\documentclass[desactivate]{aa}

\usepackage[varg]{txfonts}
\usepackage{natbib}
\usepackage{latexsym,exscale,amssymb,amsmath}
\usepackage{eurosym}
\usepackage{color}
\usepackage{graphicx}
\usepackage[percent]{overpic}
\usepackage{hyperref}

\title{Multiline observations of hydrogen, helium, and carbon radio-recombination lines toward Orion A: A detailed dynamical study and direct determination of physical conditions\thanks{Based on observations carried out with the Yebes 40m telescope (project 22A012). The 40m radio telescope at Yebes Observatory is operated by the Spanish Geographic Institute (IGN; Ministerio de Transportes, Movilidad y Agenda Urbana).}}
\titlerunning{Multiline observations of radio-recombination lines toward Orion A}
\author{C. H. M. Pabst\inst{\ref{inst1},\ref{inst2}} \and J. R. Goicoechea\inst{\ref{inst1}} \and S. Cuadrado\inst{\ref{inst1}} \and P. Salas\inst{\ref{inst3}} \and A. G. G. M. Tielens\inst{\ref{inst2},\ref{inst4}} \and N. Marcelino\inst{\ref{inst5},\ref{inst6}} }

\institute{Instituto de F\'isica Fundamental, CSIC, Calle Serrano 121-123, 28006 Madrid, Spain \label{inst1} 
\and Leiden Observatory, Leiden University, Niels Bohrweg 2, 2333 CA Leiden, Netherlands\label{inst2}
\and Green Bank Observatory, 155 Observatory Road, Green Bank, WV 24915, USA\label{inst3}
\and Department of Astronomy, University of Maryland, College Park, MD 20742, USA\label{inst4}
\and Observatorio Astron\'omico Nacional (IGN), Calle de Alfonso XII 3, 28014 Madrid, Spain\label{inst5}
\and Observatorio de Yebes (IGN), Cerro de la Palera s/n, 19141 Yebes, Guadalajara, Spain\label{inst6}
}
\date{Received 26 July 2023, Accepted 15 April 2024}

\abstract{
We present a study of hydrogen, helium, and carbon millimeter-wave radio-recombination lines (RRLs) toward ten representative positions throughout the Orion Nebula complex, using the Yebes 40m telescope in the Q band (31.3 GHz to 50.6 GHz) at an angular resolution of about $45\arcsec$ ($\sim$0.09\,pc). The observed positions include the Orion Nebula (M42) with the Orion Molecular Core 1, M43, and the Orion Molecular Core 3 bordering on NGC 1973, 1975, and 1977. While hydrogen and helium RRLs arise in the ionized gas surrounding the massive stars in the Orion Nebula complex, carbon RRLs stem from the neutral gas of the adjacent photo-dissociation regions (PDRs). The high velocity resolution ($0.3\,\mathrm{km\,s^{-1}}$) enables us to discern the detailed dynamics of the RRL emitting neutral and ionized gas. We compare the carbon RRLs with SOFIA/upGREAT observations of the [C\,{\sc ii}] $158\,\mu\mathrm{m}$ line and IRAM 30m observations of the $^{13}$CO (J=2-1) line (the complete map is presented here for the first time). We observe small differences in peak velocities between the different tracers, which cannot always be attributed to geometry but potentially to shear motions. Using the far-infrared [C\,{\sc ii}] and [$^{13}$C\,{\sc ii}] intensities with the carbon RRL intensities, we can infer physical conditions (electron temperature $T_{\rm e}$ and electron density $n_{\rm e}$, converted to hydrogen nuclei density $n_{\rm H}$ by dividing by the carbon gas-phase abundance $\mathcal{A}_{\rm C}\simeq 1.4\times 10^{-4}$) in the PDR gas using nonlocal thermal equilibrium excitation models. For positions in OMC1, we infer $n_{\rm e} \simeq 20\text{--}40\,\mathrm{cm^{-3}}$ and $T_{\rm e}\simeq 210\text{--}240\,\mathrm{K}$. On the border between OMC1 and M43, we observe two gas components with $n_{\rm e} \simeq 2\,\mathrm{cm^{-3}}$ and $n_{\rm e} \simeq 8\,\mathrm{cm^{-3}}$, and $T_{\rm e}\simeq 100\,\mathrm{K}$ and $T_{\rm e}\simeq 150\,\mathrm{K}$. In M43, we infer $n_{\rm e} \simeq 2\text{--}3\,\mathrm{cm^{-3}}$ and $T_{\rm e}\simeq 140\,\mathrm{K}$. The Extended Orion Nebula southeast of OMC1 is characterized by $n_{\rm e} \simeq 2\,\mathrm{cm^{-3}}$ and $T_{\rm e}\simeq 180\,\mathrm{K}$, while OMC3 has $n_{\rm e} \simeq 1\,\mathrm{cm^{-3}}$ and $T_{\rm e}\simeq 130\,\mathrm{K}$. Our observations are sensitive enough to detect faint lines toward two positions in OMC1, in the BN/KL PDR and the PDR close to the Trapezium stars, that may be attributed to RRLs of C$^+$ or O$^+$. In general, the RRL line widths of both the ionized and neutral gas, as well as the [C\,{\sc ii}] and $^{13}$CO line widths, are broader than thermal, indicating significant turbulence in the interstellar medium, which transitions from super-Alfv\'enic and subsonic in the ionized gas to sub-Alfv\'enic and supersonic in the molecular gas. At the scales probed by our observations, the turbulent pressure dominates the pressure balance in the neutral and molecular gas, while in the ionized gas the turbulent pressure is much smaller than the thermal pressure.
}


\begin{document}

\maketitle

\section{Introduction}

The processing of the interstellar medium (ISM) is an important driver of the evolution of galactic ecosystems. Massive stars profoundly shape the ISM, on small and large scales, by their radiation, stellar winds and at the end of their lives as supernovae. The concerted action of many massive stars leads to the formation of various phases of the ISM \citep[e.g.,][]{Field1969, McKee1977, Wolfire2003, Haffner2009}: cold diffuse clouds -- the cold neutral medium (CNM) --, a warmer intercloud phase -- the warm neutral medium (WNM) and warm ionized medium (WIM) --, compact H{\sc ii} regions, and photodissociation regions (PDRs). While the CNM, WNM, and WIM fill large volumes, compact H{\sc ii} regions and PDRs are found either deeply embedded in or on the surfaces of dense molecular clouds in the vicinity of massive stars. Only in the latter case can they be readily observed at optical wavelengths. The study of H{\sc ii} regions and PDRs reveals insights into the fate of molecular clouds, the birth sites of stars, which may be compressed or destroyed. This allows estimates of the efficiency with which new stars are formed \citep[for a review see][]{Wolfire2022}.

The extreme ultraviolet (EUV, $E>13.6\,\mathrm{eV}$) radiation of massive stars creates an H\,{\sc ii} region around the star, where hydrogen (ionization potential IP = 13.6 eV) is ionized. The most massive stars can ionize helium (IP = 24.6 eV) and doubly ionize carbon (IP = 24.4 eV). The H\,{\sc ii} region borders on a molecular cloud and the far-ultraviolet radiation (FUV, $6\,\mathrm{eV}<E<13.6\,\mathrm{eV}$) of the star creates a PDR on the surface of the molecular cloud, where hydrogen is neutral, but carbon is still ionized (IP = 11.3 eV). The radiation being attenuated in the cloud, hydrogen transitions from atomic to molecular, and carbon transitions from singly ionized to neutral to being locked up in CO. The electron density in the cloud controls the chemistry, for instance, the formation of many molecules, and the coupling of matter to magnetic fields. The ionization fraction, $x_{\rm e}\approx n_{\rm e}/n_{\rm H}$, decreases from the PDR surface ($x_{\rm e} \sim 10^{-4}$) to the shielded molecular layers ($x_e \sim 10^{-8}$) \citep[e.g.,][]{Caselli1998, Maret2007, Goicoechea2009}.

In the past, many studies have used the far-infrared $^2P_{3/2}$-$^2P_{1/2}$ fine-structure line of ionized carbon at 158\,$\mu$m to study the surface layers of PDRs and the dynamics of the ISM \citep[in Orion][]{Stacey1993, Herrmann1997, Goicoechea2015b, Pabst2019, Pabst2020, Pabst2021, Pabst2022}. Recent developments of broadband receivers have made it possible to observe multiple radio recombination lines (RRLs) simultaneously and at angular and spectral resolutions comparable to that of the [C\,{\sc ii}] 158\,$\mu$m line. RRLs are emitted when a free electron recombines with an ion and cascades down through Rydberg states to the ground state. While hydrogen and helium RRLs stem from the H\,{\sc ii} region, carbon RRLs are bright at the C$^+$/C/CO transition in the PDR\citep{Natta1994, Wyrowski1997a, Wyrowski1997b, Wyrowski2000}. While the [C\,{\sc ii}] line is often moderately optically thick toward PDRs, the carbon RRLs remain optically thin and can be used to infer electron densities and temperatures, which were previously largely unknown. In bright regions such as the Orion Bar additional RRLs of sulfur (IP = 10.4 eV) can be detected \citep{Goicoechea2021}.

The most prominent PDR for our eyes is the PDR created by the massive Trapezium stars in the Orion Nebula (M42). The Orion Nebula is the nearest site of massive star formation \citep[$d\simeq 414\,\mathrm{pc}$][]{Menten2007, Kounkel2018, Grossschedl2018} and therefore studied readily at any wavelength in great detail. Besides the central nebula, also called the Huygens Region \citep[e.g.,][]{ODell2020}, the Orion Nebula complex encompasses the two H\,{\sc ii} regions M43 and NGC 1973, 1975, and 1977. The Orion Nebula complex hosts four star-forming dense molecular cores, of which the Orion Molecular Core 1 (OMC1) is strongly irradiated by the Trapezium stars in front, of which the O7V star $\theta^1$ Ori C is the most massive. OMC1 contains several smaller regions of interest, among which the prototypical edge-on PDR of the Orion Bar and the Beckmann-Neugebauer/Kleinman-Low (BN/KL) object with massive star formation and strong molecular outflows \citep[e.g.,][]{Gomez2005, Bally2011, Morris2016, Bally2017}. Enshrouding the Trapezium stars with associated H\,{\sc ii} region in front of OMC1, several layers of ionized (emitting in tracers of ionized gas such as [O\,{\sc iii}], [O\,{\sc ii}], [O\,{\sc i}], [S\,{\sc iii}], and [N\,{\sc ii}] optical forbidden lines) and neutral gas (emitting in [C\,{\sc ii}] and the optical [O\,{\sc i}] forbidden lines, and absorbing in the H\,{\sc i} hyperfine-structure line) have been detected throughout the years \citep[e.g.,][]{VanderWerf1989, WenODell1993, ODell1993, Abel2004, Abel2006, Abel2019}. On a scale of $4\,\mathrm{pc}$, the Trapezium stars and the O9.5IV star $\theta^2$ Ori A have created and are illuminating an expanding bubble filled with hot gas, the Extended Orion Nebula (EON) with the foreground neutral Veil Shell \citep{Guedel2008,Pabst2019}.

The H\,{\sc ii} region of M43, excited by the B0.5V star NU Ori, is surrounded by an expanding half-shell of neutral gas \citep{Pabst2020}, which abuts the OMC2. The H\,{\sc ii} region of NGC 1973, 1975, and 1977 also drives the expansion of a neutral shell \citep{Pabst2020}. NGC 1977 borders on the molecular core OMC3, where the stellar radiation of the B1V star 42 Ori creates an edge-on PDR. However, the physical conditions in the various parts of the cloud complex are not well-constrained. Also the dynamic relation between the ionized, neutral and molecular phases of the H\,{\sc ii} region with adjacent PDR is not understood in many cases, the different gas phases often being observed at slightly different velocities. In both cases, RRLs are the link from the ionized to the neutral gas (hydrogen and helium RRLs), and from the neutral to the molecular phase (carbon RRLs).

We here present a broadband study of mm-wave RRLs toward the Orion Nebula complex, including M43 and NGC 1973, 1975, and 1977, in a series of observations of 10 representative positions using the 40m Yebes antenna. Earlier studies of carbon RRLs focussed on the brightest part of the Orion Nebula \citep{Balick1974, Wyrowski1997b, Tsivilev2014, Salas2019, Cuadrado2019}, while we include fainter regions that are representative for the extended outskirts of the Orion Nebula complex. Our fainter observed positions are situated in dynamic regions of interest, previously studied in other tracers, such as the EON southeast of the Orion Bar, the border between OMC1 and M43, M43 itself, and OMC3, that are bright enough to be detectable in RRLs with state-of-the-art telescopes (\cite{Salas2021} include also OMC2 and OMC3). In particular, these regions have been studied in detail in [C\,{\sc ii}] line emission in \cite{Pabst2020, Pabst2021, Pabst2022} and our selection is based on the results of these studies. We discuss the dynamics of the PDR gas compared to the H\,{\sc ii} gas and the turbulent motions. Furthermore, we aim to determine the physical conditions in the respective cloud environment from a comparison of the carbon RRLs with the [C\,{\sc ii}] lines using nonlocal thermal equilibrium (non-LTE) models.

This paper is organized as follows. In Section 2 we summarize the observations used in this paper. Section 3 compares the RRL, [C\,{\sc ii}] and CO line properties. Section 4 contains a discussion of the implications of the results obtained in Section 3 and we derive physical conditions (electron temperatures and densities) from the observed lines. We conclude with a summary of our results in Section 5.

\section{Observations}

\subsection{Yebes RRL observations}

We used the 40m radiotelescope at Yebes Observatory (Guadalajara, Spain) to observe ten representative positions throughout the Orion Nebula complex (summarized in Table \ref{Tab.positions}) in the Q band (31.3 GHz to 50.6 GHz; project 22A012, PIs: C.H.M. Pabst, J.R. Goicoechea). We employ the new Nanocosmos high-electron-mobility transistor (HEMT) Q band receiver and fast Fourier transform spectrometers, covering 18 GHz of instantaneous bandwidth per polarization at a spectral resolution of 38 kHz (i.e., a typical velocity resolution of $0.3\,\mathrm{km\,s^{-1}}$) \citep{Tercero2021}. The half-power beam width (HPBW) in the Q band ranges from $36\arcsec$ to $54\arcsec$. We observed the 10 positions in position switching mode with a reference position at $(\Delta\alpha, \Delta\delta) = (-800\arcsec, 1200\arcsec)$ relative to the central position of the nebula, $\theta^1$ Ori C, $(\mbox{RA}, \mbox{Dec}) = (5\mathrm{h}35^{\rm m}16.47^{\rm s}, -5^{\circ}23^{\rm m}22.90^{\rm s})$.

We reduced the data using the GILDAS software\footnote{\texttt{http://www.iram.fr/IRAMFR/GILDAS/}}. The antenna temperature $T_A^*$ was converted to the main-beam temperature by $T_{\rm mb} = T_A^* F_{\rm eff}/\eta_{\rm mb}$ with the main-beam efficiency $\eta_{\rm mb}$ between 0.49 and 0.67 (higher at lower frequencies\footnote{\texttt{https://rt40m.oan.es/rt40m\_en.php}}) and the forward efficiency $F_{\rm eff} = 0.967$. The root-mean-square (rms) noise in the main-beam temperature achieved outside the line window is usually between 4 and 12 mK per velocity channel ($0.37\,\mathrm{km\,s^{-1}}$) in the C59$\alpha$ line and between 11 and 23 mK per velocity channel ($0.23\,\mathrm{km\,s^{-1}}$) in the C51$\alpha$ line, after total observing times of 8 to 18 hours (depending on position; on-source times in position-switching mode and resulting rms for the C59$\alpha$ and C51$\alpha$ RRLs are given in Table \ref{Tab.positions}), varying with telescope conditions.

In order to detect faint features in the spectra, we stacked RRLs of same quantum leap $\Delta n$ and similar principal quantum number $n$ weighted by the rms noise, as is traditionally done for RRL spectra \citep{Balser2006, Emig2019}. Individual spectra, i.e. before stacking, of positions toward the molecular cores in OMC1, BNKL-PDR and TRAP-PDR (cf. Section \ref{Sec.results} for a summary of the observed positions), have additional features from molecular emission, which we discard in the present study. We stacked the $n=51\text{--}59$ $\alpha$ transitions, the $n=64\text{--}74$ $\beta$ transitions, the $n=76\text{--}84$ $\gamma$ transitions, excluding $n=74$ and $n=75$ due to bad baselines, and $n=80$ (blended with 56$\alpha$ line) and $n=83$ (blended with 91$\delta$ line), the $n=81\text{--}92$ $\delta$ transitions, excluding $n=82$ (blended with 88$\epsilon$ line) and $n=91$ due to bad baselines, and the $n=87\text{--}99$ $\epsilon$ transitions, excluding $n=88$ (blended with 82$\delta$ line). We subtracted standing wave features before stacking in some spectral chunks. The resulting velocity resolution after stacking is $0.36\,\mathrm{km\,s^{-1}}$ with a rms noise per channel of 1 to 2 mK.

\subsection{SOFIA [C\,{\sc ii}] observations}

We complement the RRL observations with existing observations of the [C\,{\sc ii}] 158\,$\mu$m line toward the Orion Nebula complex \citep{Pabst2019}. These velocity-resolved observations (covering the velocity range from $-90$ to $95\,\mathrm{km\,s^{-1}}$ at a velocity resolution of 0.3\,$\mathrm{km\,s^{-1}}$) were obtained as a square-degree sized, fully sampled map by the upgraded German Receiver At Terahertz Frequencies \citep[upGREAT,][]{Risacher2016} onboard the Stratospheric Observatory for Infrared Astronomy \citep[SOFIA,][]{Young2012} with a native angular resolution of $14.1\arcsec$. Details on the observing strategy and data reduction are given in \cite{Higgins2021}. The resulting velocity range also contains the three hyperfine components of the [$^{13}$C\,{\sc ii}] line. We convolve the map to a beam of $45\arcsec$ to compare with the Yebes observations.

\subsection{IRAM CO(2-1) observations}

We also make use of $^{12}$CO \mbox{$J$\,=\,2-1} (230.5\,GHz) and $^{13}$CO \mbox{$J$\,=\,2-1} (220.4\,GHz) line maps taken with the IRAM\,30\,m radiotelescope (Pico Veleta, Spain) at a native angular resolution of $10.7\arcsec$. The central region ($1^{\circ}\times0.8^{\circ}$) around OMC1 was originally mapped in 2008 with the HERA receiver array \citep{Berne2014}. These fully sampled CO maps were enlarged using the EMIR receiver and FFTS backends, being part of the Large Program ``Dynamic and Radiative Feedback of Massive Stars''. \cite{Goicoechea2020} provide details on the observing strategy and on how the old HERA and new EMIR CO maps were merged. Here we present the final maps, that include NGC 1973, 1975, and 1977, the last observations having been taken in October 2020. The resulting maps have a velocity resolution of $0.4\,\mathrm{km\,s^{-1}}$. We convolve the maps to a beam of $45\arcsec$ to compare with the Yebes observations.

\begin{table*}[htp]
\renewcommand{\arraystretch}{1.3}
\addtolength{\tabcolsep}{-4pt}
\caption{Summary of observed positions.}
\centering
\begin{tabular}{lcclccc}
\hline\hline
source & RA (J2000) & Dec (J2000) & comment: & on-source & rms(C59$\alpha$) & rms(C51$\alpha$) \\
 & & & region; ionizing star & time [h] & [mK] & [mK] \\ \hline
BAR-INSIDE & $5\mathrm{h}35^{\rm m}21.67^{\rm s}$ & $-5^{\circ}25^{\prime}37.00^{\prime\prime}$ & OMC1; $\theta^1$ Ori C & 11.0 & 5.4 & 13 \\
BNKL-PDR & $5\mathrm{h}35^{\rm m}16.30^{\rm s}$ & $-5^{\circ}22^{\prime}14.00^{\prime\prime}$ & OMC1; $\theta^1$ Ori C & 5.7 & 9.9 & 15 \\
TRAP-PDR & $5\mathrm{h}35^{\rm m}15.00^{\rm s}$ &$ -5^{\circ}23^{\prime}34.00^{\prime\prime}$ & OMC1; $\theta^1$ Ori C & 9.7 & 12 & 12 \\
EAST-PDR & $5\mathrm{h}35^{\rm m}25.70^{\rm s}$ & $-5^{\circ}22^{\prime}44.00^{\prime\prime}$ & OMC1; $\theta^1$ Ori C & 7.5 & 5.6 & 12 \\
ORI-P1 & $5\mathrm{h}35^{\rm m}31.37^{\rm s}$ & $-5^{\circ}21^{\prime}22.59^{\prime\prime}$ & OMC1 (dark lane); $\theta^1$ Ori C & 8.5 & 5.7 & 13 \\
ORI-P2 & $5\mathrm{h}35^{\rm m}31.37^{\rm s}$ & $-5^{\circ}18^{\prime}42.59^{\prime\prime}$ & M43 (south); NU Ori & 11.1 & 4.4 & 12 \\
ORI-P3 & $5\mathrm{h}35^{\rm m}31.37^{\rm s}$ & $-5^{\circ}16^{\prime}02.59^{\prime\prime}$ & M43 (center); NU Ori & 6.5 & 7.1 & 18 \\
ORI-P4 & $5\mathrm{h}35^{\rm m}31.37^{\rm s}$ & $-5^{\circ}13^{\prime}22.59^{\prime\prime}$ & M43 (north); NU Ori & 4.8 & 10 & 23 \\
ORI-P5 & $5\mathrm{h}35^{\rm m}29.80^{\rm s}$ & $-5^{\circ}27^{\prime}32.90^{\prime\prime}$ & EON (southeast of Orion Bar); $\theta^2$ Ori A & 5.1 & 5.7 & 15 \\
ORI-P6 & $5\mathrm{h}35^{\rm m}13.16^{\rm s}$ & $-4^{\circ}55^{\prime}08.09^{\prime\prime}$ & OMC3; 42 Orionis & 11.4 & 4.3 & 11 \\ \hline
\end{tabular}
\label{Tab.positions}
\end{table*}

\begin{figure*}[htp]
\centering
\begin{minipage}{\textwidth}
\includegraphics[width=\textwidth, height=0.514\textwidth]{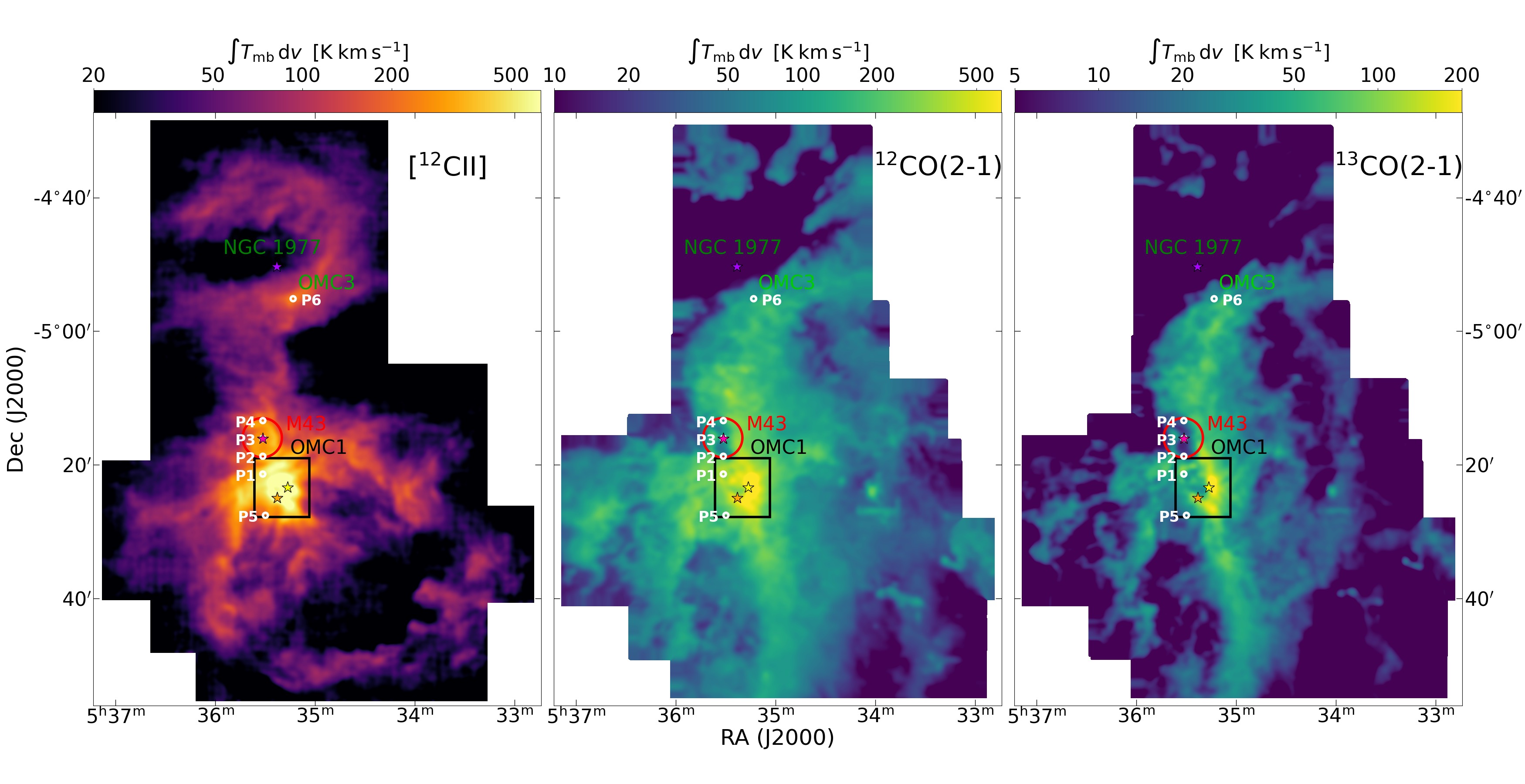}
\caption{Observed positions (except positions in OMC1) as white circles with a FWHM of $45\arcsec$ on [$^{12}$C\,{\sc ii}], $^{12}$CO(2-1), and $^{13}$CO(2-1) maps at $45\arcsec$. The black rectangle outlines the region of OMC1, shown as a zoom-in in Fig. \ref{Fig.positions_map_center_cii_co}. The red circle outlines M43. The yellow and orange stars mark the positions of $\theta^1$ Ori C and $\theta^2$ Ori A, respectively, the pink star marks NU Ori, the purple star marks 42 Orionis.}
\vspace{2.5ex}
\label{Fig.positions_map_cii_co}
\end{minipage}
\begin{minipage}{\textwidth}
\includegraphics[width=\textwidth, height=0.340\textwidth]{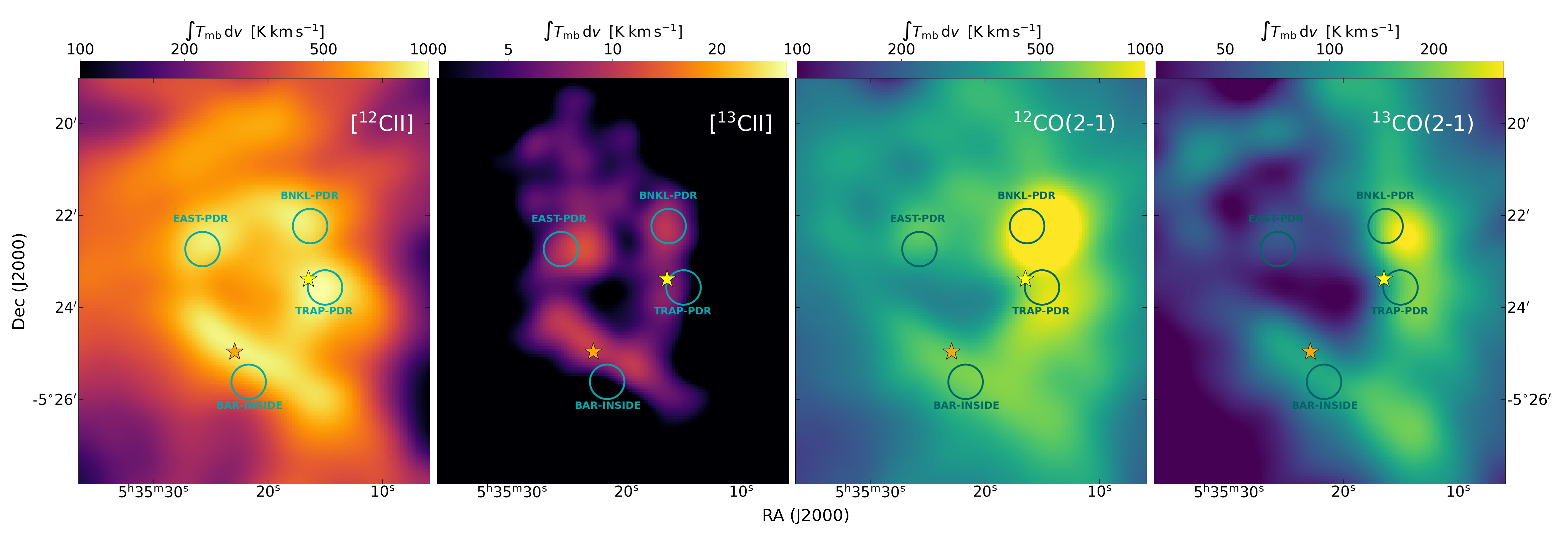}
\caption{Zoom into OMC1. Observed positions in OMC1 as turquois circles with a FWHM of $45\arcsec$ on [$^{12}$C\,{\sc ii}], [$^{13}$C\,{\sc ii}], $^{12}$CO(2-1), and $^{13}$CO(2-1) maps at $45\arcsec$. The yellow and orange stars mark the positions of $\theta^1$ Ori C and $\theta^2$ Ori A, respectively.}
\label{Fig.positions_map_center_cii_co}
\end{minipage}
\end{figure*}

\begin{figure*}[htp]
\centering
\includegraphics[width=0.9\textwidth, height=0.6\textwidth]{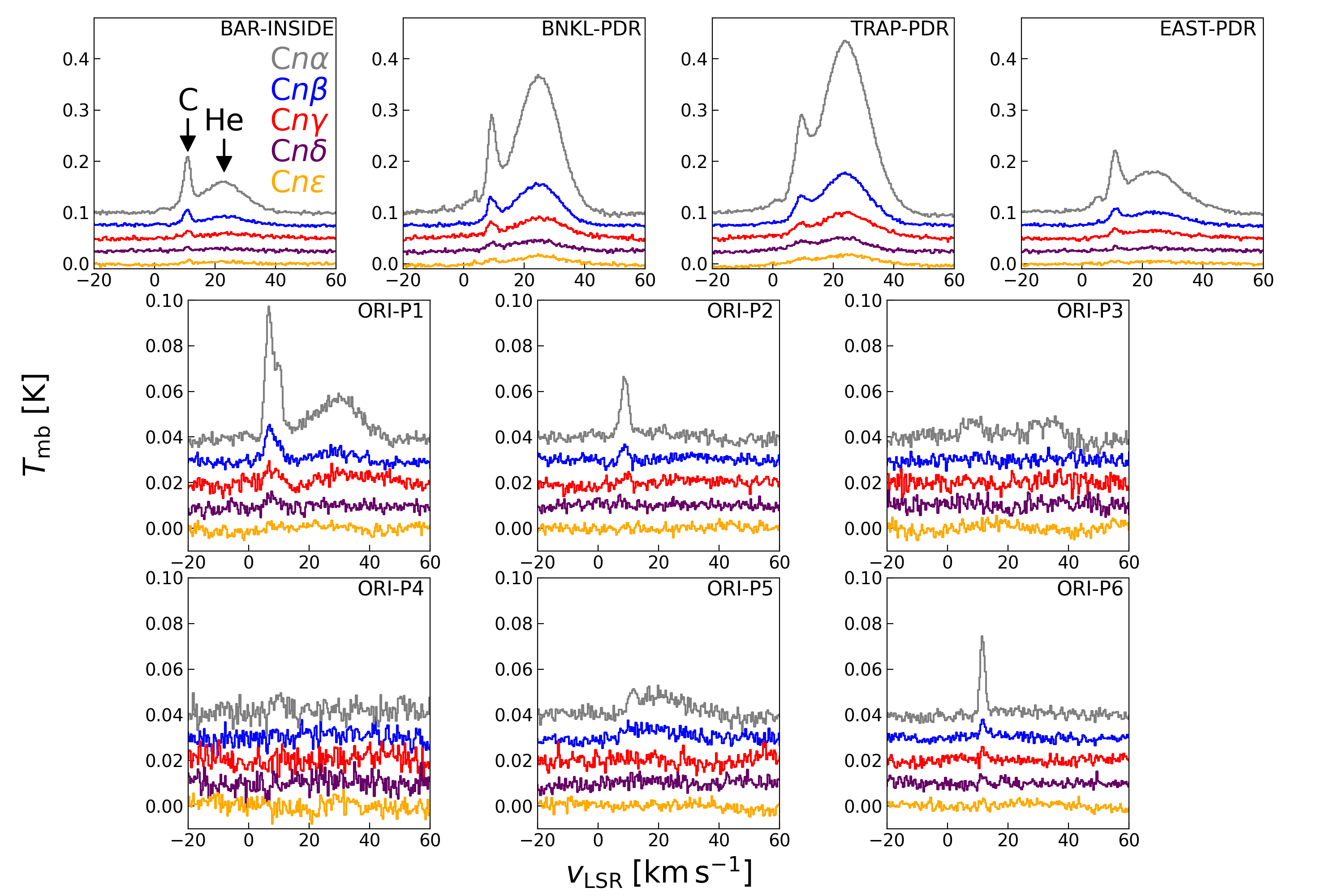}
\caption{Stacked C$n\alpha$, C$n\beta$, C$n\gamma$, C$n\delta$, and C$n\epsilon$ radio-recombination lines with He$n\alpha$, He$n\beta$, He$n\gamma$, He$n\delta$, and He$n\epsilon$ radio-recombination lines toward the ten targeted positions. In the position BAR-INSIDE the component $7.5\,\mathrm{km\,s^{-1}}$ blue of the main component may be a radio-recombination line from sulfur, while detection of the sulfur RRL in other positions is hampered by the carbon RRL component corresponding to the Veil (cf. Sec. \ref{sec.discussion-velocities}). Spectra in in the upper top are offset from the C/He$n\epsilon$ spectrum in steps of 0.025\,K, sand pectra in the lower two rows are offset in steps of 0.01\,K.}
\label{Fig.CRRLs_spectra}
\end{figure*}

\begin{figure*}[htp]
\centering
\includegraphics[width=0.9\textwidth, height=0.6\textwidth]{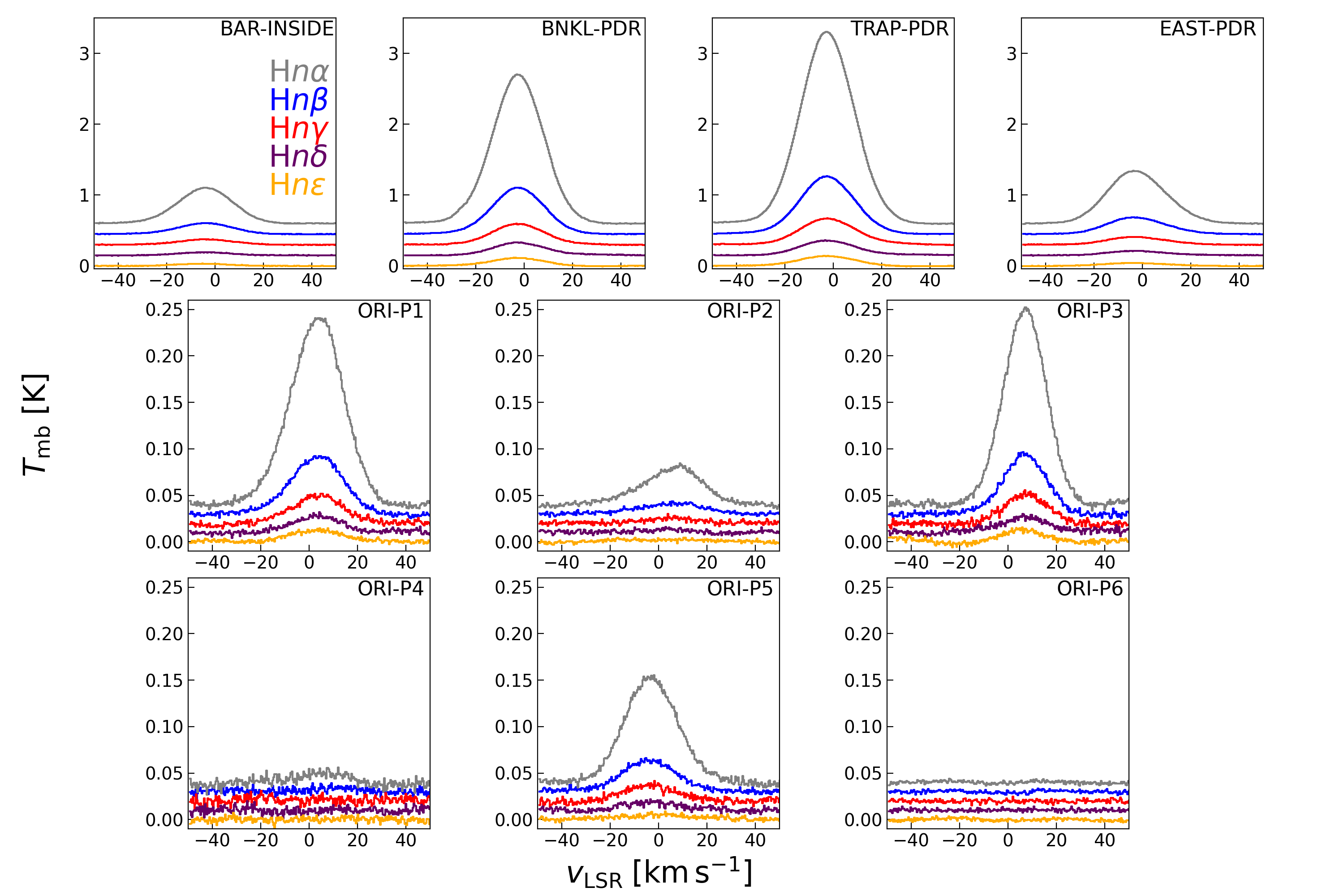}
\caption{Stacked H$n\alpha$, H$n\beta$, H$n\gamma$, H$n\delta$, and H$n\epsilon$ radio-recombination lines toward the ten targeted positions. Spectra in in the top row are offset from the H$n\epsilon$ spectrum in steps of 0.15\,K, and spectra in the lower two rows are offset in steps of 0.01\,K.}
\label{Fig.HRRLs_spectra}
\end{figure*}

\section{Results}
\label{Sec.results}

Table \ref{Tab.positions} gives the definitions of the positions observed in RRLs with the Yebes telescope with coordinates and a classification of the respective exciting star. Figures \ref{Fig.positions_map_cii_co} and \ref{Fig.positions_map_center_cii_co} show the locations of those positions on the SOFIA [C\,{\sc ii}] and IRAM $^{12}$CO(2-1) and $^{13}$CO(2-1) maps of the Orion Nebula complex at an angular resolution of $45\arcsec$. We observed four positions toward the bright OMC1, BAR-INSIDE, BNKL-PDR, TRAP-PDR, and EAST-PDR. The position BAR-INSIDE covers the CO-bright part of the Orion Bar and the background molecular cloud southeast of the Orion Bar within our beam (average of $45\arcsec$). BNKL-PDR is aimed at the BN/KL object and its surface PDR. TRAP-PDR is pointed at the Trapezium cluster and covers the bright PDR/clump west of the Trapezium stars, that is the surface of the Orion South cloud. EAST-PDR covers the bright PDR east of OMC1, toward the (optically) Dark Lane. In addition, we observed six positions toward fainter regions outside OMC1. Pointing ORI-P1 lies on the northeastern border of OMC1 at the far side of the Dark Lane, ORI-P2, ORI-P3, and ORI-P4 are situated in M43, ORI-P5 lies southeast of the Orion Bar in the cavity of the EON, and ORI-P6 lies toward OMC3. Appendix \ref{App.positions} contains zoom-ins toward OMC1, M43, and OMC3, that show the structure averaged within each beam on the SOFIA [C\,{\sc ii}] intensity map and the {\it Spitzer} 8\,$\mu$m intensity map in their original resolution.

Figures \ref{Fig.CRRLs_spectra} and \ref{Fig.HRRLs_spectra} show the stacked spectra of the observed carbon and helium RRLs and hydrogen RRLs, respectively, of the $\alpha$, $\beta$, $\gamma$, $\delta$, and $\epsilon$ transitions (with mean $\bar{n}=55$ for the $\alpha$ transitions, $\bar{n}=69$ for the $\beta$ transitions, $\bar{n}\approx 80$ for the $\gamma$ transitions, $\bar{n}\approx 87$ for the $\delta$ transitions, and $\bar{n}\approx 93$ for the $\epsilon$ transitions). The velocity separation due to different atomic weight of the Rydberg atoms between helium RRLs and carbon RRLs is $27.4\,\mathrm{km\,s^{-1}}$, and between hydrogen RRLs and carbon RRLs the velocity separation is $149.4\,\mathrm{km\,s^{-1}}$. Hence, the helium RRLs appear as a broad component with the carbon RRLs on its shoulder, while the carbon and helium RRLs are well-separated from the hydrogen RRLs.

In the brightest positions (all four positions in OMC1) we detect the $\alpha$, $\beta$, $\gamma$, $\delta$, and $\epsilon$ transitions of the hydrogen, helium and carbon RRLs. In ORI-P1 and ORI-P5 we detect the hydrogen, helium and carbon RRLs in higher transitions than the $\alpha$ transition. ORI-P2 has significant emission of hydrogen and carbon RRLs, but no helium RRL emission. ORI-P3 has weak carbon RRLs, but strong hydrogen RRL emission and we detect the $\alpha$ helium RRL. ORI-P4 has weak hydrogen and carbon RRL emission, but no significant helium RRL. In ORI-P6 we detect only the strong carbon RRL and no hydrogen or helium RRL emission. Tables \ref{tab.SN_CRRLs}, \ref{tab.SN_HRRLs} and \ref{tab.SN_HeRRLs} give the observed integrated signal-to-noise ratios from Gaussian line fits for the stacked carbon, hydrogen, and helium RRLs, respectively. Extrapolating from the tabulated signal-to-noise ratios, we would be able to detect CRRLs after stacking up to $\Delta n =7$, HRRLs up to $\Delta n = 37$ and HeRRLs up to $\Delta n = 10$ in the brightest position TRAP-PDR. Tables \ref{tab.CRRL_fits}, \ref{tab.HRRL_fits}, and \ref{tab.HeRRL_fits} give the results of the Gaussian fits for the carbon, hydrogen and helium $\alpha$ RRLs. Appendices \ref{App.CRRL-fits}, \ref{App.HRRL-fits}, and \ref{App.HeRRL-fits} contain tables giving the results of Gaussian fits to the detected $\beta$, $\gamma$, $\delta$, and $\epsilon$ RRLs of carbon, hydrogen, and helium.

\begin{table}[htb]
\caption{Signal-to-noise of stacked CRRLs.}
\begin{tabular}{lccccc}
\hline\hline
source & C$n\alpha$ & C$n\beta$ & C$n\gamma$ & C$n\delta$ & C$n\epsilon$ \\\hline
BAR-INSIDE & 44 & 15 & 5.3 & 2.8 & 3.8 \\
BNKL-PDR & 60 & 20 & 8.4 & 4.5 & 4.3 \\
TRAP-PDR & 55 & 22 & 6.9 & 8.8 & 5.8 \\
EAST-PDR-1 & 8.4 & 2.1 & -- & -- & -- \\
EAST-PDR-2 & 48 & 15 & 6.0 & 2.9 & 3.2 \\
ORI-P1-1 & 29 & 9.7 & 2.7 & 1.4 & -- \\
ORI-P1-2 & 16 & 3.8 & 2.6 & 1.1 & -- \\
ORI-P2 & 17 & 4.2 & 2.0 & -- & -- \\
ORI-P3 & 4.5 & -- & -- & -- & -- \\
ORI-P4 & 3.1 & -- & -- & -- & -- \\
ORI-P5 & 3.5 & 2.0 & -- & -- & -- \\
ORI-P6 & 20 & 5.3 & 2.6 & 2.1 & 1.9 \\ \hline
\end{tabular}
\tablefoot{Integrated signal-to-noise from Gaussian line fit. Indices of EAST-PDR and ORI-P1 refer to the two velocity components of the line. In some cases a Gaussian fit was possible at low S/N because the noise was estimated higher far outside the line window.}
\label{tab.SN_CRRLs}
\end{table}

\begin{table}[htb]
\addtolength{\tabcolsep}{-2pt}
\caption{Signal-to-noise of stacked HRRLs.}
\begin{tabular}{lccccc}
\hline\hline
source & H$n\alpha$ & H$n\beta$ & H$n\gamma$ & H$n\delta$ & H$n\epsilon$ \\ \hline
BAR-INSIDE & 496 & 166 & 73 & 50 & 37 \\
BNKL-PDR & 1322 & 489 & 184 & 130 & 89 \\
TRAP-PDR & 1745 & 665 & 240 & 166 & 104 \\
EAST-PDR & 810 & 270 & 107 & 67 & 55 \\
ORI-P1 & 208 & 74 & 28 & 20 & 17 \\
ORI-P2 & 52 & 18 & 6.4 & 4.3 & 3.3 \\
ORI-P3 & 149 & 55 & 20 & 14 & 11 \\
ORI-P4 & 6.8 & -- & -- & -- & -- \\
ORI-P5 & 105 & 34 & 13 & 8.5 & 7.4 \\
ORI-P6 & -- & -- & -- & -- & --  \\ \hline
\end{tabular}
\tablefoot{Integrated signal-to-noise from Gaussian line fit.}
\label{tab.SN_HRRLs}
\end{table}

\begin{table}[htb]
\caption{Signal-to-noise of stacked HeRRLs.}
\begin{tabular}{lccccc}
\hline\hline
source & He$n\alpha$ & He$n\beta$ & He$n\gamma$ & He$n\delta$ & He$n\epsilon$ \\ \hline
BAR-INSIDE & 47 & 15 & 7.5 & 4.4 & 3.6 \\
BNKL-PDR & 137 & 51 & 22 & 13 & 9.7 \\
TRAP-PDR & 178 & 71 & 28 & 16 & 10 \\
EAST-PDR & 78 & 27 & 12 & 5.9 & 4.6 \\
ORI-P1 & 14 & 2.6 & 3.1 & -- & -- \\
ORI-P2 & -- & -- & -- & -- & -- \\
ORI-P3 & 3.3 & -- & -- & -- & -- \\
ORI-P4 & -- & -- & -- & -- & --  \\
ORI-P5 & 6.5 & 2.7 & -- & -- & -- \\
ORI-P6 & -- & -- & -- & -- & --  \\ \hline
\end{tabular}
\tablefoot{Integrated signal-to-noise from Gaussian line fit.}
\label{tab.SN_HeRRLs}
\end{table}

\begin{figure*}[htbp]
\centering
\includegraphics[width=0.9\textwidth, height=0.6\textwidth]{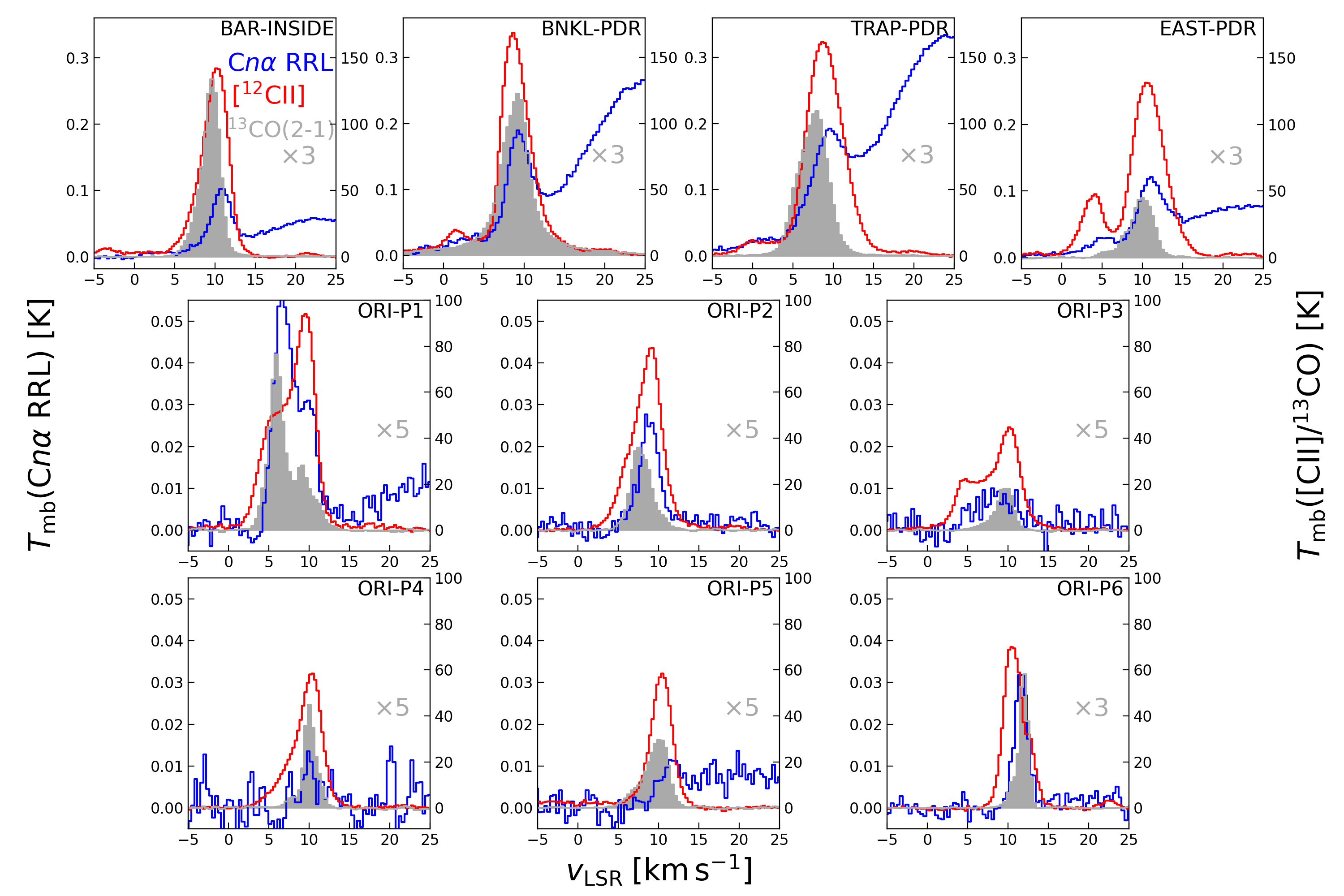}
\caption{Stacked C$n\alpha$ RRL, [$^{12}$C\,{\sc ii}], and $^{13}$CO(2-1) lines toward the ten targeted positions. The $^{13}$CO(2-1) line is scaled by the factor indicated in light gray in each panel.}
\label{Fig.Ca_12CII_13CO_spectra}
\end{figure*}

\begin{figure*}[htbp]
\centering
\includegraphics[width=0.9\textwidth, height=0.6\textwidth]{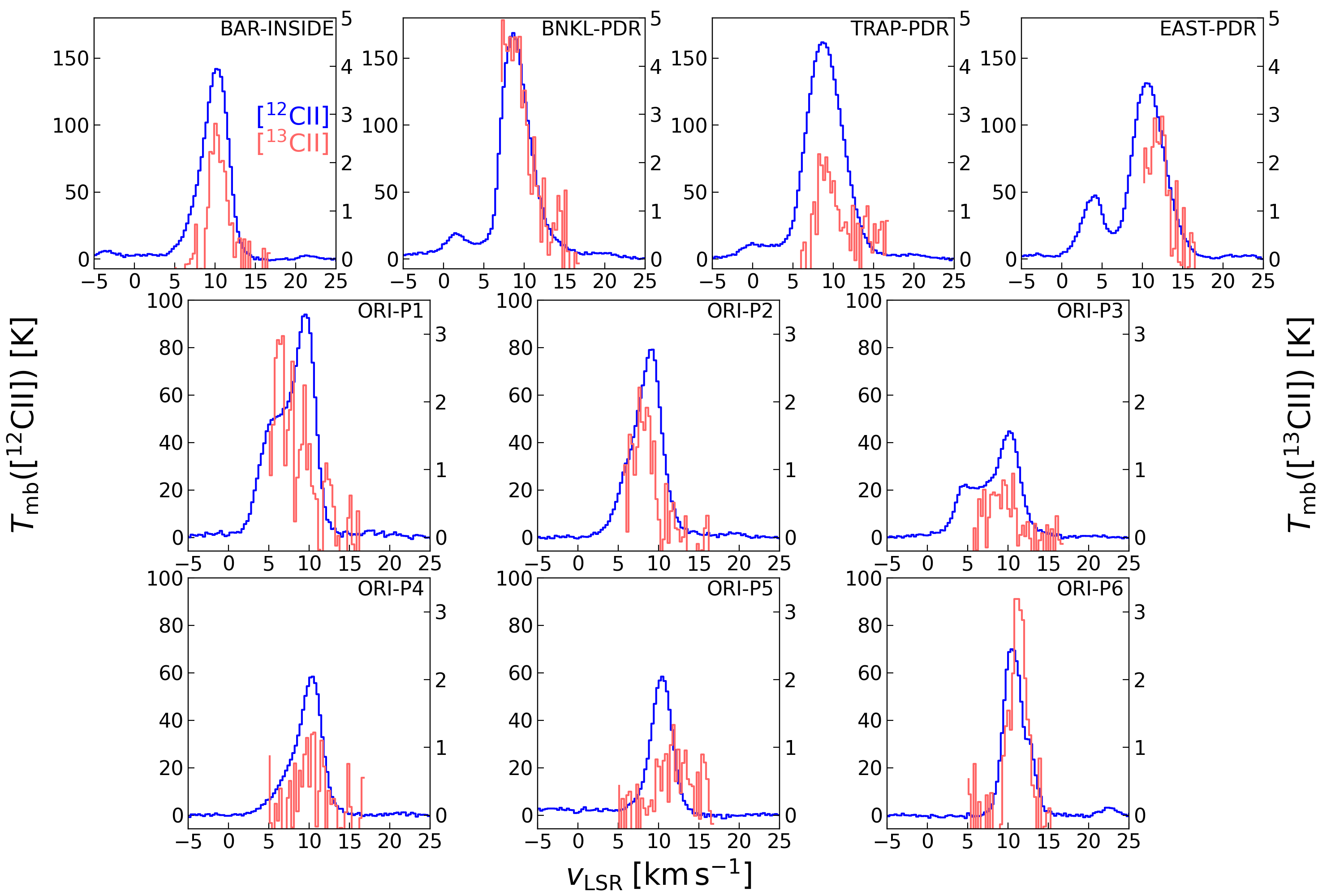}
\caption{[$^{12}$C\,{\sc ii}] and [$^{13}$C\,{\sc ii}] $F$=2-1 lines toward the ten targeted positions.}
\label{Fig.12CII_13CII_spectra}
\end{figure*}

\begin{figure*}[htbp]
\centering
\includegraphics[width=0.9\textwidth, height=0.6\textwidth]{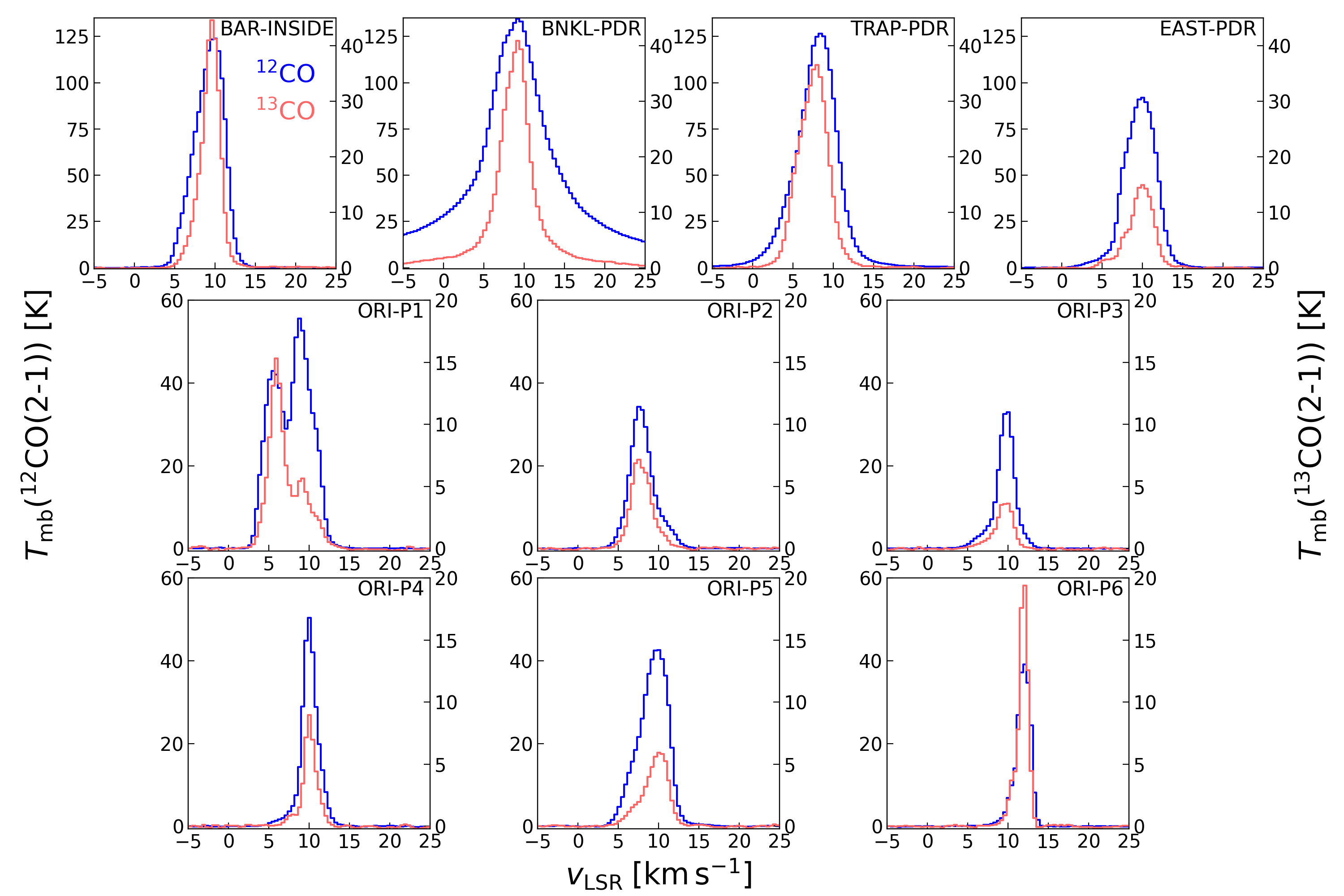}
\caption{$^{12}$CO(2-1) and $^{13}$CO(2-1) lines toward the ten targeted positions.}
\label{Fig.12CO_13CO_spectra}
\end{figure*}

\begin{figure*}[htbp]
\centering
\includegraphics[width=0.9\textwidth, height=0.6\textwidth]{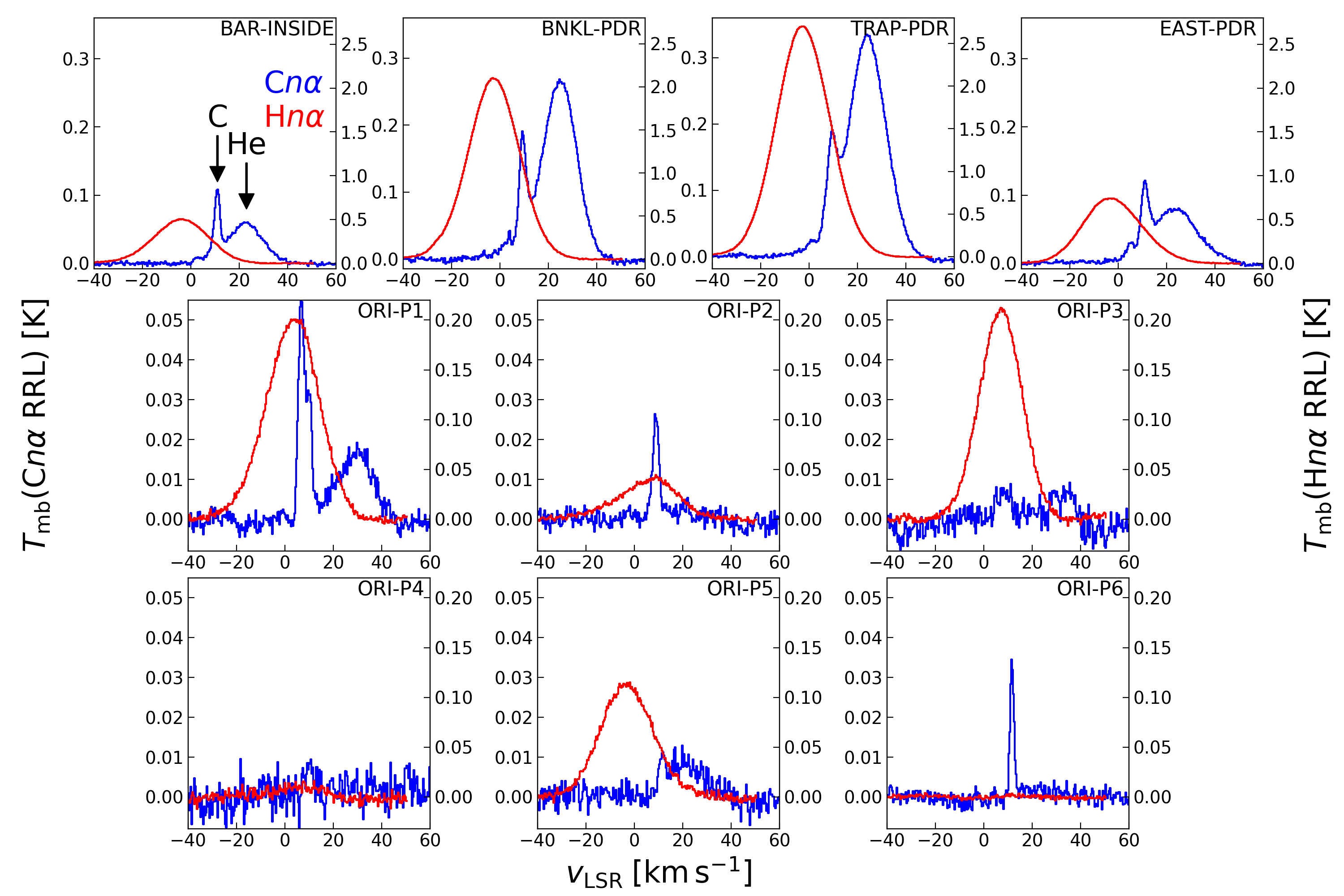}
\caption{Stacked C$n\alpha$ (blended with He$n\alpha$) with respect to the C$n\alpha$ rest frequency and stacked H$n\alpha$ RRLs with respect to the H$n\alpha$ rest frequency toward the ten targeted positions.}
\label{Fig.Ca_Ha_spectra}
\end{figure*}

Figure \ref{Fig.Ca_12CII_13CO_spectra} shows the stacked C$n\alpha$ RRL, [$^{12}$C\,{\sc ii}] and $^{13}$CO(2-1) lines toward the observed positions in the Orion A molecular cloud. We use the $^{13}$CO(2-1) line for the analysis and discussion throughout the paper as it traces the depth of the CO-emitting layer in the PDR, while the $^{12}$CO(2-1) line is optically thick and the C$^{18}$O layer traces only the more FUV-shielded and densest parts of the molecular cloud. Figure \ref{Fig.12CII_13CII_spectra} shows the [$^{12}$C\,{\sc ii}] and [$^{13}$C\,{\sc ii}] lines comparatively, while Fig. \ref{Fig.12CO_13CO_spectra} shows the $^{12}$CO(2-1) and $^{13}$CO(2-1) spectra in each observed position. In the following sections we describe the velocity structure of the C$n\alpha$ RRL, [C\,{\sc ii}] and $^{13}$CO emission lines and the intensity ratios of the carbon RRLs and the [$^{12}$C\,{\sc ii}] line. We also compare the velocity profiles of the carbon RRLs with those of the hydrogen RRLs. Fig. \ref{Fig.Ca_Ha_spectra} shows the stacked C$n\alpha$ RRLs and the stacked H$n\alpha$ RRLs together, the velocities centered on their respective rest frequency.

\begin{table}[htbp]
\caption{[C\,{\sc ii}] line fits.}
\begin{tabular}{lccc}
\hline\hline
source & $T_{\mathrm{mb}}$ & $v_{\mathrm{LSR}}$ & $\Delta v_{\mathrm{FWHM}}$ \\
 & [K] & $[\mathrm{km\,s^{-1}}]$ & $[\mathrm{km\,s^{-1}}]$ \\ \hline
BAR-INSIDE\tablefootmark{a} & $140\pm 1$ & $10.2\pm 0.1$ & $3.8\pm 0.1$ \\
BNKL-PDR & $164\pm 2$ & $9.1\pm 0.1$ & $4.1\pm 0.1$ \\
TRAP-PDR & $161\pm 1$ & $9.1\pm 0.1$ & $5.3\pm 0.1$ \\
EAST-PDR-1 & $47\pm 1$ & $3.9\pm 0.1$ & $3.2\pm 0.1$ \\
EAST-PDR-2 & $131\pm 1$ & $10.9\pm 0.1$ & $4.7\pm 0.1$ \\
ORI-P1-1 & $52\pm 1$ & $6.2\pm 0.1$ & $4.6\pm 0.1$ \\
ORI-P1-2 & $85\pm 1$ & $9.8\pm 0.1$ & $2.8\pm 0.1$ \\
ORI-P2 & $73\pm 1$ & $8.8\pm 0.1$ & $4.4\pm 0.1$ \\
ORI-P3-1 & $23\pm 1$ & $5.7\pm 0.1$ & $4.6\pm 0.1$ \\
ORI-P3-2 & $42\pm 1$ & $10.3\pm 0.1$ & $3.4\pm 0.1$ \\
ORI-P4 & $54\pm 1$ & $10.1\pm 0.1$ & $4.0\pm 0.1$ \\
ORI-P5 & $57\pm 1$ & $10.5\pm 0.1$ & $3.2\pm 0.1$ \\
ORI-P6\tablefootmark{b} & $76\pm 1$ & $11.3\pm 0.1$ & $2.6\pm 0.1$ \\ \hline
\end{tabular}
\tablefoot{\tablefoottext{a}{BAR-INSIDE can be better fitted by two components.}
\tablefoottext{b}{Result from self-absorption fit, background component.}}
\label{tab.CII_fits}
\end{table}

\begin{table}[htbp]
\caption{[$^{13}$C\,{\sc ii}] $F$=2-1 line fits.}
\begin{tabular}{lccc}
\hline\hline
source & $T_{\mathrm{mb}}$ & $v_{\mathrm{LSR}}$ & $\Delta v_{\mathrm{FWHM}}$ \\
 & [K] & $[\mathrm{km\,s^{-1}}]$ & $[\mathrm{km\,s^{-1}}]$ \\ \hline
BAR-INSIDE & $2.7\pm 0.8$ & $10.5\pm 0.2$ & $2.2\pm 0.3$ \\
BNKL-PDR\tablefootmark{a} & $3.8\pm 1.0$ & $8.0\pm 0.2$ & $1.6\pm 0.5$ \\
TRAP-PDR & $1.9\pm 0.7$ & $9.4\pm 0.2$ & $3.0\pm 0.5$ \\
EAST-PDR-1 & -- & -- & -- \\
EAST-PDR-2 & $2.8\pm 1.6$ & $11.5\pm 0.8$ & $3.6\pm 1.7$ \\
ORI-P1-1 & $2.7\pm 0.8$ & $6.6\pm 0.2$ & $2.7\pm 0.7$ \\
ORI-P1-2 & $1.6\pm 0.9$ & $9.7\pm 0.3$ & $1.7\pm 0.7$ \\
ORI-P2 & $1.6\pm 0.7$ & $7.8\pm 0.3$ & $4.3\pm 0.7$ \\
ORI-P3 & $0.6\pm 0.6$ & $8.6\pm 0.7$ & $5.5\pm 1.7$ \\
ORI-P4 & $0.9\pm 0.7$ & $10.2\pm 0.4$ & $2.9\pm 0.9$ \\
ORI-P5 & $0.8\pm 0.7$ & $12.5\pm 0.5$ & $4.9\pm 1.1$ \\
ORI-P6 & $3.2\pm 0.7$ & $11.5\pm 0.1$ & $2.2\pm 0.2$ \\ \hline
\end{tabular}
\tablefoot{\tablefoottext{a}{We use the [$^{13}$C\,{\sc ii}] $F$=1-0 and $F$=1-1 satellites. The added intensities of the $F$=1-0 and $F$=1-1 satellites are scaled to the intensity of the $F$=2-1 satellite.}}
\label{tab.13CII_fits}
\end{table}

\begin{table}[htbp]
\caption{$^{13}$CO line fits.}
\begin{tabular}{lccc}
\hline\hline
source & $T_{\mathrm{mb}}$ & $v_{\mathrm{LSR}}$ & $\Delta v_{\mathrm{FWHM}}$ \\
 & [K] & $[\mathrm{km\,s^{-1}}]$ & $[\mathrm{km\,s^{-1}}]$ \\ \hline
BAR-INSIDE\tablefootmark{a} & $43.0\pm 0.4$ & $9.7\pm 0.1$ & $2.6\pm 0.1$ \\
BNKL-PDR\tablefootmark{b} & $41.9\pm 0.3$ & $9.2\pm 0.1$ & $3.5\pm 0.1$ \\
TRAP-PDR & $35.6\pm 0.2$ & $7.7\pm 0.1$ & $4.3\pm 0.1$ \\
EAST-PDR-1 & $1.6\pm 0.1$ & $6.0\pm 0.2$ & $2.4\pm 0.3$ \\
EAST-PDR-2 & $15.2\pm 0.1$ & $10.3\pm 0.1$ & $3.0\pm 0.1$ \\
ORI-P1-1 & $11.7\pm 0.2$ & $6.0\pm 0.1$ & $1.7\pm 0.1$ \\
ORI-P1-2 & $4.9\pm 0.1$ & $8.3\pm 0.1$ & $5.7\pm 0.2$ \\
ORI-P2 & $7.1\pm 0.1$ & $8.0\pm 0.1$ & $2.9\pm 0.1$ \\
ORI-P3-1 & $0.7\pm 0.2$ & $8.2\pm 0.9$ & $3.5\pm 1.1$ \\
ORI-P3-2 & $3.5\pm 0.5$ & $10.0\pm 0.1$ & $2.0\pm 0.2$ \\
ORI-P4 & $8.3\pm 0.1$ & $10.3\pm 0.1$ & $1.8\pm 0.1$ \\
ORI-P5 & $4.9\pm 0.5$ & $10.5\pm 0.1$ & $2.4\pm 0.1$ \\
ORI-P6 & $19.5\pm 0.3$ & $12.2\pm 0.1$ & $1.3\pm 0.1$ \\ \hline
\end{tabular}
\tablefoot{\tablefoottext{a}{BAR-INSIDE can be better fitted by two components, similar to those of the [C\,{\sc ii}] line.}
\tablefoottext{b}{Lorentzian fit, wings are smaller than in $^{12}$CO, but almost Lorentzian.}}
\label{tab.13CO_fits}
\end{table}

\begin{table}[htbp]
\caption{C$n\alpha$ RRL line fits.}
\begin{tabular}{lccc}
\hline\hline
source & $T_{\mathrm{mb}}$ & $v_{\mathrm{LSR}}$ & $\Delta v_{\mathrm{FWHM}}$ \\
 & [mK] & $[\mathrm{km\,s^{-1}}]$ & $[\mathrm{km\,s^{-1}}]$ \\ \hline
BAR-INSIDE & $88\pm 2$ & $10.6\pm 0.1$ & $2.8\pm 0.1$ \\
BNKL-PDR & $156\pm 3$ & $9.2\pm 0.1$ & $3.8\pm 0.1$ \\
TRAP-PDR & $126\pm 3$ & $9.2\pm 0.1$ & $4.9\pm 0.2$ \\
EAST-PDR-1 & $16\pm 2$ & $4.7\pm 0.2$ & $2.8\pm 0.3$ \\
EAST-PDR-2 & $82\pm 2$ & $10.8\pm 0.1$ & $3.4\pm 0.1$ \\
ORI-P1-1 & $56\pm 2$ & $6.5\pm 0.1$ & $2.7\pm 0.1$ \\
ORI-P1-2 & $31\pm 2$ & $9.8\pm 0.1$ & $2.8\pm 0.2$ \\
ORI-P2 & $25\pm 1$ & $8.7\pm 0.1$ & $3.2\pm 0.2$ \\
ORI-P3 & $7\pm 1$ & $8.2\pm 0.5$ & $7.4\pm 1.1$ \\
ORI-P4 & $8\pm 2$ & $10.1\pm 0.5$ & $5.2\pm 1.1$ \\
ORI-P5 & $8\pm 2$ & $11.2\pm 0.2$ & $2.6\pm 0.6$ \\
ORI-P6 & $34\pm 1$ & $11.4\pm 0.1$ & $1.9\pm 0.1$ \\ \hline
\end{tabular}
\label{tab.CRRL_fits}
\end{table}

\begin{table}[htbp]
\caption{H$n\alpha$ RRL line fits.}
\begin{tabular}{lccc}
\hline\hline
source & $T_{\mathrm{mb}}$ & $v_{\mathrm{LSR}}$ & $\Delta v_{\mathrm{FWHM}}$ \\
 & [mK] & $[\mathrm{km\,s^{-1}}]$& $[\mathrm{km\,s^{-1}}]$ \\ \hline
BAR-INSIDE & $500\pm 1$ & $-4.2\pm 0.1$ & $26.4\pm 0.1$ \\
BNKL-PDR & $2088\pm 3$ & $-2.8\pm 0.1$ & $24.8\pm 0.1$ \\
TRAP-PDR & $2692\pm 3$ & $-2.7\pm 0.1$ & $26.0\pm 0.1$ \\
EAST-PDR & $739\pm 2$ & $-2.7\pm 0.1$ & $28.2\pm 0.1$ \\
ORI-P1 & $201\pm 1$ & $3.1\pm 0.1$ & $25.4\pm 0.1$ \\
ORI-P2 & $40\pm 1$ & $5.9\pm 0.2$ & $28.3\pm 0.3$ \\
ORI-P3 & $211\pm 1$ & $6.8\pm 0.1$ & $20.5\pm 0.1$ \\
ORI-P4 & $10\pm 1$ & $5.0\pm 0.8$ & $21.0\pm 1.8$ \\
ORI-P5 & $113\pm 1$ & $-3.3\pm 0.1$ & $25.4\pm 0.2$ \\
ORI-P6 & -- & -- & --  \\ \hline
\end{tabular}
\label{tab.HRRL_fits}
\end{table}

\begin{table}[htbp]
\caption{He$n\alpha$ RRL line fits.}
\begin{tabular}{lccc}
\hline\hline
source & $T_{\mathrm{mb}}$ & $v_{\mathrm{LSR}}$ & $\Delta v_{\mathrm{FWHM}}$ \\
 & [mK] & $[\mathrm{km\,s^{-1}}]$ & $[\mathrm{km\,s^{-1}}]$ \\ \hline
BAR-INSIDE & $57\pm 1$ & $-5.4\pm 0.1$ & $17.9\pm 0.3$ \\
BNKL-PDR & $265\pm 2$ & $-2.9\pm 0.1$ & $16.5\pm 0.2$ \\
TRAP-PDR & $330\pm 1$ & $-3.5\pm 0.1$ & $17.9\pm 0.1$ \\
EAST-PDR & $79\pm 1$ & $-4.6\pm 0.1$ & $23.0\pm 0.3$ \\
ORI-P1 & $17\pm 1$ & $1.4\pm 0.2$ & $17.0\pm 0.5$ \\
ORI-P2 & -- & -- & -- \\
ORI-P3 & $7\pm 1$ & $4.7\pm 0.6$ & $9.8\pm 1.3$ \\
ORI-P4 & -- & -- & --  \\
ORI-P5 & $9\pm 1$ & $-6.7\pm 0.6$ & $17.2\pm 1.3$ \\
ORI-P6 & -- & -- & --  \\ \hline
\end{tabular}
\vspace{0.8cm}
\label{tab.HeRRL_fits}
\end{table}


\subsection{Velocity structure}

In order to obtain line parameters (peak temperature, peak velocity, line width) we perform Gaussian fits to the observed stacked $\alpha$ RRLs, the [C\,{\sc ii}] lines and the $^{13}$CO lines. Results of the Gaussian fits are given in Tables \ref{tab.CII_fits} to \ref{tab.HeRRL_fits}. Where we fit multiple components, we append a number to the name of the position. The position BAR-INSIDE is special in this regard, as the main components of the $\alpha$ CRRL, [$^{12}$C\,{\sc ii}] and $^{13}$CO lines seems to consist of two close Gaussian components, rather than one. However, the [$^{13}$C\,{\sc ii}] line is too weak to extract two components and we opt to use a one-component fit for the velocity structure and line intensity analysis, where we need the [$^{13}$C\,{\sc ii}] line. The peak velocity of the [$^{13}$C\,{\sc ii}] line very nearly matches the peak velocity of the one-component fit to the [$^{12}$C\,{\sc ii}] line in BAR-INSIDE, and we assume that the emission is weighted toward the higher-column density region of the CO-bright Orion Bar within the beam. Furthermore, Kabanovic et al. (in prep.) observed a velocity shift between the peaks of the [$^{12}$C\,{\sc ii}] and [$^{13}$C\,{\sc ii}] line in OMC3, and we follow them in performing a fit taking into account self-absorption in the [C\,{\sc ii}] line in ORI-P6, using the recipe of \cite{Guevara2020}. We give the peak velocity of the lines relative to the respective rest frequency in the local-standard-of-rest (LSR) frame. While the statistical errors (given in the tables) are small, we estimate that the systematic errors of the measured intensities with the Yebes 40m telescope are about 10\%.

Studying the obtained line parameters, the FIR [C\,{\sc ii}] lines tend to be somewhat broader than the mm-wave C$n\alpha$ lines, apart from ORI-P3 and ORI-P4 (associated with M43). The [$^{12}$C\,{\sc ii}] line in PDRs tends to be marginally optically thick and is therefore expected to be opacity broadened. However, as observed by \cite{Ossenkopf2013}, opacity broadening in itself is not sufficient to explain the difference in line width between the [$^{12}$C\,{\sc ii}] and the [$^{13}$C\,{\sc ii}] lines. Comparing the line widths of the [$^{13}$C\,{\sc ii}] lines with the C$n\alpha$ lines, reveals different behaviors: Some [$^{13}$C\,{\sc ii}] lines are much narrower than the C$n\alpha$ lines (e.g., BNKL-PDR, TRAP-PDR, ORI-P1-2), while others are similar (e.g., BAR-INSIDE, EAST-PDR-2, and ORI-P6). In BNKL-PDR, ORI-P1, ORI-P3, and ORI-P5 line parameters of the [$^{13}$C\,{\sc ii}] line are very different when using the $F$=2-1 component compared to when adding all three hyperfine components. Due to the faintness of the two far hyperfine satellites ($F$=1-0 and $F$=1-1), we opt to fit only the $F$=2-1 satellite, apart from BNKL-PDR, where the broad line wings heavily contaminate the $F$=2-1 satellite and the $F$=1-0 and $F$=1-1 satellites are strong enough to be detected individually. We observe velocity shifts between the [$^{12}$C\,{\sc ii}] and the [$^{13}$C\,{\sc ii}] line in ORI-P2, ORI-P3 and ORI-P5.

Also the $^{12}$CO line is optically thick and thus opacity broadened, while the $^{13}$CO line is usually optically thin. Adopting $T_{\rm P}(^{13}\text{CO})/T_{\rm P}(^{12}\text{CO}) \simeq 1-\exp\left[-\tau\left(^{13}\text{CO}\right)\right]$, the optical depth of the $^{13}$CO line is estimated to be less than 0.7 in our sample. Hence, we use the $^{13}$CO lines to extract velocity information. The $^{13}$CO lines are somewhat narrower than the C$n\alpha$ lines in most positions, a notable exception being the second component of ORI-P1 (ORI-P1-2). In ORI-P3 and ORI-P4 the main $^{13}$CO lines seem to trace the molecular background (cold, with narrow lines), while the C$n\alpha$ lines in these positions are much broader. Many $^{13}$CO lines have wings or a second component. Most notably, the $^{13}$CO lines in BNKL-PDR have very broad wings from high-velocity gas in BN/KL outflows, that are also present to a lesser degree in the [C\,{\sc ii}] line, but not pronounced in the C$n\alpha$ line. In TRAP-PDR, ORI-P1-2, and ORI-P5 the $^{12}$CO and $^{13}$CO lines are observed at somewhat different velocities, which might be an opacity effect.

In BAR-INSIDE, the C$n\alpha$ line has a second faint component at $v_{\mathrm{LSR}} = 3.0\pm 0.2\,\mathrm{km\,s^{-1}}$ besides the main component at $10.6\,\mathrm{km\,s^{-1}}$, but the [C\,{\sc ii}] line has a second fainter component at $-2.3\pm 0.2\,\mathrm{km\,s^{-1}}$. The C$n\alpha$ line in TRAP-PDR has a faint component at $0.9\pm 0.4\,\mathrm{km\,s^{-1}}$ besides the main component at $9.2\,\mathrm{km\,s^{-1}}$, while the [C\,{\sc ii}] line has a weak component at $0.5\pm 0.1\,\mathrm{km\,s^{-1}}$. The lines in EAST-PDR and ORI-P1 have two components, separated by $6\,\mathrm{km\,s^{-1}}$ and $3\,\mathrm{km\,s^{-1}}$, respectively. The lines in ORI-P2, ORI-P4, ORI-P5 and ORI-P6 are approximately fitted with a single emission component, with a peak velocity increasing from $8.7\,\mathrm{km\,s^{-1}}$ to $11.4\,\mathrm{km\,s^{-1}}$. In ORI-P3 neither the peak velocity ($8.2\,\mathrm{km\,s^{-1}}$) nor the line width of the C$n\alpha$ line match the profile of the [$^{12}$C\,{\sc ii}] line, that has two peaks (at $5.7\,\mathrm{km\,s^{-1}}$ and $10.3\,\mathrm{km\,s^{-1}}$) and a broad component between the two peaks. The [$^{13}$C\,{\sc ii}] line is almost coincident with the red component of the [$^{12}$C\,{\sc ii}] line. The $^{13}$CO line in ORI-P3 exhibits non-Gaussian wings.

\begin{figure}[htb]
\includegraphics[width=0.5\textwidth, height=0.375\textwidth]{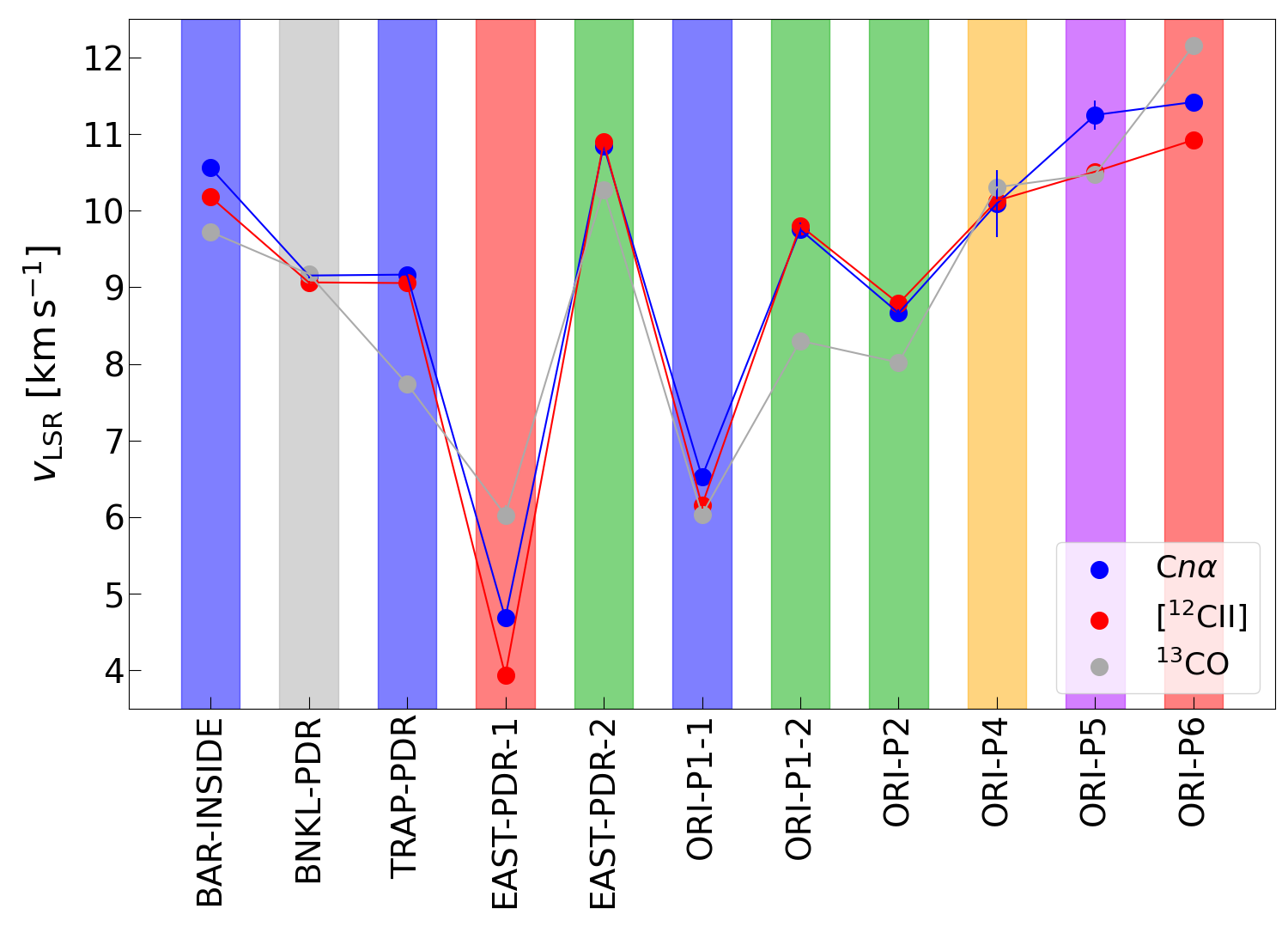}
\caption{Peak velocities in the observed positions of the [C\,{\sc ii}], C$n\alpha$, and $^{13}$CO lines obtained from Gaussian fits. The background shading indicates the velocity-ordering class, defined in the main text, the position and component belongs to: Red is class 1, green is class 2, blue is class 3, yellow is class 4, and purple is class 5.}
\label{Fig.velocities}
\end{figure}

In summary, the peak velocities of the [C\,{\sc ii}], $^{13}$CO, and C$n\alpha$ lines differ slightly in all positions, except in BNKL-PDR where they are all similar. Fig. \ref{Fig.velocities} shows the peak velocities obtained from Gaussian fits to the lines. We observe five different behaviors (or classes) in the layering of the peak velocities of the main components, from lower to higher LSR velocity (excluding ORI-P3):
\begin{enumerate}
\item Velocity layering [C\,{\sc ii}]-C$n\alpha$-CO (EAST-PDR-1, ORI-P6)
\item Velocity layering CO-C$n\alpha$/[C\,{\sc ii}] (EAST-PDR-2, ORI-P1-2, ORI-P2)
\item Velocity layering CO-[C\,{\sc ii}]-C$n\alpha$ (BAR-INSIDE, TRAP-PDR, ORI-P1-1)
\item Velocity layering [C\,{\sc ii}]/C$n\alpha$-CO (ORI-P4)
\item Velocity layering [C\,{\sc ii}]/CO-C$n\alpha$ (ORI-P5)
\end{enumerate}

Comparing the C$n\alpha$ and H$n\alpha$ RRLs toward the observed positions, we note that toward OMC1 the neutral gas and the ionized gas move at very different velocities, as has been known for a long time and repeatedly studied in increasing detail \citep[e.g.,][]{Balick1974, ODell2009, Goicoechea2015b, Abel2019, ODell2020}. This behavior is not linked to the classification above. The ionized gas is blue-shifted by $14.8\,\mathrm{km\,s^{-1}}$ toward the Orion Bar and by $12.0\,\mathrm{km\,s^{-1}}$ toward BNKL-PDR and TRAP-PDR. Toward EAST-PDR the velocity shift between the main component of the neutral gas and the ionized gas is $13.6\,\mathrm{km\,s^{-1}}$. Also in ORI-P5, southeast of the Orion Bar, the velocity shift is $14.8\,\mathrm{km\,s^{-1}}$. Opposed to that, the ionized gas is blueshifted by only $1.8\,\mathrm{km\,s^{-1}}$ toward ORI-P3, the center of M43. Here, however, the CRRL is blueshifted from the [C\,{\sc ii}] line that traces the background PDR and we do not observe a CRRL component that corresponds to the background velocity. In ORI-P1 the ionized gas is blueshifted by $3.4\,\mathrm{km\,s^{-1}}$ with respect to the brighter neutral component and by $6.6\,\mathrm{km\,s^{-1}}$ with respect to the fainter component. Toward ORI-P6, OMC3, we do not detect ionized gas.

We do not observe systematic velocity differences between the C$n\alpha$, C$n\beta$ and C$n\gamma$ lines. Also in the H$n\alpha$, H$n\beta$ and H$n\gamma$ lines there is no systematic ordering of the peak velocities. The velocity shifts are usually less than $1\,\mathrm{km\,s^{-1}}$ and do not increase with quantum leap $\Delta n$, but scatter about zero velocity shift.

\subsection{Intensity ratios}

\begin{figure}[htb]
\includegraphics[width=0.5\textwidth, height=0.357\textwidth]{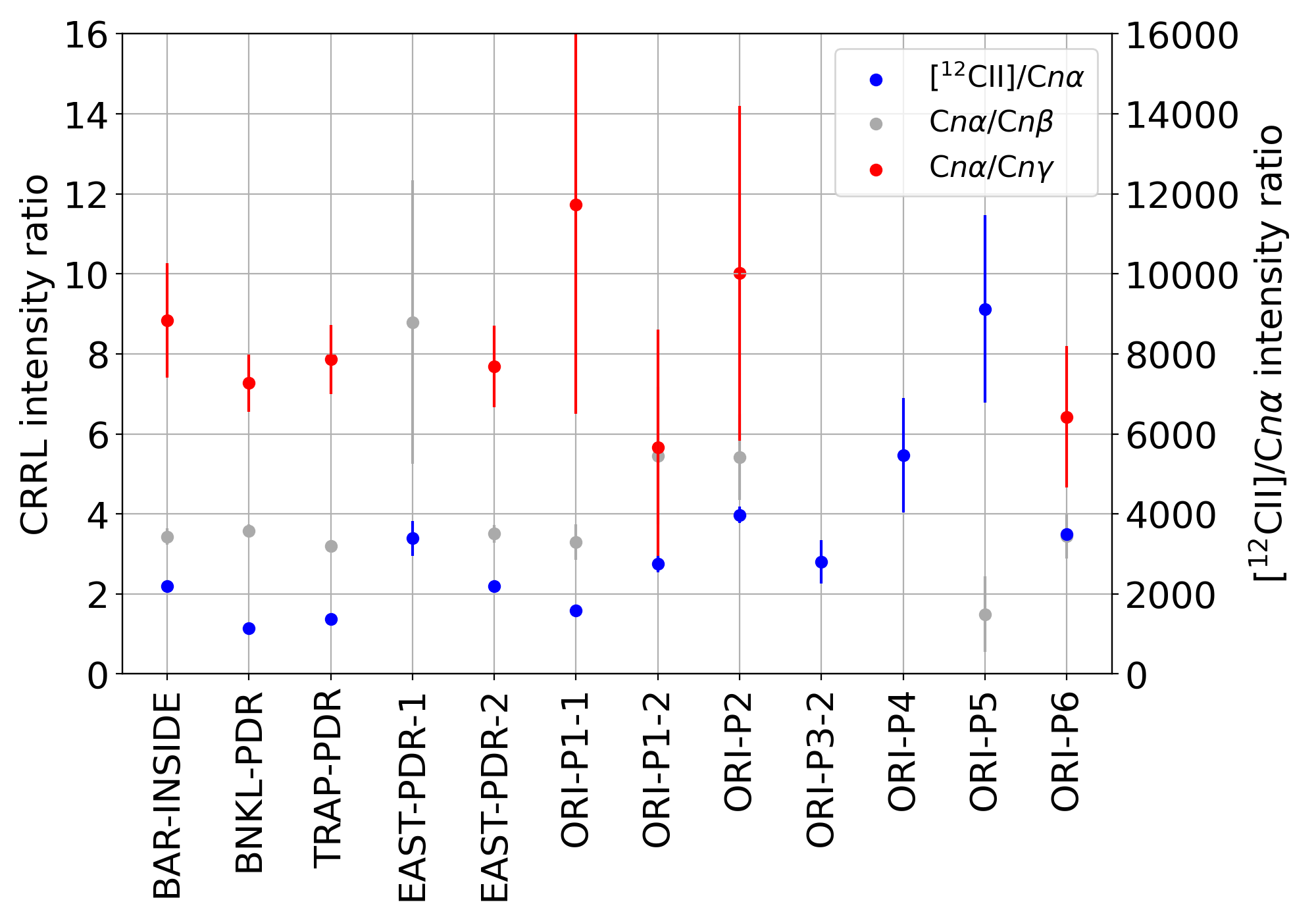}
\caption{Intensity ratios of C$n\alpha$ over C$n\beta$ and C$n\gamma$, respectively (left axis) and [$^{12}$C\,{\sc ii}] over C$n\alpha$ (right axis).}
\label{Fig.CRRL_CII_intensity_ratios}
\end{figure}

Figure \ref{Fig.CRRL_CII_intensity_ratios} shows the intensity ratios of the carbon RRLs and the intensity ratio of the [C\,{\sc ii}] line over the C$n\alpha$ line (intensity units are $\mathrm{K\,km\,s^{-1}}$). Of all positions, BNKL-PDR and TRAP-PDR have the lowest [C\,{\sc ii}]/C$n\alpha$ intensity ratio with $1.1\times 10^3$ and $1.3\times 10^3$, respectively. ORI-P1 also has a low [C\,{\sc ii}]/C$n\alpha$ intensity ratio with $1.6\times 10^3$ The majority of the regions has [C\,{\sc ii}]/C$n\alpha$ ratios between $2\times 10^3$ and $4\times 10^3$. Only in ORI-P4 and ORI-P5 the ratios are elevated at about $5.5\times 10^3$ and $9.1\times 10^3$, respectively, but the errors are large. The blue component of EAST-PDR has an elevated [C\,{\sc ii}]/C$n\alpha$ ratio compared to the other PDRs in OMC1 and much elevated CRRL intensity ratios, but again the errors are large.

The mean of the C$n\alpha$/C$n\beta$ intensity ratios of all positions is $3.6\pm 1.2$, without the outlier EAST-PDR-1 with a ratio of $9\pm 4$. ORI-P1-2 and ORI-P2 have elevated ratios of $5.5\pm 1.8$ and $5.4\pm 1.1$, respectively, while ORI-P5 has a lower ratio of $1.5\pm 1.0$.

The C$n\alpha$/C$n\gamma$ intensity ratios in OMC1 lie close to the mean ratio of $8.0\pm 0.7$. ORI-P1-1 and ORI-P2 are outliers with C$n\alpha$/C$n\gamma$ ratios of $12\pm 5$ and $10\pm 4$, respectively, but due to their faintness the errors are also largest. In ORI-P1-2 the C$n\gamma$ intensity appears to be equal to the C$n\beta$ intensity (C$n\alpha$/C$n\gamma$ intensity ratio of $6\pm 3$), but the errors in the fits allow for the 'normal' intensity ordering. The double-peaked structure of ORI-P1 is less pronounced in the C$n\beta$ line, but very pronounced in the C$n\gamma$ line. ORI-P6 has a somewhat lower ratio of $6.4\pm 1.8$.

\begin{figure}[htb]
\includegraphics[width=0.5\textwidth, height=0.34\textwidth]{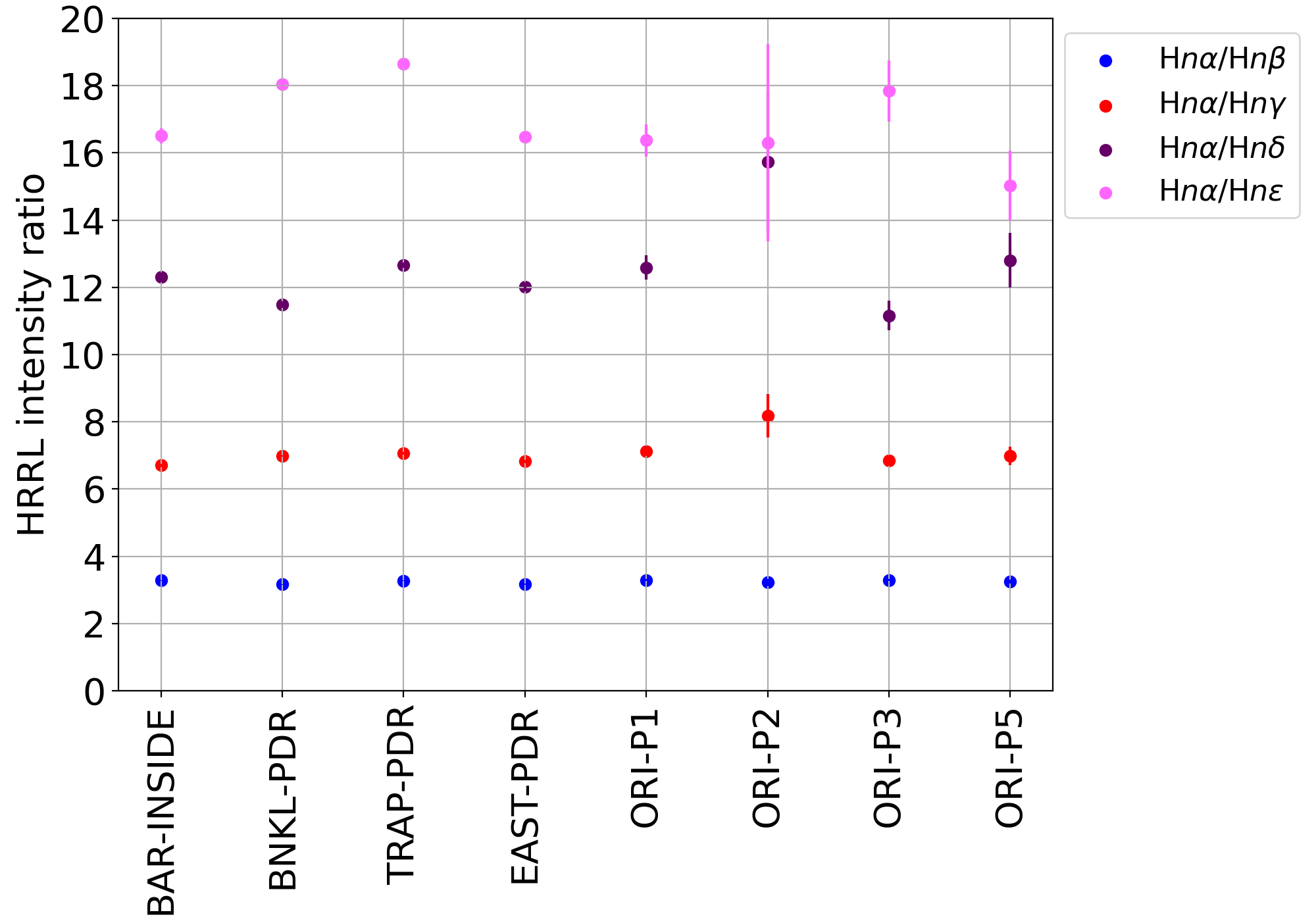}
\caption{Intensity ratios of H$n\alpha$ over H$n\beta$, H$n\gamma$, and H$n\delta$, respectively.}
\label{Fig.HRRL_intensity_ratios}
\end{figure}

Fig. \ref{Fig.HRRL_intensity_ratios} shows the integrated intensity ratios of the hydrogen RRLs, where detected. We observe little variation in the HRRL line ratios between the positions. Mean values with standard deviations are $3.24\pm 0.06$ for the H$n\alpha$/H$n\beta$ intensity ratio, $7.1 \pm 0.5$ for the H$n\alpha$/H$n\gamma$ intensity ratio, $12.6\pm 1.3$ for the H$n\alpha$/H$n\delta$ intensity ratio, and $16.9\pm 1.1$ for the H$n\alpha$/H$n\epsilon$ intensity ratio. Where the line are fainter and errors are larger (ORI-P2, ORI-P5), we observe the most deviation from the mean values.

\begin{figure*}[htb]
\includegraphics[width=\textwidth, height=0.35\textwidth]{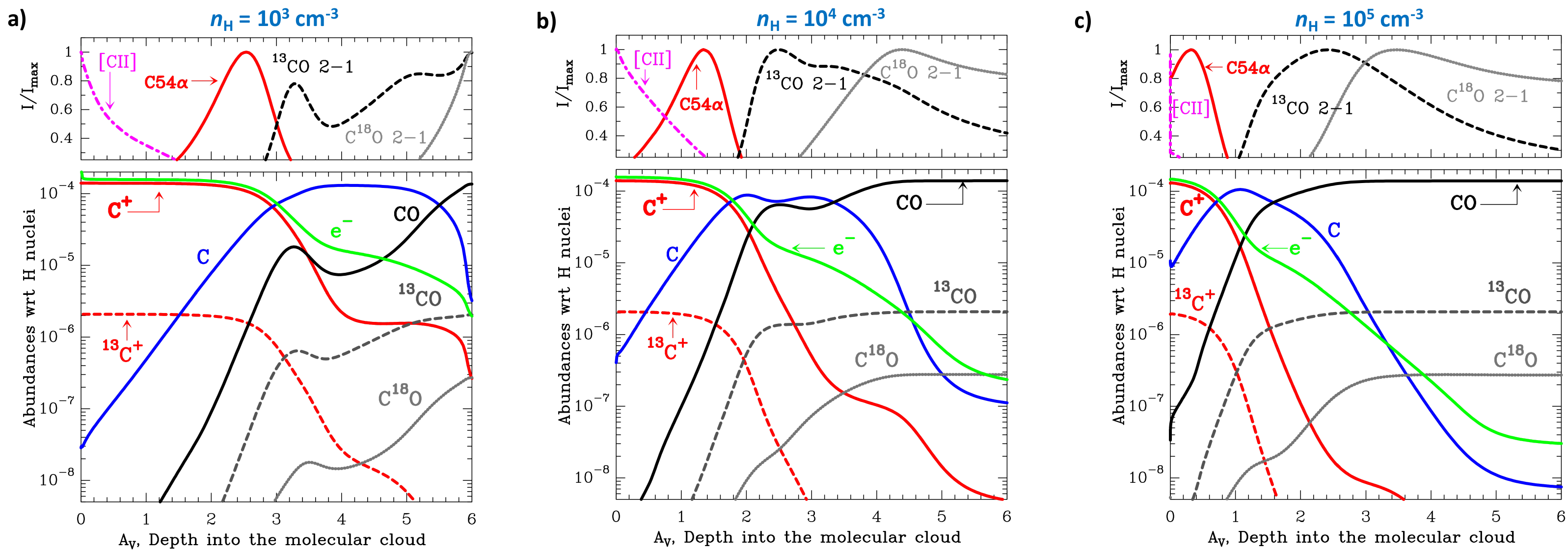}
\caption{PDR model outputs from the Meudon PDR code for an incident radiation field of $G_0=100$ and three different densities ($n=10^3,\,10^4,\,10^5\,\mathrm{cm^{-3}}$). The upper panel shows the normalized intensity, assuming LTE for the CRRL emission, the lower panel shows the abundances of the relevant species.}
\label{Fig.PDR-model}
\end{figure*}

\section{Discussion}

\subsection{Velocities of the [C\,{\sc ii}], C$n\alpha$ and CO-emitting gas}
\label{sec.discussion-velocities}

In order to interpret the velocity structure of our observations, we first run representative PDR models. Figure \ref{Fig.PDR-model} shows the stratification of emission layers in a PDR as computed by the Meudon PDR code \citep{LePetit2006} for an incident radiation field of $G_0=100$ (appropriate for regions far from OMC1) and three different densities ($n=10^3,\,10^4,\,10^5\,\mathrm{cm^{-3}}$), including $^{13}$C fractionation. We include the emission lines and abundances relevant for this discussion. Typically, the [C\,{\sc ii}] line is bright at the surface of a PDR, where the gas is warmest, but extends to the cooler gas at $A_{\rm V} \sim 2$. The C54$\alpha$ line, which we observe with the Yebes 40m telescope, peaks somewhat deeper into the PDR, as the optical depth is weighted ($\propto T^{-2.5}$) toward cooler C$^+$-rich gas ($A_{\rm V} \simeq 1\text{--}2$). Beyond $A_{\rm V} \simeq 2$ carbon is locked up in CO and low-J CO rotational lines are excited. In lower-density models (e.g., $n=10^3\,\mathrm{cm^{-3}}$), the regions of bright emission of the [C\,{\sc ii}] and C54$\alpha$ line are slightly broader in $A_{\rm V}$ space than at higher density (e.g. $n=10^5\,\mathrm{cm^{-3}}$), while CO emission sets on closer to the surface, as the C$^+$/C/CO transition is pulled toward the lower $A_{\rm v}$ for higher densities.

Considering the typical stratification of a PDR, as expected from the models, the velocity layering expected for a face-on PDR, with a photoevaporative flow from the surface, is [C\,{\sc ii}]-C$n\alpha$-CO (blue to red, class 1 in our classification). Photoevaporative flows from PDRs are expected to have velocities of about $1\,\mathrm{km\,s^{-1}}$ \citep{Bertoldi1989, Stoerzer1999, Bron2018, Maillard2021}. However, the layering expected for face-on PDRs (such as TRAP-PDR and ORI-P5) we encounter only in two edge-on PDRs, ORI-P4 and ORI-P6. For perfectly edge-on PDRs, we do not expect a radial velocity shift. ORI-P1 likely has one face-on and one edge-on component (cf. Sec. \ref{physical-conditions-neutral}), while ORI-P2 on the edge of M43 is likely observed edge-on and ORI-P3 in the center of M43 face-on. With the geometries of the cloud in the positions we observe it is unclear what causes the observed velocity differences between the lines. In most positions, we can only speculate about the origins. The velocity ordering of class 1 (EAST-PDR-1, ORI-P6) can be explained by assuming that the illuminating star is located somewhat in front of the molecular cloud, while on the other hand the class-2 behavior (EAST-PDR-2, ORI-P1-2, ORI-P2) would mean that the cloud is situated closer to the observer than the illuminating star.

Furthermore, it is likely that the cloud parts move with respect to one another due to dynamics induced by stellar feedback from the most massive stars at earlier evolutionary stages, among which fossil outflows \citep{Kavak2022a}, but also gravitational infall \citep{Hacar2017}. We may thus observe the remnants of these earlier dynamics as velocity differences between the layers of the PDR. Also the dynamics currently induced by the winds and radiation of the massive stars in the Orion Nebula complex \citep{Pabst2019, Pabst2020} and protostellar jets of less massive objects \citep[e.g.,][]{Mendez-Delgado2021, Kavak2022b} can possibly lead to shear in the cloud parts due to the irregular shapes of the clouds and their relative orientation with respect to the stars. It is also possible that velocity gradients within our beam get weighted differently in the different emission lines. This might be the case in particular in BAR-INSIDE, where we average emission from the bright Orion Bar with emission from the background molecular cloud southeast of the Orion Bar, which is known to be slightly blue-shifted with respect to the Orion Bar \citep{Tauber1994, Cuadrado2017}.

The observed velocity differences between the [$^{12}$C\,{\sc ii}] and the C$n\alpha$ lines are less than $1\,\mathrm{km\,s^{-1}}$, except in ORI-P3, with a median of the absolute values of all positions of $0.25\,\mathrm{km\,s^{-1}}$. In ORI-P3 the observed C$n\alpha$ line does not have line parameters that match either the foreground or the background neutral component, but it could also be a blend of multiple components that we are unable to identify due to the low S/N ratio. The observed velocity differences between the $^{13}$CO and the C$n\alpha$ lines scatter between 0 and $2\,\mathrm{km\,s^{-1}}$ with a median of $0.75\,\mathrm{km\,s^{-1}}$. The spectrum of ORI-P3 does have a weak $^{13}$CO component with a peak velocity matching the C$n\alpha$ line, but with very different line widths; it might also be a wing of the main component. The main $^{13}$CO component corresponds to the main [$^{12}$C\,{\sc ii}] component of the background PDR.

The lines in BAR-INSIDE show slight deviations from Gaussianity, which can be leveraged by fitting the line with two components. The Orion Bar is the prototypical edge-on PDR close to the Trapezium stars. Our beam averages emission from the narrow Orion Bar and emission from the background molecular cloud southeast of the Orion Bar. The line is dominated by the bright edge-on PDR. Also in other positions the velocity shifts may be caused by averaging multiple components with varying strengths in the beam (without leading to much deviation from Gaussianity, however). Given our relatively large beam and the systematic uncertainties in performing Gaussian fits to the observed lines, we refrain from drawing strong conclusions from the small (but significant by eye inspection) observed peak velocity differences between the different tracers.

The broad wings in BNKL-PDR are the result of molecular outflows and shocks produced by proto-stellar activity in the molecular core, likely the dynamical decay of a multiple system \citep{Gomez2005, Bally2011}, visible in both molecular lines \citep{Rosenthal2000, Gonzales-Alfonso2002, Goicoechea2015a, Bally2017} and the [C\,{\sc ii}] line \citep{Goicoechea2015b, Morris2016}. Relatively little C$^+$ is produced in shocks \citep{HollenbachMcKee1989}, hence the [C\,{\sc ii}] line and the CRRLs are relatively unaffected and mostly trace the PDR gas (main peak) and the FUV-illuminated surface of the shocked gas (wings). In contrast, the [O\,{\sc i}] 63\,$\mu$m line and CO lines are heavily affected by the shocks and outflows \citep{Stacey1993, Berne2014}, consisting of a PDR component with moderate temperature and a hot component likely produced from a J-type shock \citep{Rosenthal2000, Gonzales-Alfonso2002, Goicoechea2015a}.

Position TRAP-PDR lies in the region that has been extensively studied in optical wavelengths \citep[e.g.,][]{ODell1993, Abel2019, ODell2020}. The velocity components now known as Orion's Veil were first detected in H\,{\sc i} 21\,cm absorption \citep{VanderWerf1989}. In [C\,{\sc ii}] and CRRL emission, we observe only the PDR behind the Trapezium stars ($v_{\rm LSR} \simeq 9.1\,\mathrm{km\,s^{-1}}$) and the Veil component B ($v_{\rm LSR} \simeq 0.5\,\mathrm{km\,s^{-1}}$). The CRRL Veil component, however, seems to lie on a broader line (wing), while the [C\,{\sc ii}] Veil component is approximately Gaussian.

In EAST-PDR and ORI-P1 we observe two pronounced velocity components. While the weaker (blue) component of EAST-PDR probably corresponds to the Dark Lane in the foreground of the molecular cloud with its main PDR, the position ORI-P1 lies close or on the border between OMC1 and M43. Possibly this part of the cloud is a dense wall, through which the hot plasma created by the Trapezium stars cannot escape. The presence of two velocity components may give rise to the assumption that this part of the cloud is created by a cloud-cloud collision between the clouds containing OMC1 and the cloud containing M43. However, the molecular cores OMC1 and OMC2 close to M43 both belong to the continuous structure of the Integral-Shaped Filament (ISF), the dense molecular spine of the Orion Nebula complex.

In each position in OMC1 we observe a faint blue-shifted component in the [C\,{\sc ii}] line at small positive or negative velocities, that most likely corresponds to the Veil in positions other than BAR-INSIDE. In the C$n\alpha$ line toward BAR-INSIDE we observe a weak second component, blue-shifted from the main component by $7.5\pm 0.5\,\mathrm{km\,s^{-1}}$, that does not have a corresponding [C\,{\sc ii}] component. While the velocity observed here does not match exactly the velocity of a sulfur RRL (the velocity shift between C and S should be $-8.6\,\mathrm{km\,s^{-1}}$), an earlier study of the Orion Bar edge has reported the detection of sulfur RRLs toward this position \citep{Goicoechea2021}. The intensity ratio between the main component and the blue-shifted component is about 10 and the line widths are similar. Hence, we tentatively conclude that this component is indeed a sulfur RRL with a similar intensity ratio compared to the carbon RRL even somewhat deeper inside the cloud. We do not detect such a faint component without [C\,{\sc ii}] counterpart in the other positions: In TRAP-PDR and BNKL-PDR the sulfur RRL may be hidden in the component corresponding to the [C\,{\sc ii}] Veil component at $v_{\rm LSR} \simeq 0.5\,\mathrm{km\,s^{-1}}$ and $v_{\rm LSR} \simeq 1.7\,\mathrm{km\,s^{-1}}$, respectively, while in EAST-PDR the sulfur RRL may be hidden in the blue strong component (EAST-PDR-1) of the carbon RRL. In positions outside OMC1, the S/N ratio is not sufficient to detect the potential sulfur lines with $\sim$10 times lower intensity in the C$n\alpha$ spectra.

\subsection{Physical conditions in the neutral gas}
\label{physical-conditions-neutral}

The comparison of the carbon RRLs with the [C\,{\sc ii}] line emission allows for a first-order estimate of the physical conditions in the neutral, not yet fully molecular PDR, in particular the electron density $n_{\rm e}$ and the electron temperature $T_{\rm e}$. Since the CRRL intensities and the [C\,{\sc ii}] intensities scale differently with density and temperature, the two intensities can be used to derive the physical conditions in the gas, assuming that the emissions stem from the same regions in the PDR. This latter assumption is not exactly valid as the CRRLs come from closer to the molecular gas, offsetting the derived values from the true physical conditions. The necessary theoretical framework to compute CRRL intensities given physical conditions was developed by \cite{Natta1994, Smirnov1995, Tsivilev2014, Salgado2017a, Salgado2017b, Salas2021}. The non-LTE excitation model computes the C$n$ integrated intensity for $\Delta n = 1,2,3,4,5$, and the [$^{12}$C\,{\sc ii}] and [$^{13}$C\,{\sc ii}] integrated intensities, taking into account optical depth effects, on a grid of $T_{\rm e} = 20 - 1000\,\mathrm{K}$, $n_{\rm e} = 0.01 - 200\,\mathrm{cm^{-3}}$, and C$^+$ column densities $\log_{10} N(\mathrm{C}^+\,\mathrm{[cm^{-2}]}) = 17.5 - 19.0$. For $n$ we take the mean principal quantum number of the stacked lines. The model assumes that hydrogen nuclei are equally distributed in atomic hydrogen and molecular hydrogen (i.e., the hydrogen nuclei fractions are $f(\mathrm{HI}) = f(\mathrm{H}_2) = 0.5$ with $f(\mathrm{HI}) = n_{\rm HI}/n_{\rm H}$ and $f(\mathrm{H_2}) = 2n_{\rm H_2}/n_{\rm H}$) and that the hydrogen nuclei density is given by $n_{\rm H} = n_{\rm e}/\mathcal{A}_{\rm C}$ with $\mathcal{A}_{\rm C} = 1.4\times 10^{-4}$ the carbon gas-phase abundance in Orion \citep{Sofia2004}, i.e. all electrons stem from carbon ionization. We also assume that the gas kinetic temperature equals the electron temperature. We interpolate the model grid using a spline interpolation.

\begin{table}[ht]
\renewcommand{\arraystretch}{1.2}
\addtolength{\tabcolsep}{2pt}
\centering
\caption{Dust properties and hydrogen nuclei column density along each line of sight computed from SED fits.}
\begin{tabular}{lccccccc}
\hline\hline
source & $T_{\rm d}$ & $\tau_{\rm d, 160\,\mu m}$ & $N_\mathrm{H}$ \\
 & [K] & $ [10^{-3}]$ & [$10^{22}$ cm$^{-2}$] & \\ \hline
BAR-INSIDE & 38 & 35 & 21 \\
BNKL-PDR & 45 & 150 & 87 \\
TRAP-PDR & 44 & 120 & 71 \\
EAST-PDR & 41 & 23 & 14 \\
ORI-P1 & 36 & 24 & 14  \\
ORI-P2 & 36 & 13 & 7.7  \\
ORI-P3 & 38 & 5.8 & 3.5  \\
ORI-P4 & 36 & 7.8 & 4.7  \\
ORI-P5 & 36 & 5.4 & 3.3 \\
ORI-P6 & 30 & 14 & 8.6 \\ \hline
\end{tabular}
\tablefoot{Uncertainties in the dust temperature and dust optical depth are about 50\%, see discussion in \cite{Pabst2020}. We convert from dust optical depth $\tau_{\rm d, 160\,\mu m}$ to hydrogen nuclei column density $N_\mathrm{H}$ following \cite{Weingartner2001}.}
\label{tab.dust_properties}
\end{table}

\begin{table*}[!tb]
\centering
\caption{Results of Bayesian analysis using results from the non-LTE excitation code.}
\renewcommand{\arraystretch}{1.3}
\addtolength{\tabcolsep}{5pt}
\begin{tabular}{lcccccc}
\hline\hline
source & $n_{\mathrm{e}}$ & $T_{\mathrm{e}}$ & $N(\mathrm{C}^+)$ & $\tau_{\mathrm{[CII]}}$ & $T_{\mathrm{ex}}(\mathrm{[CII]})$ & $p_{\rm th}/k_{\rm B}$ \\
 & [cm$^{-3}$] & [K] & [$10^{18}$ cm$^{-2}$] &  & [K] & [K cm$^{-3}$] \\ \hline
BAR-INSIDE & $20\substack{+20 \\ -10}$ & $230\substack{+50 \\ -30}$ & $3.5\substack{+1.5 \\ -1.0}$ & $1.5\substack{+0.5 \\ -0.5}$ & $220\substack{+40 \\ -20}$ & $3\times 10^7$ \\
BNKL-PDR\tablefootmark{a} & $40\substack{+100 \\ -20}$ & $240\substack{+140 \\ -30}$ & $5.0\substack{+3.9 \\ -2.5}$ & $2.0\substack{+1.5 \\ -1.3}$ & $230\substack{+140 \\ -20}$ & $6\times 10^7$ \\
TRAP-PDR\tablefootmark{b} & $140\substack{+60 \\ -50}$ & $360\substack{+80 \\ -60}$ & $2.8\substack{+0.7 \\ -0.6}$ & $0.7\substack{+0.3 \\ -0.2}$ & $350\substack{+80 \\ -50}$ & $3\times 10^8$ \\
EAST-PDR-2 & $20\substack{+90 \\ -10}$ & $210\substack{+200 \\ -30}$ & $3.5\substack{+2.1 \\ -1.8}$ & $1.6\substack{+1.0 \\ -1.2}$ & $200\substack{+200 \\ -20}$ & $3\times 10^7$ \\
ORI-P1-1 & $2\substack{+1 \\ -1}$ & $100\substack{+10 \\ -10}$ & $5.0\substack{+2.1 \\ -2.0}$ & $5.4\substack{+1.4 \\ -0.6}$ & $90\substack{+10 \\ -10}$ & $1\times 10^6$ \\
ORI-P1-2 & $8\substack{+100 \\ -3}$ & $150\substack{+190 \\ -10}$ & $1.6\substack{+0.6 \\ -0.9}$ & $1.8\substack{+0.9 \\ -1.5}$ & $140\substack{+190 \\ -10}$ & $8\times 10^6$ \\
ORI-P2 & $2\substack{+2 \\ -1}$ & $140\substack{+10 \\ -10}$ & $4.5\substack{+0.5 \\ -1.2}$ & $2.6\substack{+0.1 \\ -1.0}$ & $120\substack{+10 \\ -10}$ & $2\times 10^6$ \\
ORI-P3-2\tablefootmark{c} & -- & -- & $1.6\substack{+3.4 \\ -0.8}$ & $1.2\substack{+3.6 \\ -1.0}$ & $130\substack{+300 \\ -30}$ & -- \\
ORI-P4 & $3\substack{+62 \\ -2}$ & $140\substack{+330 \\ -20}$ & $1.4\substack{+1.4 \\ -0.7}$ & $1.2\substack{+1.7 \\ -1.0}$ & $120\substack{+330 \\ -30}$ & $3\times 10^6$ \\
ORI-P5 & $2\substack{+12 \\ -1.5}$ & $180\substack{+220 \\ -50}$ & $1.8\substack{+1.4 \\ -0.7}$ & $0.9\substack{+1.3 \\ -0.7}$ & $140\substack{+190 \\ -40}$ & $1\times 10^6$ \\
ORI-P6 & $1\substack{+1 \\ -0.1}$ & $130\substack{+10 \\ -10}$ & $5.0\substack{+0.6 \\ -0.5}$ & $5.2\substack{+0.1 \\ -0.1}$ & $120\substack{+10 \\ -10}$ & $8\times 10^5$ \\ \hline
\end{tabular}
\tablefoot{Errors are given at $1\sigma$. The electron densities may be converted to hydrogen nuclei densities by dividing by the carbon gas-phase abundance in Orion, $\mathcal{A}_{\rm C} = 1.4\times 10^{-4}$.
\tablefoottext{a}{We used [$^{13}$C\,{\sc ii}] $F$=1-0 and $F$=1-1 satellites.}
\tablefoottext{b}{[$^{13}$C\,{\sc ii}] line in TRAP-PDR weaker than expected given the isotopic ratio of Orion.}
\tablefoottext{c}{CRRL in ORI-P3 does not match PDR component, hence only [C\,{\sc ii}] properties.}}
\label{tab.ne-Te}
\end{table*}

\begin{figure*}[htb]
\includegraphics[width=\textwidth, height=0.67\textwidth]{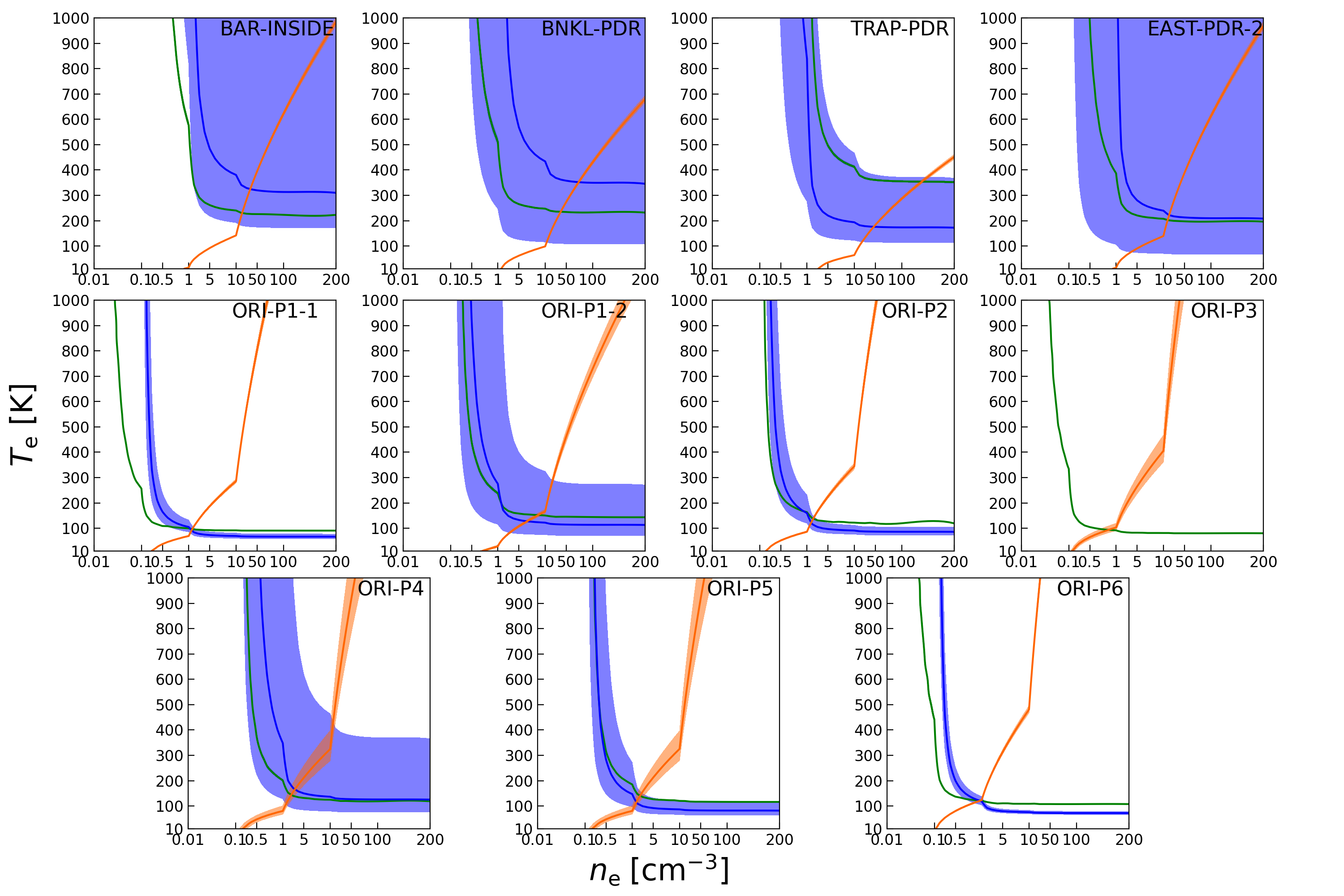}
\caption{Observed CRRL, [$^{12}$C\,{\sc ii}], and [$^{13}$C\,{\sc ii}] intensities constrain electron temperature $T_{\rm e}$ and electron density $n_{\rm e}$ (orange, C55$\alpha$; green, [$^{12}$C\,{\sc ii}]; blue, [$^{13}$C\,{\sc ii}]). The shaded area gives the $1\sigma$ error of each observation. Each panel uses the respective model grid with the column density inferred from the Bayesian analysis.}
\label{Fig.CRRL_models_sources}
\end{figure*}

The peak temperature of the [$^{12}$C\,{\sc ii}] line for an extended source is given by
\begin{align}
T_{\rm P} &= \left(J(T_{\rm ex}) - J(T_{\rm bg})\right) [1-e^{-\tau_{\mathrm{[CII]}}}] \\
&= \left(\frac{91.2\,\mathrm{K}}{e^{91.2\,\mathrm{K}/T_{\rm ex}} - 1} - \frac{91.2\,\mathrm{K}}{e^{91.2\,\mathrm{K}/T_{\rm bg}} - 1} \right) [1-e^{-\tau_{\mathrm{[CII]}}}] ,\label{eq.TP}
\end{align}
where 91.2\,K is the energy level separation of the [$^{12}$C\,{\sc ii}] fine-structure transition, $T_{\rm ex}$ is the excitation temperature of the [$^{12}$C\,{\sc ii}] transition, $T_{\rm bg}$ is the temperature of a background permeating the [C\,{\sc ii}] emitting region, and $\tau_{\mathrm{[CII]}}$ is the optical depth of the transition \citep{Goldsmith2012}. The background temperature $T_{\rm bg}$ generally arises from the cosmic microwave background (CMB) and the dust FIR field, external or internal to the [C\,{\sc ii}] emitting region. In the case of the [C\,{\sc ii}] line, the CMB is unimportant and the dust FIR field with temperature $T_{\rm d}$ is attenuated by the dust optical depth $\tau_{\rm d}$, such that $J(T_{\rm bg}) = J(T_{\rm d})(1-e^{-\tau_{\rm d}})$. In our case, the warm dust is intermingled with the [C\,{\sc ii}] emitting region in the PDR. In OMC1, the continuum brightness temperature at 158\,$\mu$m $T_{\rm c} = J(T_{\rm bg})$ has been observed to range between 1 and 7 K, thus also being negligible \citep{Goicoechea2015b}. The [$^{13}$C\,{\sc ii}] peak temperature can be computed from the same equation, where the optical depth is scaled with the $^{12}$C/$^{13}$C isotopic ratio. In Orion, we assume a $^{12}$C/$^{13}$C the isotopic ratio of 67 \citep{BoreikoBetz1996}.

The [$^{12}$C\,{\sc ii}] excitation temperature is related to the kinetic gas temperature $T_{\rm kin}$ by
\begin{align}
e^{91.2\,\mathrm{K}/T_{\rm ex}} = \left(1+\frac{n_{\rm cr}}{n} \right) e^{91.2\,\mathrm{K}/T_{\rm kin}} ,
\end{align}
where $n_{\rm cr}$ is the critical density of the transition and $n$ the gas density \citep[cf. eq. 13 in][with $G=0$]{Goldsmith2012}. If there are multiple collision partners, that collisionally excite the transition (electrons, hydrogen atoms, and hydrogen molecules), we define the gas density as the hydrogen nuclei density $n = n_{\rm H} + 2n_{\rm H_2}$, and
\begin{align}
\frac{n_{\rm cr}}{n} = \beta(\tau_{\mathrm{[CII]}})\frac{A_{\rm ul}}{C_{\rm ul}},
\end{align}
where $\beta(\tau_{\mathrm{[CII]}}) = (1-e^{-\tau_{\mathrm{[CII]}}})/\tau_{\mathrm{[CII]}}$ is the 158\,$\mu$m photon escape probability, $A_{\rm ul}$ is the Einstein coefficient for spontaneous radiative de-excitation, and $C_{\rm ul} = \gamma_{\rm ul,e}n_{\rm e} + \gamma_{\rm ul,H}n_{\rm H} + \gamma_{\rm ul,H_2}n_{\rm H_2}$ is the combined collision de-excitation rates with coefficients $\gamma_i$ \citep[see][for parametric expressions]{Goldsmith2012}.

The peak temperature of a high-frequency carbon RRL ($\nu > 2\,\mathrm{GHz}$), under the assumption that the line is optically thin and there is no background, is given by \citep[cf.][]{Salas2019}
\begin{align}
T_{\rm P} = \tau_{\rm CRRL} b_{\rm u} T_{\rm e} ,
\end{align}
where $\tau_{\rm CRRL}$ is the optical depth of the CRRL in LTE and $b_u = N_{\rm u}/N_{\rm u, LTE}$ is the departure coefficient of the upper level of the CRRL transition, defined as the ratio of the level population $N_{\rm u}$ over the equilibrium level population $N_{\rm u, LTE}$ \citep{Shaver1975}. Departure coefficients were computed using the equations of \cite{Salgado2017a}. The peak temperature of the CRRL scales approximately with $n_{\rm e}^2 T_{\rm e}^{-1.5}$ \citep{Salgado2017b}. The line-integrated optical depth is given by
\begin{align}
\tau_{\rm CRRL}\Delta\nu = 1.069\times 10^7\Delta n M T_{\rm e}^{-2.5} e^{\chi_n} n_{\rm e} N(\mathrm{C^+}) \;\mathrm{Hz},
\end{align}
with the oscillator strength of the transition $M$, $\chi_n=157800 n^{-2} T_{\rm e}^{-1}$, and the C$^+$ column density $N(\mathrm{C^+})$ \citep{Menzel1968, Salgado2017b, Salas2019}.

We use the [$^{12}$C\,{\sc ii}], [$^{13}$C\,{\sc ii}] $F$=2-1 and the C$n\alpha$ intensities to constrain the gas density and temperature in the PDR in a Bayesian analysis. The errors are estimated using the 0.68 confidence interval, i.e. at $1\sigma$. In the model we use the line width of the [$^{13}$C\,{\sc ii}] $F$=2-1 line to compute the [C\,{\sc ii}] intensities. The model computes the opacity broadening of the optically thick [$^{12}$C\,{\sc ii}] line.

We make use of the dust spectral energy distributions (SEDs) by \cite{Pabst2019}, to compute the dust properties in our positions. They combine the far-IR and submillimeter {\it Herschel}/PACS and SPIRE bands to obtain the dust temperature $T_{\rm d}$ and dust optical depth $\tau_{\rm d, 160\,\mu m}$ at $160\,\mu\mathrm{m}$ in a modified blackbody fit at $36\arcsec$ resolution. We extract dust properties with a beam of $45\arcsec$. To calculate dust properties in BNKL-PDR and TRAP-PDR, where the SPIRE bands are saturated, we use a PACS-only SED (also saturated in a few pixels around BNKL-PDR). Since the FUV-heated dust FIR radiation in Orion is dominated by emission in the PACS bands, especially in warmer regions, values do not vary greatly between the PACS+SPIRE and the PACS-only SEDs. Table \ref{tab.dust_properties} summarizes the dust properties toward the observed positions. We note that the derived dust parameters render the $J(T_{\rm bg})$ term in Eq. \ref{eq.TP} negligible: The equivalent continuum brightness temperature $T_{\rm c}$ ranges between 10 and 20\,K. To estimate the total gas column density in the PDR, we convert the dust optical depth to a hydrogen column density, $N_{\rm H} = N(\mathrm{HI}) + 2N(\mathrm{H}_2)$ using \citep{Weingartner2001, Pabst2020},
\begin{align}
N_{\mathrm{H}} \simeq 6\cdot 10^{24}\,\mathrm{cm}^{-2}\;\tau_{\rm d, 160\,\mu m}.
\end{align}

Table \ref{tab.ne-Te} summarizes the results of the CRRL Bayesian analysis. Fig. \ref{Fig.CRRL_models_sources} illustrates how the observed intensities constrain $T_{\rm e}$ and $n_{\rm e}$. The sharp breaks in some of the plots are artifacts from the interpolation of the sparsely sampled original grid, on which the departure coefficients were computed. In general, the observations constrain the physical conditions well. We note that the temperature within a PDR really decreases from the surface, which affects the [C\,{\sc ii}] and C$n\alpha$ intensities differently. Hence, the inferred $n_{\rm e}$ and $T_{\rm e}$ are an ill-defined average over the [C\,{\sc ii}] and C$n\alpha$ emitting layers along the line of sight and within our beam. With the physical conditions in our sample, $T_{\rm e}$ is predominantly set by the [C\,{\sc ii}] line intensity, while the C$n\alpha$ intensity largely defines $n_{\rm e}$. Hence, in general, we expect the derived temperature to reflect more the temperature at the PDR surface, where the [C\,{\sc ii}] emission is strongest, while the density will reflect more the deeper denser layers, where the CRRL emission is strongest. On the other hand, CRRLs will be enhanced compared to the [C\,{\sc ii}] intensity in clumpy PDRs (increased $n_{\rm e}$). How these different conditions affect our type of line intensity analysis is difficult to judge as it strongly depends on the adopted PDR model with its detailed incorporated physical and chemical processes (e.g., constant density versus constant pressure, or homogeneous versus clumpy). More investigation is needed in this regard, which is beyond the scope of this paper.

The results in TRAP-PDR are not to be trusted, since the [$^{13}$C\,{\sc ii}] intensity is much weaker than the [$^{12}$C\,{\sc ii}] intensity scaled with the $^{12}$C/$^{13}$C isotopic ratio (assumed to be 67; earlier studies used a $^{12}$C/$^{13}$C isotopic ratio of 43 at $\theta^1$ Ori C \citep[][and references therein]{Stacey1993}). The results in BAR-INSIDE, where the bright PDR lines seem to consist of two close Gaussian components, are most likely weighted toward the high-column density region of the Orion Bar rather than the background molecular cloud southeast of the Orion Bar.

Positions in OMC1 (except for TRAP-PDR) have high electron densities (ranging from 20 to 40\,$\mathrm{cm^{-3}}$) and moderate temperatures (ranging from 210 to 240 K). Although ORI-P1 is close to OMC1, the electron densities and temperatures of the two components are much less than in OMC1, 2 and 8\,$\mathrm{cm^{-3}}$ and 100 and 150 K. Possibly we observe ORI-P1-1 edge-on (though moving away from Trapezium), and ORI-P1-2 face-on (same systemic velocity as Trapezium), judging from the C$^+$ column density. This is also reflected in the relative strength of the $^{12}$CO and $^{13}$CO lines: The blue component appears to be cooler and with higher column density, while the red component appears warmer with lower column density. This two-layer model, with a warm layer behind a cooler layer, for the Dark Lane, in which ORI-P1 lies, was first proposed by \cite{Herrmann1997}, derived from their [C\,{\sc ii}] and [O\,{\sc i}] 63\,$\mu$m and 145\,$\mu$m fine-structure FIR line observations. The gas density in the warmer layer derived in this work, assuming that all electrons result from carbon ionization and a carbon gas-phase abundance of $\mathcal{A}_{\rm C}\simeq 1.4\times 10^{-4}$ \citep{Sofia2004}, is on the order of $10^5\,\mathrm{cm^{-3}}$, in agreement with the assumption of \cite{Herrmann1997} for the molecular component, while the gas density in the cooler layer is somewhat lower with about $3\times 10^4\,\mathrm{cm^{-3}}$.

In M43 the electron densities are much lower, 2 to 3\,$\mathrm{cm^{-3}}$, and the temperatures are 140 K. The non-detection of a CRRL component from the background molecular cloud in ORI-P3 suggests a low density and/or high temperature and/or a low column density, but from the [C\,{\sc ii}] line properties we derive an optical depth of 1.2 and an excitation temperature of 130 K, values consistent with a moderately dense, moderately illuminated cloud surface with a significant C$^+$ column density. On the west border of M43, where the H\,{\sc ii} region presses against the molecular ridge, \cite{Herrmann1997} derive a density of $3\times 10^5\,\mathrm{cm^{-3}}$ for the molecular gas. Judging from the [O\,{\sc i}] 63\,$\mu$m and 145\,$\mu$m intensities the density toward the center of M43 must be somewhat lower, but still within the typical range of PDRs \citep{Herrmann1997}. This is in agreement with the gas densities calculated from the electron densities derived in this work, $n_{\rm H} \simeq 2\times 10^4\,\mathrm{cm^{-3}}$.

ORI-P5, associated with the background PDR in the EON, has an electron density of 2\,$\mathrm{cm^{-3}}$ and a temperature of 180 K. The electron density in ORI-P6 is the lowest of all the sources, 1\,$\mathrm{cm^{-3}}$, and the region is cool with 130 K. ORI-P6 lies in an edge-on PDR, giving rise to the high C$^+$ optical depth and column density.

Our findings indicate that the PDRs in OMC1 have gas densities above $10^5\,\mathrm{cm^{-3}}$ ($n_{\rm H} \simeq n_{\rm e}/\mathcal{A}_{\rm C} \simeq 1\text{--}3\times 10^5\,\mathrm{cm^{-3}}$), while PDRs outside OMC1 have gas densities lower than $10^5\,\mathrm{cm^{-3}}$. The PDRs in OMC1 are known to have high densities, and our estimates are in good agreement with earlier estimates \citep{TielensHollenbach1985b, Wolfire1990, Stacey1993, Herrmann1997, Goicoechea2015b, Cuadrado2019}. Our estimates for the BNKL-PDR yield a somewhat higher [C\,{\sc ii}] excitation temperature than the average position toward BN/KL of \cite{Morris2016} with similar optical depth. Our temperature estimates in OMC1 are consistent with earlier findings from [C\,{\sc ii}] observations \citep{Goicoechea2015b}, but lower than results from CRRL analysis at the edge of the Bar \citep{Cuadrado2019}. The density estimates for M43 are somewhat higher than in \cite{Pabst2022}, while the estimate for OMC3 is consistent. The derived gas temperatures for M43 and OMC3 are somewhat higher than estimates in \cite{Pabst2022}. Our position ORI-P6 in OMC3 is equivalent to the position EFF5 of \cite{Salas2021}, who derive lower temperatures ($T_{\rm ex}\mathrm{[C\,\textsc{ii}]} \simeq 63\pm6\,\mathrm{K}$ and $T\simeq 55\pm2\,\mathrm{K}$), but a similar electron density from C110$\alpha$, C109$\alpha$, and C102$\alpha$ lines.

In highly irradiated regions the ratio of the C$^+$ column density divided by $\mathcal{A}_{\rm C}$ (i.e., the hydrogen nucleus column density in the gas containing ionized carbon), inferred from the previous analysis, and the total H nucleus column density, derived from the dust optical depth, is low ($N(\mathrm{C^+})/\mathcal{A}_{\rm C}/N_{\rm H}$ below 0.25). In M43 and OMC3 the ratio is significantly elevated in comparison (between 0.3 and 0.6). This is in agreement with the higher density inferred for the highly irradiated regions in OMC1 and the EON, where the [C\,{\sc ii}] emission is more concentrated on the PDR surface (cf. Fig \ref{Fig.PDR-model}).

There are some issues with the adopted approach above. The position of TRAP-PDR gives special riddles. Here, the [$^{13}$C\,{\sc ii}] line is weaker than expected assuming the standard isotopic ratio of carbon and relative strengths of the hyperfine components. Fractionation, reducing the $^{13}$C$^+$ abundance by locking up carbon-13 in $^{13}$CO, should be a minor effect, according to our PDR models. We observe a curious velocity shift between the $^{12}$CO and $^{13}$CO line in TRAP-PDR -- the $^{13}$CO line is blue-shifted --, which we do not observe at all in other positions. Both CO lines are blue-shifted from the [C\,{\sc ii}] line in this position. We also observe velocity shifts between the [$^{12}$C\,{\sc ii}] and [$^{13}$C\,{\sc ii}] lines in some positions. While ORI-P3 has several components with possible self-absorption that might explain the apparent shift, the reason for the shift in ORI-P2 is not clear. It is curious that we do not observe a CRRL component from the background PDR in ORI-P3. It is unlikely that the density is too low or the temperature too high, since we observe both [C\,{\sc ii}] and CO as fairly strong lines. The [$^{13}$C\,{\sc ii}] line in ORI-P5 is noisy, but still we observe a extreme shift, while the [$^{12}$C\,{\sc ii}] profile seems symmetric.

Ideally one would like to use the [$^{13}$C\,{\sc ii}] line intensity together with the C$n\alpha$, $\beta$, and $\gamma$ intensities to constrain the physical properties from the line ratios of these optically thin lines, as \cite{Cuadrado2019} do for the edge of the Orion Bar. However, the predicted intensity ratios for the CRRLs of different $\Delta n$ are very close to each other and the comparison with the observed intensity ratio therefore very sensitive to the observed intensity ratio. The C$n\beta$/C$n\alpha$ ratio only varies from 0.22 to 0.28 in the temperature-density range of $(T_{\rm e}, n_{\rm e}) = (20-1000\,\mathrm{K}, 0.01-100\,\mathrm{cm^{-3}})$. Owing to the low S/N ratio, the errors in our observations are actually larger (due to baseline issues, multiple components within the line of sight and the helium RRL being blended with the CRRL) and often observed values lie outside this range. For the positions in OMC1, the errors in the ratios are less than 10\%, but the inferred temperatures and densities lie outside the grid. In ORI-P6, where the CRRLs are strong, as well, and there is no ionized gas, the ratios also exhibit high values that indicate higher temperatures and/or densities than the grid range, which we deem unlikely considering the quiescent nature of this region and previous estimates (Kabanovic et al., in prep.).

\subsection{The pressure equilibrium}

\begin{figure}[htb]
\includegraphics[width=0.5\textwidth, height=0.367\textwidth]{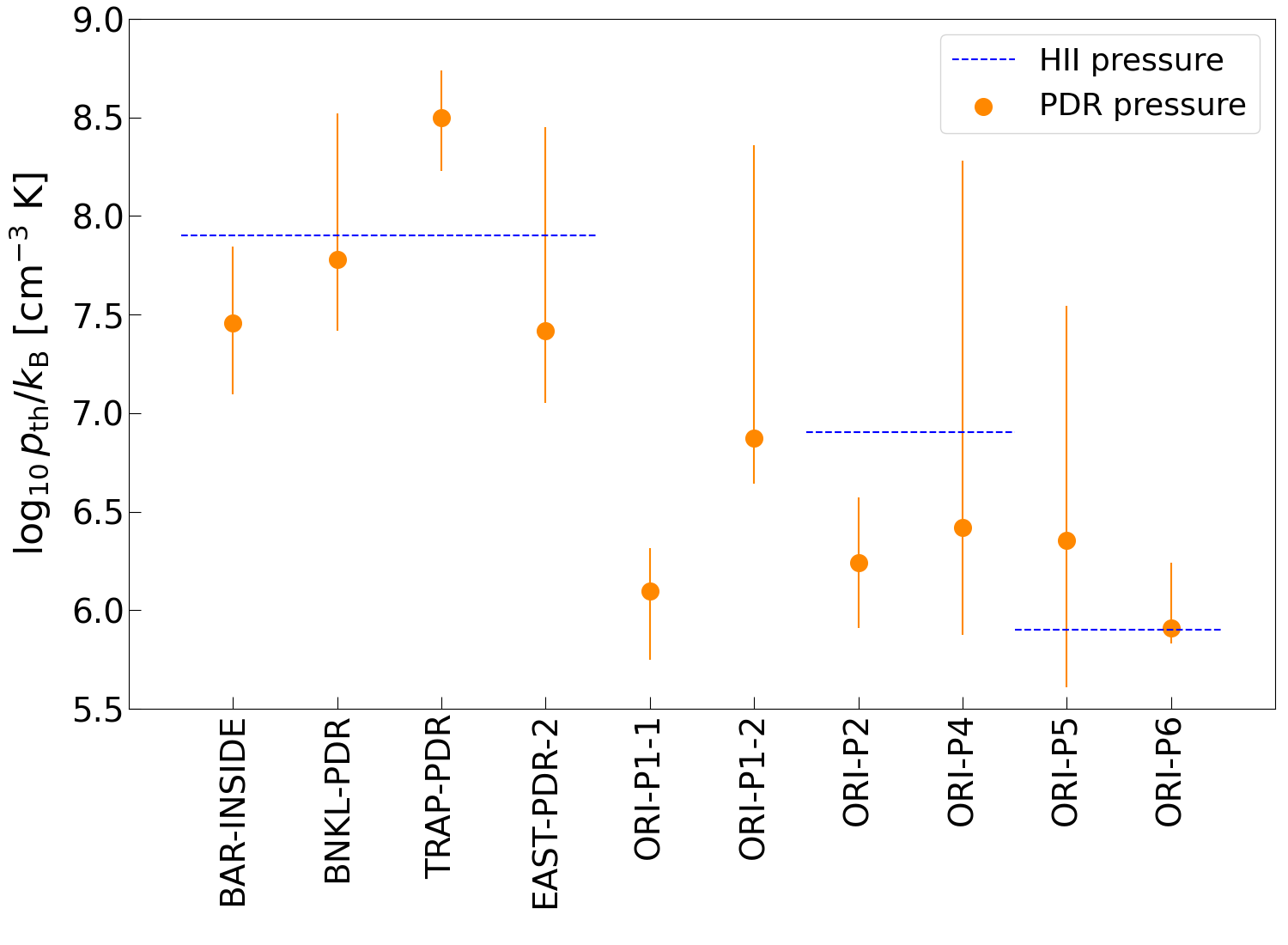}
\caption{Thermal pressure in the neutral gas derived in this work. Dashed lines indicates the thermal pressure in the adjacent H\,{\sc ii} region taken from the literature.}
\label{Fig.CRRL_analysis_pth}
\end{figure}

Figure \ref{Fig.CRRL_analysis_pth} shows the thermal pressures computed from the physical conditions in the neutral PDR gas derived in this work, $p_{\mathrm{th}}/k_{\rm B} = n_{\rm e}/\mathcal{A}_{\rm C} T_{\rm e}$, compared to the thermal pressure in the ionized gas, in the observed positions. The ionized gas in the Orion Nebula was analyzed by \cite{ODellHarris2010}, the ionized gas in M43 was analyzed by \cite{SimonDiaz2011}, and physical conditions of the ionized gas in NGC 1977 were estimated by \cite{Pabst2020}.

The total pressure in a PDR is the sum of several pressure terms: thermal, turbulent, magnetic, radiation and line pressure (cf. discussion in Sec. 4.1 in \cite{Pabst2020}). We should usually expect the PDR to be governed by equipartition of the thermal, turbulent and magnetic pressures, while radiation pressure and line pressure are generally unimportant in Orion. However, as we shall see below (Sec. \ref{line-broadening}), at the spatial scales we probe the turbulent pressure is usually somewhat larger than the thermal pressure in our PDRs. The magnetic pressure in OMC1 is on the order of the thermal pressure \citep{Chuss2019}.

With the physical conditions derived in this work, the two gas phases, neutral and ionized, are in approximate pressure equilibrium in OMC1, albeit values in OMC1 are somewhat lower than earlier findings, that indicate high thermal pressures at the edge of the Orion Bar \citep[$p_{\rm th}/k_{\rm B} \sim 10^8\,\mathrm{K\,cm^{-3}}$,][from mm-wave CRRLs and molecular tracers]{Cuadrado2019,Goicoechea2019}. CRRLs may be enhanced in dense clumps, while the [C\,{\sc ii}] line arises in the inter-clump medium. Other high-density tracers, such as high-J CO lines, also imply much higher pressures \citep{Joblin2018}. \cite{Pellegrini2009} find a maximum of $p_{\rm th}/k_{\rm B} \simeq 8\times 10^7\,\mathrm{K\,cm^{-3}}$ (close to the transition from H to H$_2$) from modeling various emission lines from the H$^+$/H/H$_2$ transition in the Orion Bar. The thermal pressures in the PDR gas of the Orion Bar, M43 and OMC3 are consistent with those cited in \cite{Pabst2020}, derived from analysis of the [C\,{\sc ii}] line alone. The thermal pressure of M43 is short of the thermal pressure of the ionized gas by a factor of 3, possibly reflecting the equipartition of the thermal, turbulent and magnetic pressure in the PDR. The thermal pressure in the EON, measured in ORI-P5, is higher than the thermal pressure in the ionized gas \citep[$n\sim 10^2\,\mathrm{cm^{-3}}$,][]{ODellHarris2010}, suggesting that the ionized gas is insignificant in driving the expansion of the Veil Shell.

Overall, RRLs are useful tracers of the physical conditions and pressures in the respective gas components. Higher angular resolution observations will allow us to resolve finer structures and potentially clumps in the PDRs to fully characterize the properties of the different gas phases.

\subsection{Physical conditions in the ionized gas}

If ionized gas can freely expand against a background of neutral (PDR) gas, the ionization front moves through the neutral gas at a velocity of $v_{\rm if} \sim \frac{c_{\rm s,I}^2}{2c_{\rm s,II}}$, where $c_{\rm s,I}$ is the sound speed of the neutral gas and $c_{\rm s,II}$ is the sound speed of the ionized gas. Depending on the temperature of the neutral gas, $v_{\rm if} \sim 0.3\text{--}0.5 \,\mathrm{km\,s^{-1}}$. The ionized gas will be accelerated away from the background along the pressure gradient reaching values of $v_{\rm i}  \sim  2c_{\rm s,II}$ for free expansion \citep{Spitzer1968, Kahn1969}.

In OMC1 the ionized gas is enshrouded in a multilayer system of ionized and neutral components \citep[e.g.,][]{Abel2019, ODell2020} in front of the background PDR. The ionized gas traced by our hydrogen RRLs streams away from the background PDR vary rapidly ($v_{\rm i} \simeq 12-15\,\mathrm{km\,s^{-1}}$) due to the strong pressure gradient from the interface to the dilute foreground interstellar medium. The expansion velocity is on the order of the sound speed in the ionized gas, $v_{\rm i} \sim c_{\rm s,II}$. We note that the peak velocity of the HRRL emitting gas does not match any of the components \cite{Abel2019} identify, but \cite{ODell2020} comment that the H41$\alpha$ emission of \cite{Goicoechea2015b} should arise from the [O{\sc iii}] emitting layer, which is shifted from the PDR velocity by $v_{\rm evap,[OIII]} = 9\pm 3\,\mathrm{km\,s^{-1}}$ (their observations). The main component of our observed HRRLs is a few km\,s$^{-1}$ bluer than the [O{\sc iii}] component, and it is unclear why. Especially the position TRAP-PDR samples part of the Orion-South cloud, where several velocity systems have been identified \citep{ODell2021a, ODell2021b}. This is however not clearly reflected in our HRRL spectrum in TRAP-PDR, where the line is still very nearly Gaussian. All HRRL seem to be broadened by other effects than thermal motion of the atoms alone (cf. Sec. \ref{line-broadening} for a discussion on extra line broadening).

We note that we do not observe [C\,{\sc ii}] line and CRRL components from the ionized gas toward OMC1 (nor in ORI-P1). In principle, ionized gas emits in the [C\,{\sc ii}] line and CRRLs, but the radiation from the close Trapezium cluster is energetic enough to doubly ionize carbon and the emission from singly ionized carbon will be weak. The EON, to which our position ORI-P5 corresponds, contains largely hot plasma and little ionized gas, and we do not observe the weak [C\,{\sc ii}] line and CRRL components from the ionized gas either. In M43 and NGC 1977, in contrast, the illuminating stars are less massive and therefore their radiation is less energetic, and a significant fraction of carbon is expected to be singly ionized. We observe those [C\,{\sc ii}] lines from the ionized gas as broad components ($\Delta v_{\rm FWHM} \sim 5\text{--}10\,\mathrm{km\,s^{-1}}$) in spectra toward the center of M43 and NGC 1977 \citep{Pabst2017, Pabst2020}.

In M43 the ionized gas is contained within an expanding shell. The HRRL in ORI-P3, the center of M43, is blueshifted from the CRRL by $1.5\,\mathrm{km\,s^{-1}}$. However, the CRRL in ORI-P3 is blue-shifted by $2\,\mathrm{km\,s^{-1}}$ from the background PDR traced by [C\,{\sc ii}] and CO. Possibly this CRRL also stems from the ionized gas, where carbon is singly ionized. The ionized gas is streaming away from the background PDR with $3.5\,\mathrm{km\,s^{-1}}$. The HRRL is significantly broader than the CRRL due to the different atomic weight ($20.5\,\mathrm{km\,s^{-1}}$ compared to $7.4\,\mathrm{km\,s^{-1}}$), but the CRRL is broader than in the other positions. In the rim of the M43 shell, in positions ORI-P2 and ORI-P4, the ionized gas is blue-shifted from the [C\,{\sc ii}] and CO-traced PDR gas by 3 and 5\,km\,s$^{-1}$, respectively, and the CRRLs peak velocities and line widths are close to the peak velocities and line widths of the [C\,{\sc ii}] line.

Plotting the peak intensities of the individual hydrogen transitions in the observed positions reveals that the ionized gas is in local thermal equilibrium (LTE). This is characterized by the intensities being proportional to frequency. The nearly constant ratios between the different orders of RRLs suggest that the gas has similar electron temperatures in all positions. The ionized gas in the Huygens Region, however, is denser \citep[$n_{\rm e}\simeq 5\times 10^3\,\mathrm{cm^{-3}}$ close to $\theta^1$ Ori C][]{ODellHarris2010} than the ionized gas in M43 \citep[$n_{\rm e}\simeq 500\,\mathrm{cm^{-3}}$][]{SimonDiaz2011}

\begin{figure}[htb]
\includegraphics[width=0.5\textwidth, height=0.405\textwidth]{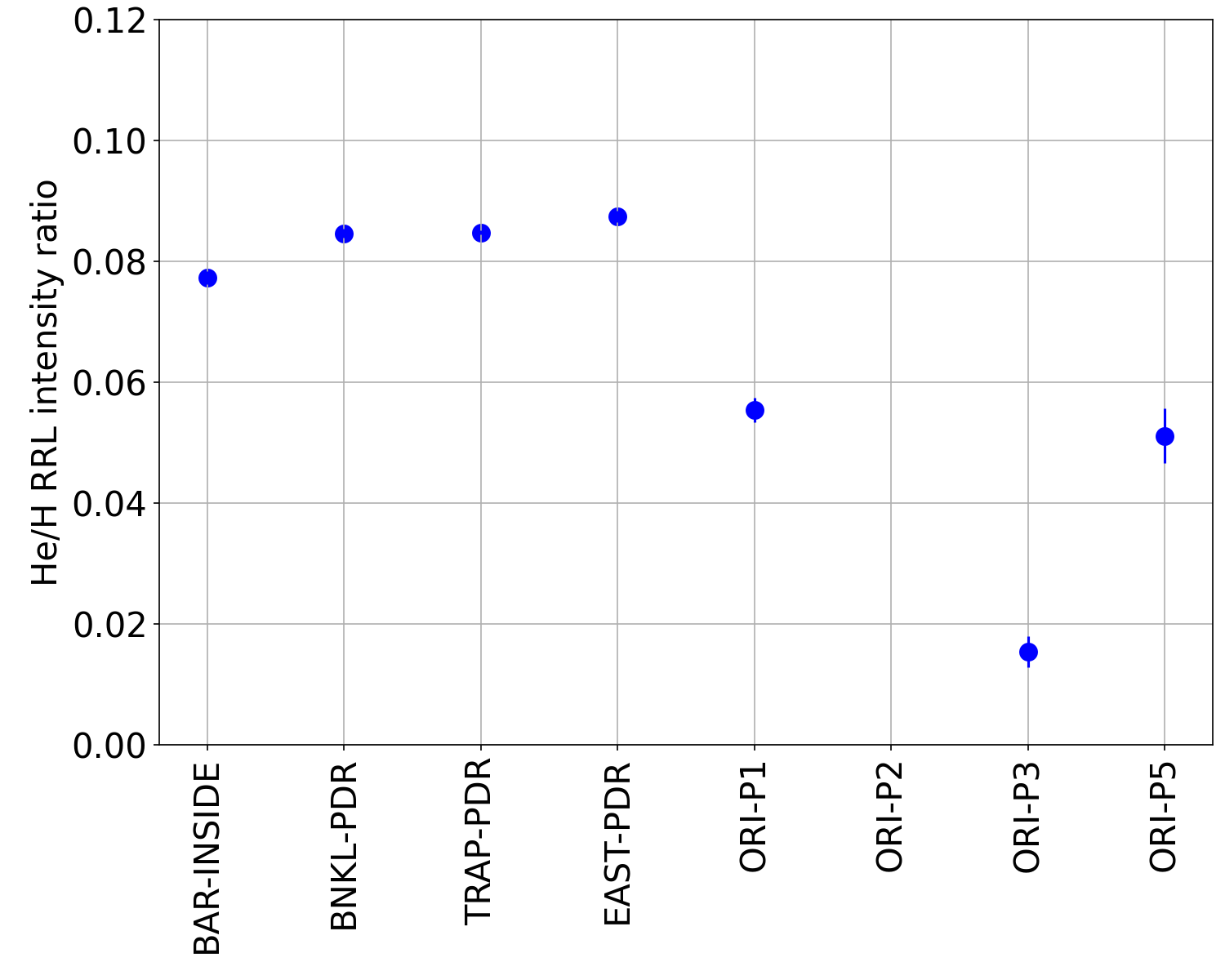}
\caption{Intensity ratio of He$n\alpha$ over H$n\alpha$ RRLs.}
\label{Fig.He_H_ratio}
\end{figure}

Fig. \ref{Fig.He_H_ratio} shows the integrated intensity ratios of the helium RRLs over the hydrogen RRLs. Using $I_{\mathrm{He}n\alpha}/I_{\mathrm{H}n\alpha} = n(\mathrm{He}^+)/n(\mathrm{H}^+)$ \citep[e.g.,][]{Roshi2017}, where we assume the H and He RRLs coincide spatially, we derive a helium ionization fraction. The He$n\alpha$/H$n\alpha$ ratio is high in OMC1 ($0.070\pm 0.002$, $0.0824\pm 0.0009$, $0.0826\pm 0.0006$, $0.084\pm 0.001$ for BAR-INSIDE, BNKL-PDR, TRAP-PDR, and EAST-PDR, respectively) and significantly lower in the center of M43 ($0.014\pm 0.003$). ORI-P1 and ORI-P5 take intermediate values ($0.051\pm 0.003$ and $0.049\pm 0.006$). In ORI-P2 we do not observe the He lines, meaning a He$^+$/H$^+$ ratio close to zero.

In OMC1 the ionizing radiation is energetic enough to ionize helium (IP = 24.6 eV) and to doubly ionize carbon (IP = 24.4 eV). In M43 there still is some ionized helium, but only about a quarter of the fraction in OMC1. ORI-P1 and ORI-P5 are still subject to radiation from the Trapezium stars and have a fair amount of ionized helium. The He RRL in ORI-P2, associated with M43, is too weak to be detected, but the noise level is consistent with the helium fraction in ORI-P3.

\subsection{Detection of heavy-ion RRLs toward BNKL-PDR and TRAP-PDR}

\begin{figure}[htb]
\includegraphics[width=0.5\textwidth, height=0.5\textwidth]{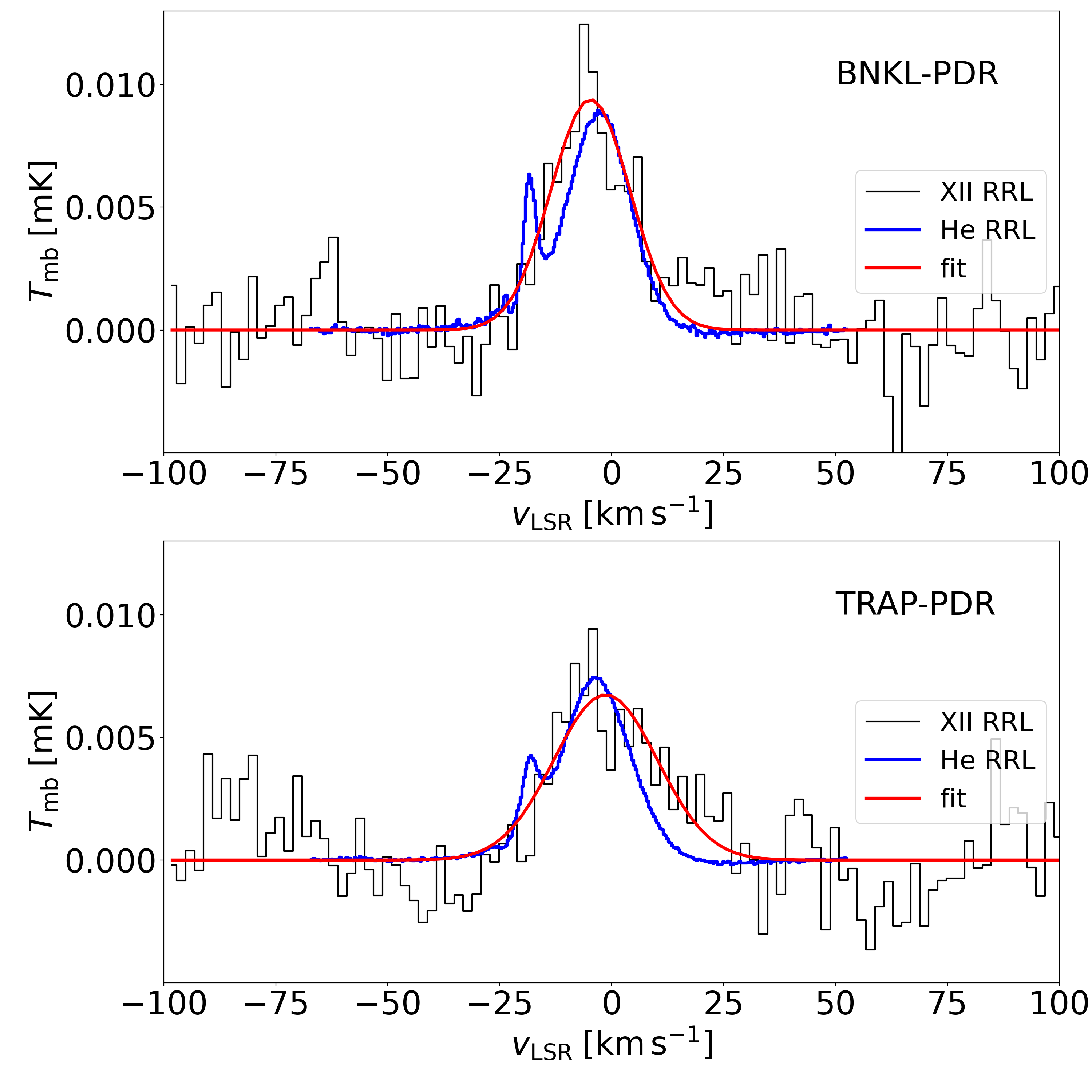}
\caption{Stacked X\,{\sc ii}$n\alpha$ RRLs toward BNKL-PDR (upper panel) and TRAP-PDR (lower panel) in the velocity frame of the C$^+$ RRL, together with the carbon and helium RRLs (scaled to the intensity of the X\,{\sc ii}$n\alpha$ RRL) in the velocity frame of the helium RRL for reference.}
\label{Fig.XII-RRL}
\end{figure}

In the H\,{\sc ii} region of OMC1, the Huygens Region, carbon is doubly ionized and we do not observe the ionized gas in [C\,{\sc ii}] or CRRLs. Toward the positions BNKL-PDR and TRAP-PDR, the helium RRLs are strongest and, given the sensitivity of our observations, we can potentially detect $\alpha$ RRLs from C$^+$ and/or O$^+$, that arise in regions where helium is ionized. The first detection of RRLs of C$^+$ and/or O$^+$ (dubbed X\,{\sc ii}) toward OMC1 has been reported recently by \cite{Liu2023}, by matching the theoretical frequencies with highly sensitive multiband radio observations. Given the reported line width ($\Delta v_{\rm FWHM} > 10\,\mathrm{km\,s^{-1}}$) and the velocity offset between the same C$^+$ and O$^+$ RRL transitions of $\sim 3.5\,\mathrm{km\,s^{-1}}$, it is not possible to distinguish between the two ions. Using the frequencies \cite{Liu2023} give for RRLs of C$^+$, we stack the X\,{\sc ii}$n\alpha$ lines for $n=81,82,83,85,86,88,89,90$ toward BNKL-PDR and $n=81,82,83,85,86,88,89$ toward TRAP-PDR, where we apply a standing-wave baseline correction and resample to a velocity resolution of $2.0\,\mathrm{km\,s^{-1}}$. We exclude spectra that contain molecular lines and other, earlier identified, RRLs, or where a standing-wave baseline correction is not possible. We obtain a rms noise of 1.8 mK toward BNKL-PDR and 2.4 mK toward TRAP-PDR. Fig. \ref{Fig.XII-RRL} shows the resulting stacked X\,{\sc ii}$n\alpha$ line together with the carbon and helium RRLs (scaled to the intensity of the X\,{\sc ii}$n\alpha$ line) in the velocity frame of the helium RRL for reference.

From a Gaussian fit of the X\,{\sc ii}$n\alpha$ RRL, we obtain a peak temperature of $T_{\rm mb} \simeq 9.4\pm 0.8\,\mathrm{mK}$ for BNKL-PDR and $T_{\rm mb} \simeq 6.7\pm 0.8\,\mathrm{mK}$ for TRAP-PDR. The peak velocity and the line width depend critically on the baseline correction, especially for TRAP-PDR, where the line is somewhat weaker and baselines are worse. We tentatively obtain, assuming X=C, a peak velocity of $v_{\rm LSR}\simeq -5\pm 1\,\mathrm{km\,s^{-1}}$ and a line width of $\Delta v_{\rm FWHM}\simeq 21\pm 2\,\mathrm{km\,s^{-1}}$ for BNKL-PDR. For TRAP-PDR we tentatively report $v_{\rm LSR}\simeq -1\pm 2\,\mathrm{km\,s^{-1}}$ and $\Delta v_{\rm FWHM}\simeq 27\pm 4\,\mathrm{km\,s^{-1}}$. If X=O, the peak velocity is redder by $3.5\,\mathrm{km\,s^{-1}}$. In the integrated intensity we thus detect the X\,{\sc ii}$n\alpha$ RRL in BNKL-PDR at a $17\sigma$ level, and in TRAP-PDR at $12\sigma$ significance.

The X\,{\sc ii}$n\alpha$ RRLs closely resemble those of helium. Assuming X=O would shift the peak velocities further from the H and He RRL peak velocities. The ionization potential of C$^+$ is very close to that of helium, while O$^+$ has a much higher ionization potential at 35 eV, hence it is likely that C$^+$ and O$^+$ RRLs arise in different ionized layers, while C$^+$ and He RRLs may stem from the same layer. Among the many layers \cite{Abel2019, ODell2020} identify in the Huygens Region, they identify a high-ionization component in optical [O\,{\sc iii}] lines at $v_{\rm LSR} \simeq 0\,\mathrm{km\,s^{-1}}$, attributed to the main ionization front (MIF). This may indicate that O$^+$ RRLs arise in the same layer in the MIF, that cannot be distinguished from a layer of lower-ionization gas, where the C$^+$ and He RRLs arise. Our beam toward TRAP-PDR includes parts of their NE (northeast) region and the SW (southwest) dark cloud, while our BNKL-PDR region lies somewhat to the north of the region of their observations. \cite{ODell2020} also identify a high-ionization component in optical [O\,{\sc iii}] lines at $v_{\rm LSR} \simeq 9\,\mathrm{km\,s^{-1}}$, that they attribute to shocked gas moving into the MIF. We do not see that component in our observations; possibly this component is too localized to contribute significantly in our beam.

While the peak velocities and line widths we obtain are in agreement with line properties reported by \cite{Liu2023}, our line intensities seem to be somewhat smaller (a factor of $\sim 3$), while our velocity resolution is a factor of 1.5 better, but our beam is 1.5 times larger. From the intensity ratio with the H$n\alpha$ intensities, assuming that the intensity ratio equals the density ratio if they stem from the same region, we obtain $n(\mathrm{X}^{++})/n(\mathrm{H}^+) \simeq 3.8\pm 0.4\times 10^{-3}$ in BNKL-PDR and $n(\mathrm{X}^{++})/n(\mathrm{H}^+) \simeq 2.6\pm 0.4\times 10^{-3}$ in TRAP-PDR, which is 8 and 5 times larger, respectively, than the expected abundance ratio in Orion of C$^+$ and O$^+$ together, $\sim 5\times 10^{-4}$ \citep{Sofia2004}. As \cite{Liu2023} observe, dielectronic recombination may be important in enhancing the intensity of heavy-ion RRLs. Detailed photoionization modeling of the excitation conditions would shed light on the region where RRLs from C$^+$ and O$^+$ arise, but this is beyond the scope of this paper.

\subsection{Line broadening}
\label{line-broadening}

In this section we discuss the observed line widths in both the [C\,{\sc ii}] and the hydrogen, helium and carbon RRLs. While RRLs can be broadened by a number of physical processes, this does not apply to the [C\,{\sc ii}] line. Also, some of these processes are expected to only apply to higher quantum numbers (lower frequencies) than those of our observations. Collisional and radiation broadening lead to the line having a Voigt profile in general, a convolution of a Gaussian profile, produced by turbulence, with a Lorentzian profile \citep{Salgado2017b}. However, our observations cover only quantum numbers $n<100$, hence we expect the profiles to be dominated by Doppler broadening. In fact, the lines can be fitted very well by Gaussian profiles.

A discussion of the extra line broadening (in addition to thermal broadening) in optical lines can be found in \cite{ODell2023}. Our observed radio and far-infrared lines are usually broader than the thermal line width, while preserving a Gaussian profile. The extra line broadening observed in optically thin lines can either stem from turbulence or Alfv\'{e}n waves, according to \cite{ODell2023}. We follow along the lines of that discussion. While our rather large beam width can be problematic in comparing results to those from higher spatial resolution optical spectra, we note that our high velocity resolution does make it technically possible to distinguish separate velocity components averaged in our beam (as long as the separation is large enough compared to the line width), as we clearly observe in BAR-INSIDE (averaging the Orion Bar and the molecular cloud southeast of the Orion Bar), where the [C\,{\sc ii}], C$n\alpha$, and $^{13}$CO lines deviate from a single-component Gaussian profile. Especially in OMC1, where velocity shifts occur on rather small spatial scales in certain regions \citep{ODell2021a}, this is critical to check. However, we emphasize that the lines in our sample are surprisingly Gaussian, in contradiction with this given. It seems that our beam avoids these regions, e.g. toward TRAP-PDR we do not cover much of the Orion-South cloud, where spatial and spectral complexity is a known property \citep[also][]{ODell2021b}. We expect more quiescent, more extended regions to be less affected by small-scale velocity shifts. However, we can estimate the effect on the line widths of large-scale (turbulent) motions averaged within our beam with an average beam size of $45\arcsec$ (0.09\,pc at the distance of Orion) for the PDR gas.

To estimate this effect, we compare the [$^{13}$C\,{\sc ii}] and $^{13}$CO lines at $16\arcsec$ resolution with those at $45\arcsec$, that we use in the discussion, and the C51$\alpha$ line at $36\arcsec$ with the C59$\alpha$ line at $54\arcsec$. In the [$^{13}$C\,{\sc ii}] line this line-broadening effect seems to be on the order of at most 20\%, with the exception of BAR-INSIDE, where the two fits give a difference of 36\%. Most accurately the effect of the beam size can be estimated for the $^{13}$CO line, as the S/N ratio is usually high. Here we see that the line width is smaller for the $16\arcsec$ beam by at most 20\%, but in the brightest sources it is smaller by less than 10\%. For the C$n\alpha$ lines the effect is on the order of 10 to 20\%. For 10\% narrower lines, the turbulent pressures we compute in Section \ref{Sec.dots} thus would be lower by 19\%, while for 20\% narrower lines the turbulent pressures would be lower by 36\%. Where the turbulent Mach number (cf. Sec. \ref{Sec.dots}) is close to or slightly higher than 1, turbulence would then be subsonic rather than sonic or slightly supersonic. Running ahead of the discussion, however, our main conclusions would remain unchanged by this modification, but we have to keep this cautionary note in mind when drawing conclusions.

In the following, we need the equation for the total line width. The thermal line width (FWHM) for an ideal gas is given by
\begin{align}
\Delta v_{\mathrm{th}}^2 = \frac{8\ln 2 k_{\rm B}T}{m},
\end{align}
with $T$ being the gas temperature and $m$ the mass of the atom or molecule. The total line width, in the presence of turbulence, is then given by
\begin{align}
\Delta v^2 = \Delta v_{\mathrm{th}}^2 + \Delta v_{\mathrm{turb}}^2. \label{eq.delta-v}
\end{align}
The Alfv\'en line width, that can contribute to the extra line broadening, is given by \citep{Henney2005}
\begin{align}
\Delta v_{\rm A}^2 = \frac{B^2}{4\pi \rho},
\end{align}
where $B$ is the magnetic field strength and $\rho$ is the ion mass density of the gas. In addition, the Alfv\'enic Mach number $\mathcal{M}_{\rm A}$ is defined as the ratio of the line width (standard deviation) over the Alfv\'en velocity. The spectral resolution of the data contributes to the observed line width, but in our case it is negligible: With a smallest observed line width of $1.3\,\mathrm{km\,s^{-1}}$ in the $^{13}$CO line, the velocity resolution of these observations ($0.4\,\mathrm{km\,s^{-1}}$; $0.3\,\mathrm{km\,s^{-1}}$ for the [$^{13}$C\,{\sc ii}] lines, and $0.36\,\mathrm{km\,s^{-1}}$ for the CRRLs) contributes at most 5\% to the measured line width; in the vast majority of the $^{13}$CO, [$^{13}$C\,{\sc ii}], and C$n\alpha$ observations, this instrumental broadening is at most 1\%.

\subsubsection{Line broadening in the neutral gas}

For carbon in the PDR at a temperature of $T=300$ K, the thermal line width is $\Delta v_{\mathrm{FWHM}} \approx 1.1\,\mathrm{km\,s^{-1}}$. The extra line-broadening component in the CRRLs toward OMC1 is therefore between $2.5\,\mathrm{km\,s^{-1}}$ and $4.8\,\mathrm{km\,s^{-1}}$. If this was due to Alfv\'en waves, the magnetic field can have a maximum value of $23\,\mu\mathrm{G}$, assuming a density of $3\times 10^5\,\mathrm{cm^{-3}}$. This is lower than the magnetic field observed in the Veil \citep[$B_{\rm los}\simeq -50\text{--}-75\,\mu\mathrm{G}$,][]{Troland2016} and much weaker than the magnetic field observed in the Orion Bar \citep[$B\simeq 300\,\mu\mathrm{G}$,][]{Chuss2019}.

If we attributed the CRRL in ORI-P3 to the neutral gas, as we do in all other positions, the line width in ORI-P3 would indicate extreme turbulence in the CRRL emitting gas, which is however not reflected in the [C\,{\sc ii}] lines. The line width is consistent, though, with the thermal line width of gas at the temperature of ionized gas with a small contribution from turbulence. The CRRL emission in ORI-P3 would then stem from an ionized layer closer to the molecular background compared to the H and He RRLs. It is extremely puzzling, though, that we do not observe a component that stems from the PDR. In ORI-P2 and ORI-P4 the extra line broadening components are 3.1 and $5.1\,\mathrm{km\,s^{-1}}$, respectively, which results in a maximum magnetic field strength estimate of $5\,\mu\mathrm{G}$ if Alfv\'enic.

The CRRL in ORI-P6 is likely also turbulence dominated. Assuming $T=100$ K, the thermal line width is $\Delta v_{\mathrm{FWHM}} \approx 0.6\,\mathrm{km\,s^{-1}}$ and the turbulent line width would be $1.8\,\mathrm{km\,s^{-1}}$. The extra line broadening component is smaller than in OMC1, OMC3 being a more quiescent region. If the extra line broadening was dominated by Alfv\'en waves, the magnetic field in ORI-P6 could have a maximum value of $1.5\,\mu$G, assuming a density of $10^4\,\mathrm{cm^{-3}}$. 

Assuming these recent estimates of the magnetic field strength in OMC1 to be accurate and applicable to our positions, we conclude that the extra line broadening is highly sub-Alfv\'enic in the neutral gas ($\mathcal{M}_{\rm A} \ll 1$).

\subsubsection{Line broadening in the ionized gas}
\label{Line-broadening-ionized}

For hydrogen at a temperature of $T=10^4$ K, the thermal line width is $\Delta v_{\mathrm{FWHM}} \approx 21.4\,\mathrm{km\,s^{-1}}$. For helium in thermal equilibrium with the hydrogen gas, the thermal line width is $\Delta v_{\mathrm{FWHM}} \approx 10.7\,\mathrm{km\,s^{-1}}$.

If we assume that hydrogen and helium are in thermal equilibrium and that the turbulence is equal in the hydrogen and helium gas, we can solve the two equations for the total line widths of the hydrogen and helium lines, $\Delta v^2 = \Delta v_{\rm th}^2 + \Delta v_{\rm turb}^2$ (Eq. \ref{eq.delta-v}), for the gas temperature and the turbulent velocity. This way, we derive a temperature of the ionized gas and a turbulent velocity of $1.1\times 10^4\,\mathrm{K}$ and  $14.0\,\mathrm{km\,s^{-1}}$ toward the Orion Bar, $0.99\times 10^4\,\mathrm{K}$ and $12.5\,\mathrm{km\,s^{-1}}$ toward BNKL-PDR, $1.0\times 10^4\,\mathrm{K}$ and $14.2\,\mathrm{km\,s^{-1}}$ toward TRAP-PDR, $0.77\times 10^4\,\mathrm{K}$ and $21.0\,\mathrm{km\,s^{-1}}$ toward EAST-PDR. In ORI-P3, however, the observed line widths are in good agreement with sole thermal broadening in a gas with $T\approx 0.9\times 10^4\,\mathrm{K}$.

These assumptions may not be exactly valid. We observe the centroid velocities of the helium RRLs to be more blueshifted with respect to the hydrogen RRLs in the positions toward OMC1 (including ORI-P1) and ORI-P5 in the EON. The biggest blueshift is observed in ORI-P5 with $-3.4\,\mathrm{km\,s^{-1}}$. Also in ORI-P3 we observe a significant blueshift of $-2.1\,\mathrm{km\,s^{-1}}$. It can be expected from dynamic modeling of the Huygens region that helium RRLs arise from gas closer to the stars (since its ionization potential is higher than that of hydrogen, ``the High Ionization Zone'') and hence be more blueshifted \citep{Henney2003, ODell2023}. On the other hand, according to recent observations with the JWST, the He Str\"omgren sphere is observed to be slightly smaller than the H Str\"omgren sphere \citep[$5\times10^{-3}\,\mathrm{pc}$][]{Peeters2023}. As the gas density will decrease as the gas expands into the Huygens Region, and the recombination lines will be strongly weighted with density, this demonstrates that the ionized gas is rapidly accelerated once it leaves the ionization front to about the sound speed over $5\times10^{-3}\,\mathrm{pc}$ – as indicated by the velocity of the HRRL – and that further acceleration only adds about $3.4\,\mathrm{km\,s^{-1}}$ to this, as indicated by the HeRRL. Further dynamical modeling taking the full geometry of M42 H{\sc ii} region and the stellar bubble into account may be very insightful in the dynamical feedback of this region.

In the ionized gas of OMC1, the magnetic field would need to be rather strong to produce the extra line broadening, about $400\,\mu$G. In the light of this, the assumption that the extra line broadening is only due to Alf\'en waves becomes unreasonable, as the frozen-in magnetic field should rather increase in strength from the ionized gas to the denser PDR. Rather, turbulence in the ionized gas is slightly super-Alfv\'enic ($\mathcal{M}_{\rm A} \gtrsim 1$): The Alfv\'en speed in the ionized gas is at most about $15\,\mathrm{km\,s^{-1}}$, while the extra line broadening we observe in OMC1 is between 12.5 and $21.0\,\mathrm{km\,s^{-1}}$.

\subsubsection{Line broandening in the molecular gas}

The $^{13}$CO line in BNKL-PDR is characterized by very broad wings, that are due to the strong outflows from this region. The [C\,{\sc ii}] line toward this position also shows deviation from Gaussianity, with broad wings, albeit less than the $^{13}$CO lines. In contrast, the RRLs seem to be even less affected by this line-broadening effect, although we note that this is hard to judge as the carbon RRL is blended with the helium RRL and a blue weaker component at $v_{\mathrm{LSR}}\simeq 1\,\mathrm{km\,s^{-1}}$, corresponding with a [C\,{\sc ii}] component. Likely the $^{13}$CO and [C\,{\sc ii}] lines consist of broad wings from shocked gas and a narrow component from the PDR, that emits in the carbon RRL.

The temperature of the molecular gas can be estimated from the peak temperature of the optically thick $^{12}$CO line by
\begin{align}
T_{\rm P}(^{12}\mbox{CO}) = J(T_{\rm ex}) - J(T_{\rm bg}),
\end{align}
with the equivalent brightness temperature of a black body at $T$ $J(T) = (h\nu/k_{\rm B}) / (e^{h\nu/k_{\rm B}T} -1)$, the excitation temperature of the $^{12}$CO line $T_{\rm ex}$ and the cosmic background temperature $T_{\rm bg} = 2.7\,\mathrm{K}$. The excitation temperature of the CO line is a good lower limit on the kinetic temperature in the CO gas. For the $^{12}$CO(J=2-1) line, $h\nu/k_{\rm B} = 11.07\,\mathrm{K}$. If $h\nu/k_{\rm B}T \ll 1$, $T_{\rm P} \simeq T_{\rm ex} \leq T_{\rm kin}$. In this manner, we obtain estimates for the gas temperature in the CO gas of 100 to 130\,K in OMC1, and 40 to 60\,K in positions outside OMC1. We use these temperatures to determine the thermal line widths of the $^{13}$CO lines.

In molecular gas with a temperature of $T=100$ K, the thermal line width of $^{13}$CO is $\Delta v_{\mathrm{FWHM}} \approx 0.2\,\mathrm{km\,s^{-1}}$. Hence, the line widths we observe here are heavily dominated by the extra line broadening. Generally the line widths are somewhat smaller than those of the corresponding [C\,{\sc ii}] lines, between 1 and $2\,\mathrm{km\,s^{-1}}$ (the exception being BNKL-PDR, of course). Assuming an electron abundance of $x_{\rm e}\sim 10^{-5}$ and a density of $3\times 10^5\,\mathrm{cm^{-3}}$, the magnetic field in the CO gas can be at most $3\,\mu\mathrm{G}$ to produce a corresponding Alfv\'en velocity. If the density increases from the surface layer of the PDR to the molecular gas, the magnetic field can be higher. \cite{Berne2014} conclude that the magnetic field is important in stabilizing the molecular cloud against gravitational collapse, but cite higher values of the magnetic field strength \citep[50 to 250$\,\mu$G][]{Abel2004, Brogan2005}. Hence, in the molecular gas, turbulence is highly sub-Alfv\'enic ($\mathcal{M}_{\rm A} \ll 1$).

The $^{13}$CO lines in OMC1 are not consistently broader than those in regions more distance from the Trapezium stars, which suggests that the cloud is intrinsically turbulent and not much energy is transmitted deep into the cloud. The only exceptions are the observed lines in ORI-P6 that are noticeably narrower than in the other positions. OMC3 is a much more quiescent region than OMC1, M43 and the EON, with still significant star formation (\cite{Megeath2012,Megeath2016}, see Fig. 11 in \cite{Pabst2021}). This suggests that at least some turbulence is caused by the presence of massive stars, but our sample is too small to draw definite conclusions. \cite{Berne2014} conclude that only 0.01\% of the radiative energy injected by stars ($E_{\star} \simeq 5\times 10^{51}\,\mathrm{erg}$) is transferred into the molecular cloud as turbulent kinetic energy, $E_{\rm k}\simeq 3.6\times 10^{47}\,\mathrm{erg}$. A large amount, 50\%, of the injected wind energy ($E_{\rm w} \simeq 5\times 10^{48}\,\mathrm{erg}$) is transferred into the large-scale expansion of the Veil Shell, \mbox{$E_{\rm kin} \simeq 2.5\times 10^{48}\,\mathrm{erg}$} \citep{Pabst2020}.

\subsubsection{Connecting the dots}
\label{Sec.dots}

\begin{table*}[htb]
\centering
\caption{Turbulent line widths and Mach numbers of [$^{13}$C\,{\sc ii}], C$n\alpha$, and $^{13}$CO lines.}
\renewcommand{\arraystretch}{1.1}
\addtolength{\tabcolsep}{0pt}
\begin{tabular}{lcccccc}
\hline\hline
position & $\sigma_{\mathrm{turb}}$([$^{13}$C\,{\sc ii}]) & $\mathcal{M}_{\mathrm{turb}}$([$^{13}$C\,{\sc ii}]) & $\sigma_{\mathrm{turb}}(\mathrm{C}n\alpha)$ & $\mathcal{M}_{\mathrm{turb}}(\mathrm{C}n\alpha)$ & $\sigma_{\mathrm{turb}}(\mathrm{^{13}CO})$ & $\mathcal{M}_{\mathrm{turb}}(\mathrm{^{13}CO})$ \\ 
 & $[\mathrm{km\,s^{-1}}]$ &  & $[\mathrm{km\,s^{-1}}]$ &  & $[\mathrm{km\,s^{-1}}]$ &  \\ \hline
BAR-INSIDE & $0.8\pm 0.2$ & $0.9\pm 0.3$ & $1.2\pm 0.1$ & $1.3\pm 0.1$ & $1.1\pm 0.1$ & $2.2\pm 0.1$ \\
BNKL-PDR & $0.5\pm 0.5$ & $0.6\pm 0.5$ & $1.6\pm 0.1$ & $1.8\pm 0.1$ & $3.5\pm 0.1$ & $6.9\pm 0.2$ \\
TRAP-PDR & $1.2\pm 0.4$ & $1.1\pm 0.4$ & $2.0\pm 0.1$ & $1.8\pm 0.1$ & $1.8\pm 0.1$ & $3.6\pm 0.1$ \\
EAST-PDR-2 & $1.5\pm 1.5$ & $1.7\pm 1.7$ & $1.4\pm 0.1$ & $1.7\pm 0.1$ & $1.3\pm 0.1$ & $2.9\pm 0.1$ \\
ORI-P1-1 & $1.1\pm 0.5$ & $1.9\pm 0.9$ & $1.1\pm 0.1$ & $1.9\pm 0.2$ & $0.7\pm 0.1$ & $2.4\pm 0.1$ \\
ORI-P1-2 & $0.7\pm 0.6$ & $0.9\pm 0.8$ & $1.2\pm 0.2$ & $1.6\pm 0.3$ & $2.4\pm 0.1$ & $7.3\pm 0.3$ \\
ORI-P2 & $1.8\pm 0.6$ & $2.6\pm 0.8$ & $1.3\pm 0.1$ & $1.8\pm 0.2$ & $1.2\pm 0.1$ & $4.6\pm 0.1$ \\
ORI-P3-2 & $2.3\pm 1.4$ & $3.3\pm 2.0$ & $2.8\pm 0.9$ & $4.0\pm 1.2$ & $0.9\pm 0.1$ & $3.2\pm 0.4$ \\
ORI-P4 & $1.2\pm 0.7$ & $1.7\pm 1.0$ & $2.6\pm 1.3$ & $3.7\pm 1.9$ & $0.8\pm 0.1$ & $2.3\pm 0.1$ \\
ORI-P5 & $2.0\pm 0.9$ & $2.6\pm 1.2$ & $1.1\pm 0.6$ & $1.4\pm 0.7$ & $1.0\pm 0.1$ & $3.7\pm 0.3$ \\
ORI-P6 & $0.9\pm 0.2$ & $1.3\pm 0.2$ & $0.7\pm 0.1$ & $1.1\pm 0.1$ & $0.6\pm 0.1$ & $1.9\pm 0.1$ \\ \hline
\end{tabular}
\tablefoot{Temperatures for the [$^{13}$C\,{\sc ii}] and C$n\alpha$ line from Bayesian analysis in Tab. \ref{tab.ne-Te}; in ORI-P3, we use $T=140\,\mathrm{K}$ as for ORI-P2 and ORI-P4. Temperatures for the $^{13}$CO line from peak temperature of $^{12}$CO line.}
\label{Tab.v-turb}
\end{table*}

\begin{table*}[htb]
\centering
\caption{Turbulent pressures, using results of Bayesian analysis in Table \ref{tab.ne-Te}.}
\renewcommand{\arraystretch}{1.1}
\addtolength{\tabcolsep}{0pt}
\begin{tabular}{lcccccc}
\hline\hline
 & [$^{13}$C\,{\sc ii}] & & C$n\alpha$ & & $^{13}$CO & \\
position & $p_{\mathrm{turb}}/k_{\rm B}$ & $p_{\mathrm{turb}}/p_{\mathrm{th}}$ & $p_{\mathrm{turb}}/k_{\rm B}$ & $p_{\mathrm{turb}}/p_{\mathrm{th}}$ & $p_{\mathrm{turb}}/k_{\rm B}$ & $p_{\mathrm{turb}}/p_{\mathrm{th}}$ \\
 & [$\mathrm{K\,cm^{-3}}$] &  & [$\mathrm{K\,cm^{-3}}$] &  & [$\mathrm{K\,cm^{-3}}$] & \\ \hline
BAR-INSIDE & $2\times 10^7$ & 0.8 & $5\times 10^7$ & 1.6 & $4\times 10^7$ & 2.6 \\
BNKL-PDR & $2\times 10^7$ & 0.4 & $2\times 10^8$ & 3.0 & -- & -- \\
TRAP-PDR & $3\times 10^8$ & 1.1 & $1\times 10^9$ & 3.2 & $8\times 10^8$ & 7.2 \\
EAST-PDR-2 & $8\times 10^7$ & 2.9 & $7\times 10^7$ & 2.6 & $6\times 10^7$ & 4.5 \\
ORI-P1-1 & $4\times 10^6$ & 3.4 & $4\times 10^6$ & 3.3 & $2\times 10^6$ & 3.3 \\
ORI-P1-2 & $6\times 10^6$ & 0.8 & $2\times 10^7$ & 2.5 & $8\times 10^7$ & 28 \\
ORI-P2 & $1\times 10^7$ & 6.6 & $6\times 10^6$ & 3.1 & $5\times 10^6$ & 11 \\
ORI-P3-2 & -- &  -- & -- & -- & -- & -- \\
ORI-P4 & $8\times 10^6$ & 2.9 & $3\times 10^7$ & 13 & $3\times 10^6$ & 3.0 \\
ORI-P5 & $1\times 10^7$ & 6.5 & $5\times 10^6$ & 1.9 & $4\times 10^6$ & 7.5 \\
ORI-P6 & $1\times 10^6$ & 1.7 & $9\times 10^5$ & 1.1 & $5\times 10^5$ & 1.9 \\
\end{tabular}
\label{Tab.turbulent-pressure}
\end{table*}

Comparing the turbulent line widths in Table \ref{Tab.v-turb}, we conclude that the turbulent line widths of each gas phase are about equal in BAR-INSIDE, EAST-PDR-2, ORI-P1-1, and ORI-P6. In the other positions/components, the line widths in the different phases vary a great deal with respect to one another. In BNKL-PDR this can be due to the fact the molecular outflows and associated shocks affects mostly the molecular lines and lines close to the molecular phase. In TRAP-PDR, the [$^{13}$C\,{\sc ii}] line has somewhat lower turbulent line width than the C$n\alpha$ and $^{13}$CO line. In ORI-P1-2, the turbulent line widths increase from the neutral to the molecular phase, while in ORI-P1-1, ORI-P5 and ORI-P6, the turbulent line widths decrease going from the neutral to the molecular phase. ORI-P2 and ORI-P4 show unordered behavior, with the turbulent line width in the C$n\alpha$ line largest, which may be due to the faintness of the CRRLs. In ORI-P3(-2), the C$n\alpha$ line may stem not from the neutral gas, but from the ionized gas, as discussed earlier.

We can also calculate the turbulent Mach numbers $\mathcal{M}_{\mathrm{turb}}$ of the different layers. The one-dimensional turbulent Mach number $\mathcal{M}_{\mathrm{turb,z}}$ is the ratio of the velocity dispersion $\sigma_{\mathrm{turb}}$ over the sound speed. The total Mach number is given by $\mathcal{M}_{\mathrm{turb}} = \sqrt{3}\mathcal{M}_{\mathrm{turb,z}}$ if the turbulence is isotropic \citep{Orkisz2017}. For interstellar gas the sound speed is given by $c_s \simeq \sqrt{\gamma p_{\mathrm{th}}/\rho} = \sqrt{\gamma k_{\rm B} T/m}$, where $p_{\mathrm{th}}$ is the thermal pressure, $\rho$ is the mass density of the gas, $\gamma$ is the adiabatic index, which is 5/3 for a monatomic ideal gas and 7/5 for a diatomic ideal gas, $T$ is the temperature, and $m$ is the atomic or molecular weight of the gas.

In ionized gas with a temperature of $T=10^4\,\mathrm{K}$, the sound speed is about $c_s \simeq 14\,\mathrm{km\,s^{-1}}$. In neutral atomic hydrogen at a temperature of $T=300\,\mathrm{K}$, the sound speed is approximately $c_s \simeq 2.0\,\mathrm{km\,s^{-1}}$, while at $T=100\,\mathrm{K}$ it is $c_s \simeq 1.2\,\mathrm{km\,s^{-1}}$. If hydrogen is molecular with $T=100\,\mathrm{K}$, $c_s \simeq 0.8\,\mathrm{km\,s^{-1}}$. If half of the hydrogen nuclei are in atomic gas and half are locked up in molecular hydrogen, the sound speed can be computed to be effectively $c_s \simeq 1.0\,\mathrm{km\,s^{-1}}$. In molecular hydrogen gas with $T=30\,\mathrm{K}$, the sound speed is $c_s \simeq 0.4\,\mathrm{km\,s^{-1}}$.

We assume that the hydrogen nuclei fractions in atomic and molecular form, respectively, are each 0.5 in the neutral gas (i.e., $f(\mathrm{HI}) = f(\mathrm{H}_2) = 0.5$). In BAR-INSIDE, the turbulence in the neutral gas is slightly subsonic. In BNKL-PDR, the turbulence is subsonic in [$^{13}$C\,{\sc ii}], but supersonic in C$n\alpha$, while in TRAP-PDR and EAST-PDR it is slightly supersonic. In the positions outside OMC1 turbulence in the neutral gas is supersonic. In ORI-P6 the turbulent Mach number is lowest. In the molecular gas Mach numbers are significantly higher than in the neutral gas, all positions are supersonic (in BNKL-PDR the high Mach number is caused by the molecular outflows). Plotting the turbulent line widths and Mach numbers, respectively, of the three PDR tracers ([$^{13}$C\,{\sc ii}], C$n\alpha$, $^{13}$CO) against each other does not reveal an overall correlation of the turbulent properties between the different tracers/PDR layers, nor do Pearson correlation coefficients. In general, we observe that the Mach numbers increase from the neutral to the molecular gas, albeit by very different factors. In the ionized gas in OMC1, turbulence is slightly subsonic with Mach numbers between 0.7 and 1.0.

Table \ref{Tab.turbulent-pressure} summarizes the turbulent pressures and the ratio of the turbulent pressure over the thermal pressure in the observed PDRs, derived from the Bayesian analysis of our line intensities. For the $^{13}$CO lines we assume the same density as for the [C\,{\sc ii}]/CRRL layer and the excitation temperature of the $^{12}$CO line as a lower limit for the gas temperature. The turbulent pressure, $p_{\mathrm{turb}} = \rho \sigma_{\mathrm{turb}}^2$, in the [C\,{\sc ii}]/CRRL layer is usually on the order of the thermal pressure, albeit somewhat larger within a factor of 3. Only in ORI-P2 and ORI-P5 the turbulent pressure in [$^{13}$C\,{\sc ii}] is 6 to 7 times larger than the thermal pressure, and in ORI-P4 the turbulent pressure in C$n\alpha$ is 13 times larger than the thermal pressure. In the molecular gas, the turbulent pressure dominates over the thermal pressure with factors between 1.9 (ORI-P6) and 28 (ORI-P1-2) and is usually within a factor of 2 of the turbulent pressure in the CRRL layer (except in ORI-P4, where it is somewhat lower).

Turning to the ionized gas, assuming a density of $5\times 10^3\,\mathrm{cm^{-3}}$ for the ionized gas \citep{ODellHarris2010}, the turbulent pressure in the ionized gas is on the order of $5\times 10^7\,\mathrm{K\,cm^{-3}}$, thus on the order of the thermal pressure in the ionized gas and the turbulent pressure in the PDRs in OMC1. In M43, the turbulent pressure in the ionized gas is negligible, the H$n\alpha$ and He$n\alpha$ lines are in first order consistent with thermal broadening alone (cf. Sec. \ref{Line-broadening-ionized}). In NGC 1977, \cite{Pabst2020} report a gas density of the ionized gas of $n\simeq 40\,\mathrm{cm^{-3}}$ and a width of the [C\,{\sc ii}] line of $\Delta v_{\mathrm{FWHM}}\simeq 11.6\,\mathrm{km\,s^{-1}}$. With these givens, assuming a gas temperature of $T\sim 10^4\,\mathrm{K}$, the turbulent pressure can be computed to be $p_{\mathrm{turb}}\simeq 2\times 10^5\,\mathrm{K\,cm^{-3}}$, which is a factor of 5 lower than the thermal pressure of the ionized gas and the turbulent pressure in the PDR of OMC3.

Estimates from [C\,{\sc ii}] observations in \cite{Pabst2020} suggest that the thermal, turbulent and magnetic pressures in the Veil Shell and the PDR of M43 are in approximate equipartition, while the thermal pressure is an order of magnitude larger than the turbulent and magnetic pressure in the Orion Bar and the PDR of NGC 1977. The radiation pressure in the Veil Shell and M43 is about equal to the other pressure terms, while it is an order of magnitude lower in the Orion Bar and the PDR of NGC 1977. The Veil Shell and M43 are very dynamic regions that expand rather rapidly (with $13\,\mathrm{km\,s^{-1}}$ and $6\,\mathrm{km\,s^{-1}}$, respectively), while NGC 1977 expands more slowly (with $1.5\,\mathrm{km\,s^{-1}}$). Possibly the still rapid expansion of the Veil Shell and M43 injects turbulence into the interstellar medium, while the shell of NGC 1977 interacts more quiescently in a Spitzer-type expansion and has aged sufficiently to have let turbulence be dissipated away. The time scale for turbulent dissipation may be estimated by $\tau \sim l/c_{\rm s}$ \citep{Kolmogorov1941, ElmegreenScalo2004}; if the characteristic length scale is $l\sim 1\,\mathrm{pc}$ and the sound speed $c_{\rm s} \sim 1\,\mathrm{km\,s^{-1}}$, the dissipation timescale is $\tau \sim 10^6\,\mathrm{yr}$, which is on the order of the age of NGC 1977, $0.4\times 10^6\,\mathrm{yr}$. The Orion Bar is heavily irradiated by the nearby Trapezium stars and the high turbulence may be a result of flows caused by the strong radiation field and pressure gradients in this region.

This same behavior is reflected in our samples, where the ratios of the turbulent pressure over the thermal pressure in positions in OMC1 is smaller than in other positions. We note that we derive rather low temperatures of the PDRs in OMC1 compared to earlier studies: despite exhibiting high turbulent pressure, the thermal pressure in the PDRs in OMC1 may be still dominating the total pressure in the PDRs. The turbulent pressure we derive in M43 (ORI-P2 and ORI-P4) is an order of magnitude higher than in \cite{Pabst2020} and larger than the thermal pressure. Those two positions lie on the edges of M43, though, while \cite{Pabst2020} look at the central background PDR. Possibly, turbulence is increased where the gas can freely stream away. We derive a ten times higher turbulent pressure in OMC3 than \cite{Pabst2020} derive for the PDR associated with the large shell of NGC 1977, rendering the thermal pressure and the turbulent pressure in OMC3 in approximate equipartition.

In conclusion, at the spatial scales our observations probe we observe that turbulence in the Orion A molecular cloud is highly supersonic in the molecular gas, but sub-Alfv\'enic, varying between slightly sub- and supersonic and sub- and super-Alfv\'enic in the neutral gas, but slightly subsonic and highly super-Alfv\'enic in the ionized gas. The turbulent pressure dominates the pressure balance in the neutral and molecular gas, but plays a minor role compared to the thermal pressure in the ionized gas.

The data presented here may help settle the flow of turbulent energy in regions of massive star formation. Since the gas phases border each other, it is likely that turbulence is transmitted from one gas phase into the other. In our case, the extra line broadening is larger in the ionized gas by an order of magnitude, hence it is unlikely that it stems from the turbulence of the molecular gas. We cannot exclude the possibility of Alfv\'en waves being responsible for the extra line broadening in the ionized gas, but turbulent motions from gas flows are likely to contribute in this setting. How magnetic fields and turbulence are passed on between the phases of the ISM is of yet unknown. Another obvious contribution to the line width is the dynamic structure of the observed regions, blister H\,{\sc ii} regions (OMC1) and thermally expanding H\,{\sc ii} regions (M43). However, lines are nearly Gaussian, not giving away much hints on the dynamic origin of the line width.

An alternative way to study the flow of turbulent energy in star-forming regions, is to statistically analyze the full data cubes using structure functions. However, the scale at which turbulence is important is comparable to the spatial resolution of our [C\,{\sc ii}] and CO data cubes (for the ionized gas \cite{Arthur2016} gives a turbulent scale of $22\arcsec$; \cite{ODell2001} gives an average turbulent scale of $18\arcsec$, depending on emission line), while the larger-scale morphology of the region becomes important at our resolutions and the radial velocity distribution in each spectrum is complicated. For this approach higher-resolution data cubes of lines from neutral atomic PDR gas are needed, which may be available with the next generation of far-infrared and radio telescopes.

It is generally acknowledged that molecular clouds are strongly turbulent, stabilizing the cloud against collapse, usually interpreted as the result of a turbulent cascade from cloud scales to filament scales \citep[e.g.,][]{Hennebelle2012,Federrath2016, Padoan2016}. Gas flows caused by local energy injection in the interstellar medium are believed to be turbulent, while also gravitational collapse can induce turbulence \citep{Heigl2020}. On galactic scales, gravity may be the dominant source of turbulence in molecular clouds \citep{Krumholz2016}. If the line broadening is caused by turbulence, there may be two reasons for turbulence in the molecular gas: either that the $^{13}$CO lines are broadened by turbulence that is transmitted from the ionized gas through the neutral gas to the molecular gas, or else that the molecular clouds are intrinsically turbulent which is transmitted to the neutral gas and exacerbated by photoevaporation on the surface. In the case of the Orion B molecular cloud, the compression of the molecular gas caused by H\,{\sc ii} regions leads to the formation of filaments and thus increased star formation \citep{Gaudel2023}, while turbulence in the molecular gas is highly supersonic \citep{Orkisz2017}.

\section{Conclusion}

We have analyzed multiple hydrogen, helium and carbon mm-wave RRLs observed with the Yebes 40m telescope toward 10 representative positions in the Orion Nebula complex, including M43 and OMC3. Comparing with the [C\,{\sc ii}] $158\,\mu\mathrm{m}$ line as a PDR surface tracer and the $^{13}$CO (2-1) line as a tracer of the molecular gas deeper in the PDR, we observe slight velocity shifts between the main components of these two tracers and the carbon RRLs, that cannot always be attributed to geometry, but may indicate intrinsic shear. The hydrogen RRL emitting ionized gas in OMC1 is streaming away from the PDR surface in the background at a velocity of $12\text{--}15\,\mathrm{km\,s^{-1}}$, while in M43 the streaming velocity is $3.5\,\mathrm{km\,s^{-1}}$.

Combining the carbon RRL intensities with the intensities of the [$^{12}$C\,{\sc ii}] main component and its [$^{13}$C\,{\sc ii}] $F$=2-1 satellite, we infer electron temperatures and electron densities. This approach has its shortcomings, as the three lines originate from slightly different PDR layers, which is reflected in different observed line parameters (peak velocity and line width). We obtain robust estimates, that allow us to compare the thermal pressure in the observed PDRs with other pressure terms (turbulent, magnetic) and the pressure in the ionized gas. Besides in the position southeast of the Orion Bar, where the expansion of the Veil is driven by the contained hot plasma, the PDR and the ionized gas are in approximate pressure equilibrium.

In OMC1, the RRLs show that helium is ionized, while in M43 only a quarter of the amount of helium is ionized. This reflects the difference in radiation field, of a O7V star in OMC1 and a B0.5V star in M43. In OMC3, we do not observe ionized gas. Toward two positions in OMC1, BNKL-PDR and TRAP-PDR, we detect very faint lines that can be attributed to RRLs of C$^+$ and/or O$^+$ (X\,{\sc ii}$n\alpha$ RRLs), stemming from the ionized layers in OMC1.

The observed lines, whether H, He, or C RRLs, are significantly broader than thermal broadening would allow. The only exception is M43, where we attribute the observed carbon RRL to the ionized gas instead of the background PDR due to its peak velocity and line width. Here, the line widths are consistent with thermal broadening alone. In the other positions we explore the possibility that the extra line broadening is due to Alfv\'en waves. We conclude that the extra line broadening is super-Alfv\'enic and subsonic in the ionized gas, Alfv\'enic and varying between slightly sub- and supersonic in the neutral gas at the H/H$_2$ transition, but sub-Alfv\'enic and supersonic in the molecular gas. Hence, we conclude that the extra line broadening is mainly due to interstellar turbulence that is non-Alfv\'enic, while the precise origin of such turbulence remains unknown.

By observing RRLs of different elements and in different ionization states, future telescopes such as the SKA will be able to resolve much finer structures, which will result in a refinement of the estimates of the physical conditions in exquisite detail, in Orion as well in more distant massive star forming regions.

\begin{acknowledgements}
We thank the anonymous referee for valuable comments that helped improve the manuscript. We thank C.R. O'Dell for valuable discussions on the properties of OMC1.

This work is based in part on observations made with the NASA/DLR Stratospheric Observatory for Infrared Astronomy (SOFIA). SOFIA was jointly operated by the Universities Space Research Association, Inc. (USRA), under NASA contract NNA17BF53C, and the Deutsches SOFIA Institut (DSI) under DLR contract 50 OK 0901 to the University of Stuttgart.

CHMP and JRG thank the Spanish MCINN for funding support under grant PID2019-106110GB-100. CHMP thanks the Dutch NWO for funding support through a Rubicon grant No. 13257.
\end{acknowledgements}

\bibliographystyle{aa}
\bibliography{RRLs_Orion}

\begin{appendix}

\section{Observed positions on SOFIA/upGREAT [C\,{\sc ii}] and {\it Spitzer}/IRAC4 8\,$\mu$m maps}
\label{App.positions}

\begin{figure*}[htp]
\centering
\includegraphics[width=1.0\textwidth, height=0.5\textwidth]{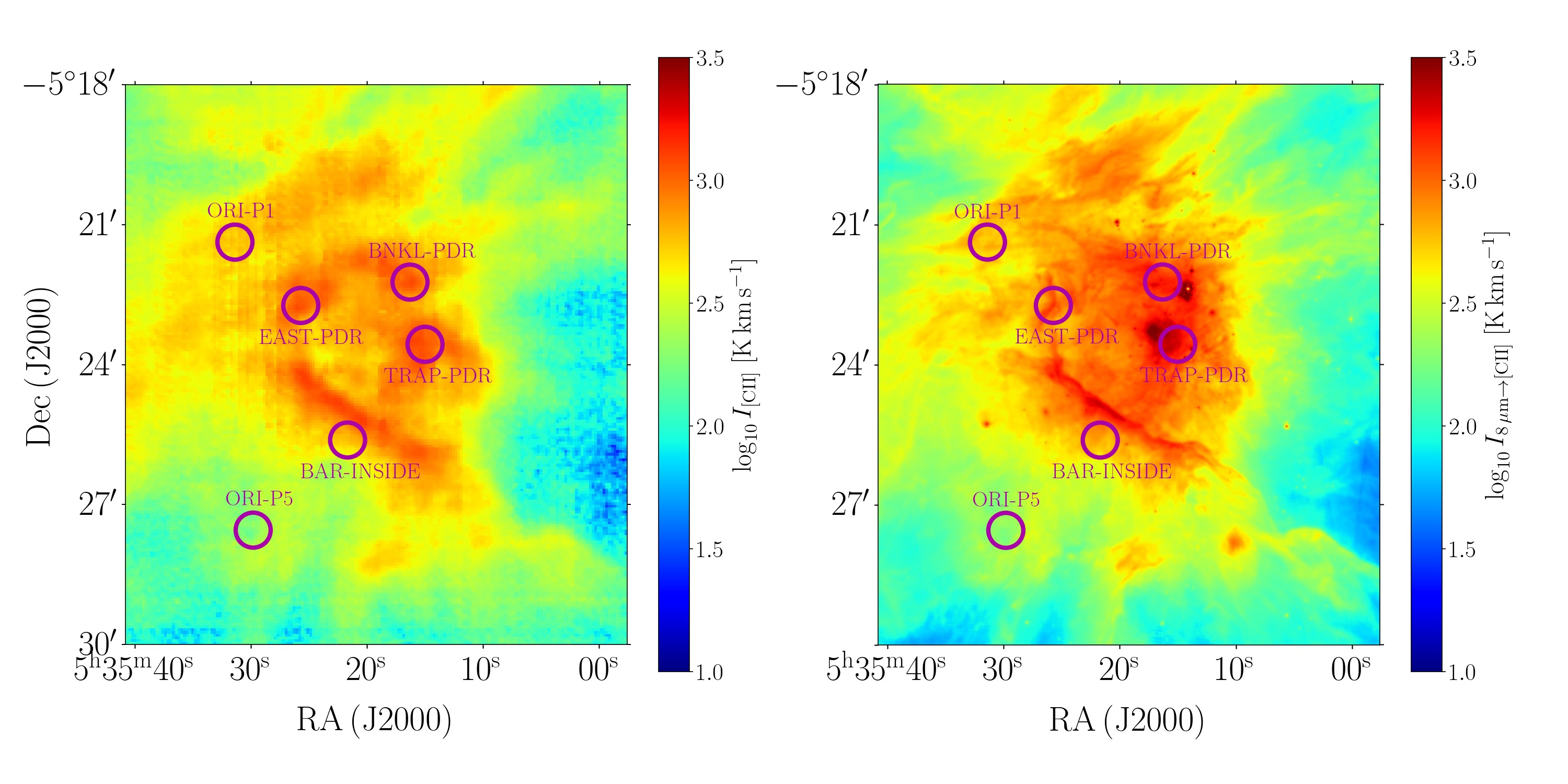}
\caption{Zoom into OMC1. {\it Left:} SOFIA/upGREAT [C\,{\sc ii}] intensity at $16\arcsec$. {\it Right:} {\it Spitzer}/IRAC4 8\,$\mu$m intensity converted to [C\,{\sc ii}] intensity at $1.9\arcsec$. Observed positions in the region of OMC1 (BAR-INSIDE, BNKL-PDR, TRAP-PDR, EAST-PDR, ORI-P1, and ORI-P5) are indicated as purple circles, that represent the average $45\arcsec$ beam size of our Yebes 40m telescope observations.}
\label{Fig.orion_omc1_sofia_irac4}
\end{figure*}

\begin{figure*}[htp]
\centering
\includegraphics[width=1.0\textwidth, height=0.5\textwidth]{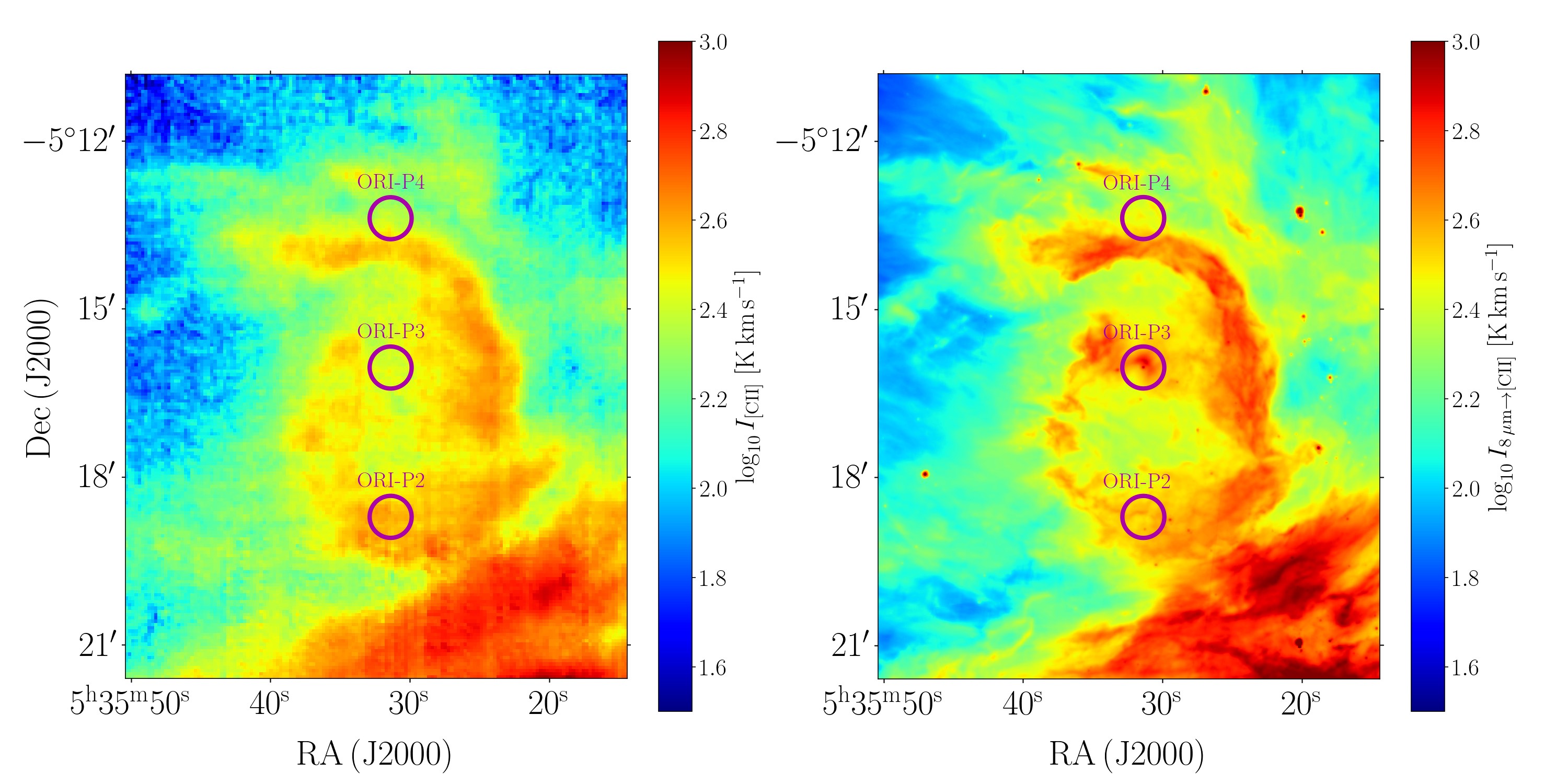}
\caption{Zoom into M43. {\it Left:} SOFIA/upGREAT [C\,{\sc ii}] intensity at $16\arcsec$. {\it Right:} {\it Spitzer}/IRAC4 8\,$\mu$m intensity converted to [C\,{\sc ii}] intensity at $1.9\arcsec$. Observed positions in the region of M43 (ORI-P2, ORI-P3, and ORI-P4) are indicated as purple circles, that represent the average $45\arcsec$ beam size of our Yebes 40m telescope observations.}
\label{Fig.orion_m43_sofia_irac4}
\end{figure*}

\begin{figure*}[htp]
\centering
\includegraphics[width=1.0\textwidth, height=0.5\textwidth]{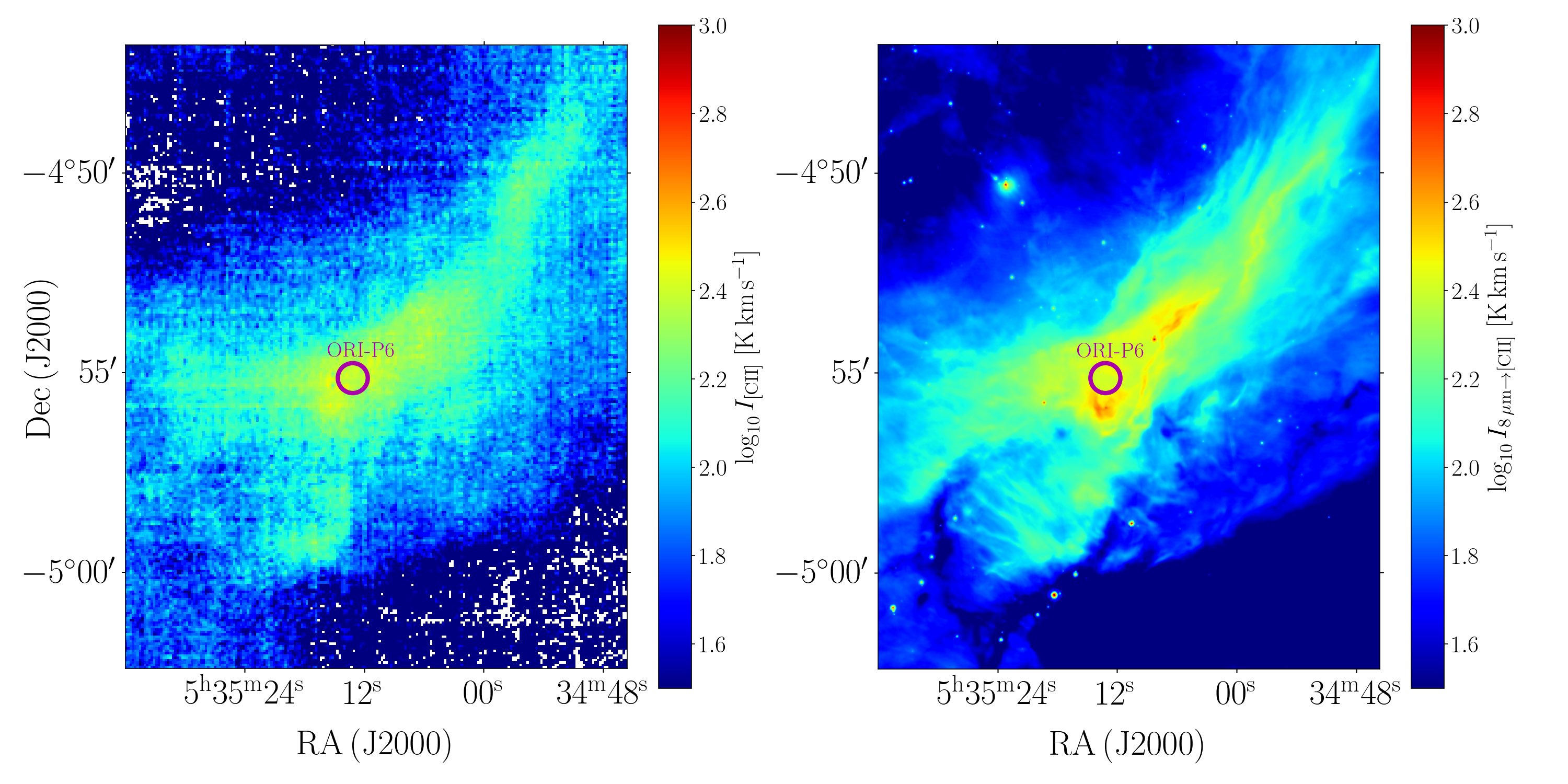}
\caption{Zoom into OMC3. {\it Left:} SOFIA/upGREAT [C\,{\sc ii}] intensity at $16\arcsec$. {\it Right:} {\it Spitzer}/IRAC4 8\,$\mu$m intensity converted to [C\,{\sc ii}] intensity at $1.9\arcsec$. Observed positions in the region of OMC3 (ORI-P6) are indicated as purple circles, that represent the average $45\arcsec$ beam size of our Yebes 40m telescope observations.}
\label{Fig.orion_omc3_sofia_irac4}
\end{figure*}

Figures \ref{Fig.orion_omc1_sofia_irac4} to \ref{Fig.orion_omc3_sofia_irac4} show the average beam (FWHM = $45\arcsec$) toward the observed positions on maps of the SOFIA/upGREAT [C\,{\sc ii}] intensity and the {\it Spitzer}/IRAC4 8\,$\mu$m intensity converted to [C\,{\sc ii}] intensity using the relations provided by \cite{Pabst2022}, zoomed in to discern the structure that is averaged in each beam, in their original resolution (SOFIA: FWHM = $16\arcsec$, {\it Spitzer}: FWHM = $1.9\arcsec$). As the IRAC4 8\,$\mu$m image reveals, each beam contains finer structure, that is also lost in the upGREAT [C\,{\sc ii}] image. Future telescopes (such as the SKA and far-infrared heterodyne receivers on large telescopes) can provide the spatial resolution necessary to resolve the structure in velocity-resolved observations, which will lead to much refined estimates of the physical conditions in the observed regions.

\clearpage

\section{Line fits to C$n\beta$, $\gamma$, $\delta$ and $\epsilon$ RRLs}
\label{App.CRRL-fits}

Tables B.1, B.2, B.3, and B.4 give the fit parameters of the C$n\beta$, $\gamma$, $\delta$ and $\epsilon$ RRLs toward the ten targeted positions.

\begin{table}[ht]
\caption{C$n\beta$ RRL line fits.}
\begin{tabular}{lccc}
\hline\hline
source & $T_{\mathrm{mb}}$ & $v_{\mathrm{LSR}}$ & $\Delta v_{\mathrm{FWHM}}$ \\
 & [$10^{-3}\mathrm{K}]$ & $[\mathrm{km\,s^{-1}}]$ & $[\mathrm{km\,s^{-1}}]$ \\ \hline
BAR-INSIDE & $26\pm 1$ & $10.5\pm 0.1$ & $2.8\pm 0.2$ \\
BNKL-PDR & $43\pm 2$ & $9.1\pm 0.1$ & $3.9\pm 0.2$ \\
TRAP-PDR & $37\pm 1$ & $9.0\pm 0.1$ & $5.2\pm 0.2$ \\
EAST-PDR-1 & $7\pm 2$ & $5.3\pm 0.1$ & $0.7\pm 0.3$ \\
EAST-PDR-2 & $22\pm 1$ & $10.9\pm 0.1$ & $3.6\pm 0.2$ \\
ORI-P1-1 & $15\pm 1$ & $6.8\pm 0.2$ & $3.1\pm 0.4$ \\
ORI-P1-2 & $7\pm 2$ & $10.0\pm 0.4$ & $2.4\pm 0.7$ \\
ORI-P2 & $6\pm 1$ & $8.6\pm 0.2$ & $2.4\pm 0.4$ \\
ORI-P3 & -- & -- & -- \\
ORI-P4 & -- & -- & -- \\
ORI-P5 & $3\pm 2$ & $10.2\pm 0.6$ & $4.5\pm 2.0$ \\
ORI-P6 & $6\pm 1$ & $11.7\pm 0.2$ & $2.8\pm 0.4$ \\ \hline
\end{tabular}
\end{table}

\begin{table}[ht]
\caption{C$n\gamma$ RRL line fits.}
\begin{tabular}{lccc}
\hline\hline
source & $T_{\mathrm{mb}}$ & $v_{\mathrm{LSR}}$ & $\Delta v_{\mathrm{FWHM}}$ \\
 & [$10^{-3}\mathrm{K}]$ & $[\mathrm{km\,s^{-1}}]$ & $[\mathrm{km\,s^{-1}}]$ \\ \hline
BAR-INSIDE & $12\pm 2$ & $10.8\pm 0.2$ & $2.4\pm 0.3$ \\
BNKL-PDR & $23\pm 2$ & $8.9\pm 0.2$ & $3.5\pm 0.3$ \\
TRAP-PDR & $16\pm 2$ & $8.5\pm 0.2$ & $4.9\pm 0.5$ \\
EAST-PDR-1 & -- & -- & -- \\
EAST-PDR-2 & $12\pm 1$ & $10.9\pm 0.2$ & $3.0\pm 0.4$ \\
ORI-P1-1 & $7\pm 2$ & $6.2\pm 0.4$ & $2.0\pm 0.8$ \\
ORI-P1-2 & $5\pm 2$ & $9.1\pm 0.7$ & $3.1\pm 1.5$ \\
ORI-P2 & $4\pm 2$ & $9.5\pm 0.3$ & $2.1\pm 0.7$ \\
ORI-P3 & -- & -- & -- \\
ORI-P4 & -- & -- & -- \\
ORI-P5 & -- & -- & -- \\
ORI-P6 & $5\pm 1$ & $11.8\pm 0.2$ & $2.0\pm 0.5$ \\ \hline
\end{tabular}
\end{table}

\begin{table}[ht]
\caption{C$n\delta$ RRL line fits.}
\begin{tabular}{lccc}
\hline\hline
source & $T_{\mathrm{mb}}$ & $v_{\mathrm{LSR}}$ & $\Delta v_{\mathrm{FWHM}}$ \\
 & [$10^{-3}\mathrm{K}]$ & $[\mathrm{km\,s^{-1}}]$ & $[\mathrm{km\,s^{-1}}]$ \\ \hline
BAR-INSIDE & $6\pm 2$ & $10.6\pm 0.2$ & $1.7\pm 0.4$ \\
BNKL-PDR & $11\pm 2$ & $9.1\pm 0.2$ & $3.4\pm 0.5$ \\
TRAP-PDR & $13\pm 2$ & $8.7\pm 0.4$ & $8.7\pm 0.9$ \\
EAST-PDR-1 & -- & -- & -- \\
EAST-PDR-2 & $9\pm 2$ & $10.9\pm 0.1$ & $1.1\pm 0.3$ \\
ORI-P1-1 & $3\pm 5$ & $6.0\pm 7.9$ & $1.4\pm 2.3$ \\
ORI-P1-2 & $3\pm 2$ & $0.7\pm 0.9$ & $1.2\pm 1.6$ \\
ORI-P2 & -- & -- & -- \\
ORI-P3 & -- & -- & -- \\
ORI-P4 & -- & -- & -- \\
ORI-P5 & -- & -- & -- \\
ORI-P6 & $3\pm 1$ & $11.3\pm 0.3$ & $1.8\pm 0.6$ \\ \hline
\end{tabular}
\end{table}

\begin{table}[ht]
\caption{C$n\epsilon$ RRL line fits.}
\begin{tabular}{lccc}
\hline\hline
source & $T_{\mathrm{mb}}$ & $v_{\mathrm{LSR}}$ & $\Delta v_{\mathrm{FWHM}}$ \\
 & [$10^{-3}\mathrm{K}]$ & $[\mathrm{km\,s^{-1}}]$ & $[\mathrm{km\,s^{-1}}]$ \\ \hline
BAR-INSIDE & $8\pm 2$ & $11.1\pm 0.2$ & $1.7\pm 0.3$ \\
BNKL-PDR & $9\pm 2$ & $9.2\pm 0.3$ & $4.1\pm 0.7$ \\
TRAP-PDR & $9\pm 2$ & $9.3\pm 0.7$ & $8.5\pm 1.6$ \\
EAST-PDR-1 & -- & -- & -- \\
EAST-PDR-2 & $4\pm 1$ & $10.6\pm 0.4$ & $3.9\pm 0.9$ \\
ORI-P1-1 & -- & -- & -- \\
ORI-P1-2 & -- & -- & -- \\
ORI-P2 & -- & -- & -- \\
ORI-P3 & -- & -- & -- \\
ORI-P4 & -- & -- & -- \\
ORI-P5 & -- & -- & -- \\
ORI-P6 & $4\pm 2$ & $11.7\pm 0.2$ & $1.1\pm 0.4$ \\ \hline
\end{tabular}
\end{table}

\clearpage

\section{Line fits to H$n\beta$, $\gamma$, $\delta$ and $\epsilon$ RRLs}
\label{App.HRRL-fits}

Tables C.1, C.2, C.3, and C.4 give the fit parameters of the H$n\beta$, $\gamma$, $\delta$ and $\epsilon$ RRLs toward the ten targeted positions.

\begin{table}[ht]
\caption{H$n\beta$ RRL line fits.}
\begin{tabular}{lccc}
\hline\hline
source & $T_{\mathrm{mb}}$ & $v_{\mathrm{LSR}}$ & $\Delta v_{\mathrm{FWHM}}$ \\
 & $[10^{-3}\mathrm{K}]$ & $[\mathrm{km\,s^{-1}}]$ & $[\mathrm{km\,s^{-1}}]$ \\ \hline
BAR-INSIDE & $154\pm 1$ & $-4.3\pm 0.1$ & $26.2\pm 0.1$ \\
BNKL-PDR & $651\pm 1$ & $-2.7\pm 0.1$ & $25.2\pm 0.1$ \\
TRAP-PDR & $809\pm 2$ & $-2.7\pm 0.1$ & $26.4\pm 0.1$ \\
EAST-PDR & $234\pm 1$ & $-2.8\pm 0.1$ & $28.1\pm 0.1$ \\
ORI-P1 & $62\pm 1$ & $3.1\pm 0.1$ & $25.0\pm 0.2$ \\
ORI-P2 & $11\pm 1$ & $5.2\pm 0.3$ & $31.2\pm 0.7$ \\
ORI-P3 & $64\pm 1$ & $6.7\pm 0.1$ & $20.5\pm 0.2$ \\
ORI-P4 & -- & -- & -- \\
ORI-P5 & $33\pm 1$ & $-4.1\pm 0.2$ & $26.3\pm 0.3$ \\
ORI-P6 & -- & -- & -- \\ \hline
\end{tabular}
\end{table}

\begin{table}[ht]
\caption{H$n\gamma$ RRL line fits.}
\begin{tabular}{lccc}
\hline\hline
source & $T_{\mathrm{mb}}$ & $v_{\mathrm{LSR}}$ & $\Delta v_{\mathrm{FWHM}}$ \\
 & $[10^{-3}\mathrm{K}]$ & $[\mathrm{km\,s^{-1}}]$ & $[\mathrm{km\,s^{-1}}]$ \\ \hline
BAR-INSIDE & $75\pm 1$ & $-4.0\pm 0.1$ & $26.3\pm 0.2$ \\
BNKL-PDR & $292\pm 1$ & $-2.6\pm 0.1$ & $25.4\pm 0.1$ \\
TRAP-PDR & $365\pm 2$ & $-2.4\pm 0.1$ & $27.1\pm 0.1$ \\
EAST-PDR & $108\pm 1$ & $-2.7\pm 0.1$ & $28.2\pm 0.2$ \\
ORI-P1 & $30\pm 1$ & $3.4\pm 0.2$ & $24.1\pm 0.4$ \\
ORI-P2 & $5\pm 1$ & $5.0\pm 0.7$ & $24.9\pm 1.5$ \\
ORI-P3 & $31\pm 1$ & $7.2\pm 0.2$ & $20.2\pm 0.4$ \\
ORI-P4 & -- & -- & -- \\
ORI-P5 & $17\pm 1$ & $-4.0\pm 0.4$ & $24.6\pm 0.8$ \\
ORI-P6 & -- & -- & -- \\ \hline
\end{tabular}
\end{table}

\begin{table}[ht]
\caption{H$n\delta$ RRL line fits.}
\begin{tabular}{lccc}
\hline\hline
source & $T_{\mathrm{mb}}$ & $v_{\mathrm{LSR}}$ & $\Delta v_{\mathrm{FWHM}}$ \\
 & $[10^{-3}\mathrm{K}]$ & $[\mathrm{km\,s^{-1}}]$ & $[\mathrm{km\,s^{-1}}]$ \\ \hline
BAR-INSIDE & $43\pm 1$ & $-4.5\pm 0.1$ & $25.2\pm 0.3$ \\
BNKL-PDR & $177\pm 1$ & $-2.2\pm 0.1$ & $25.5\pm 0.2$ \\
TRAP-PDR & $207\pm 1$ & $-2.5\pm 0.1$ & $26.6\pm 0.2$ \\
EAST-PDR & $61\pm 1$ & $-2.6\pm 0.1$ & $28.2\pm 0.3$ \\
ORI-P1 & $18\pm 1$ & $3.1\pm 0.3$ & $22.9\pm 0.5$ \\
ORI-P2 & $3\pm 1$ & $2.7\pm 1.0$ & $21.1\pm 2.2$ \\
ORI-P3 & $16\pm 1$ & $6.6\pm 0.4$ & $24.7\pm 0.8$ \\
ORI-P4 & -- & -- & -- \\
ORI-P5 & $9\pm 1$ & $-3.1\pm 0.6$ & $25.8\pm 1.3$ \\
ORI-P6 & -- & -- & -- \\ \hline
\end{tabular}
\end{table}

\begin{table}[ht]
\caption{H$n\epsilon$ RRL line fits.}
\begin{tabular}{lccc}
\hline\hline
source & $T_{\mathrm{mb}}$ & $v_{\mathrm{LSR}}$ & $\Delta v_{\mathrm{FWHM}}$ \\
 & $[10^{-3}\mathrm{K}]$ & $[\mathrm{km\,s^{-1}}]$ & $[\mathrm{km\,s^{-1}}]$ \\ \hline
BAR-INSIDE & $31\pm 1$ & $-5.5\pm 0.2$ & $26.0\pm 0.3$ \\
BNKL-PDR & $113\pm 1$ & $-3.2\pm 0.1$ & $25.5\pm 0.2$ \\
TRAP-PDR & $140\pm 1$ & $-3.0\pm 0.1$ & $26.8\pm 0.2$ \\
EAST-PDR & $42\pm 1$ & $-3.3\pm 0.1$ & $30.0\pm 0.2$ \\
ORI-P1 & $12\pm 1$ & $2.9\pm 0.3$ & $25.1\pm 0.6$ \\
ORI-P2 & $2\pm 1$ & $9.1\pm 2.1$ & $31.5\pm 5.1$ \\
ORI-P3 & $13\pm 1$ & $5.8\pm 0.4$ & $18.3\pm 0.8$ \\
ORI-P4 & -- & -- & -- \\
ORI-P5 & $5\pm 1$ & $-0.1\pm 1.0$ & $40.7\pm 2.2$ \\
ORI-P6 & -- & -- & -- \\ \hline
\end{tabular}
\end{table}

\clearpage

\section{Line fits to He$n\beta$, $\gamma$, $\delta$ and $\epsilon$ RRLs}
\label{App.HeRRL-fits}

Tables D.1, D.2, D.3, and D.4 give the fit parameters of the He$n\beta$, $\gamma$, $\delta$ and $\epsilon$ RRLs toward the ten targeted positions.

\begin{table}[ht]
\caption{He$n\beta$ RRL line fits.}
\begin{tabular}{lccc}
\hline\hline
source & $T_{\mathrm{mb}}$ & $v_{\mathrm{LSR}}$ & $\Delta v_{\mathrm{FWHM}}$ \\
 & $[10^{-3}\mathrm{K}]$ & $[\mathrm{km\,s^{-1}}]$ & $[\mathrm{km\,s^{-1}}]$ \\ \hline
BAR-INSIDE & $17\pm 1$ & $-4.4\pm 0.2$ & $17.5\pm 0.5$ \\
BNKL-PDR & $80\pm 1$ & $-3.2\pm 0.1$ & $17.0\pm 0.2$ \\
TRAP-PDR & $99\pm 1$ & $-3.8\pm 0.1$ & $18.7\pm 0.2$ \\
EAST-PDR & $25\pm 1$ & $-3.9\pm 0.2$ & $21.8\pm 0.5$ \\
ORI-P1 & $4\pm 1$ & $0.8\pm 0.7$ & $12.1\pm 1.6$ \\
ORI-P2 & -- & -- & -- \\
ORI-P3 & -- & -- & -- \\
ORI-P4 & -- & -- & -- \\
ORI-P5 & $4\pm 1$ & $-8.5\pm 1.6$ & $14.2\pm 3.2$ \\
ORI-P6 & -- & -- & -- \\ \hline
\end{tabular}
\end{table}

\begin{table}[ht]
\caption{He$n\gamma$ RRL line fits.}
\begin{tabular}{lccc}
\hline\hline
source & $T_{\mathrm{mb}}$ & $v_{\mathrm{LSR}}$ & $\Delta v_{\mathrm{FWHM}}$ \\
 & $[10^{-3}\mathrm{K}]$ & $[\mathrm{km\,s^{-1}}]$ & $[\mathrm{km\,s^{-1}}]$ \\ \hline
BAR-INSIDE & $8\pm 1$ & $-2.6\pm 0.6$ & $23.1\pm 1.4$ \\
BNKL-PDR & $38\pm 1$ & $-2.6\pm 0.2$ & $20.3\pm 0.5$ \\
TRAP-PDR & $47\pm 1$ & $-3.5\pm 0.2$ & $21.6\pm 0.4$ \\
EAST-PDR & $14\pm 1$ & $-4.1\pm 0.4$ & $22.5\pm 0.8$ \\
ORI-P1 & $4\pm 1$ & $7.0\pm 1.1$ & $18.3\pm 2.6$ \\
ORI-P2 & -- & -- & -- \\
ORI-P3 & -- & -- & -- \\
ORI-P4 & -- & -- & -- \\
ORI-P5 & -- & -- & -- \\
ORI-P6 & -- & -- & -- \\ \hline
\end{tabular}
\end{table}

\begin{table}[ht]
\caption{He$n\delta$ RRL line fits.}
\begin{tabular}{lccc}
\hline\hline
source & $T_{\mathrm{mb}}$ & $v_{\mathrm{LSR}}$ & $\Delta v_{\mathrm{FWHM}}$ \\
 & $[10^{-3}\mathrm{K}]$ & $[\mathrm{km\,s^{-1}}]$ & $[\mathrm{km\,s^{-1}}]$ \\ \hline
BAR-INSIDE & $4\pm 1$ & $-5.2\pm 0.9$ & $21.5\pm 2.1$ \\
BNKL-PDR & $20\pm 1$ & $-3.3\pm 0.3$ & $20.5\pm 0.8$ \\
TRAP-PDR & $25\pm 1$ & $-4.0\pm 0.4$ & $17.7\pm 0.9$ \\
EAST-PDR & $6\pm 1$ & $-2.5\pm 0.8$ & $26.8\pm 1.8$ \\
ORI-P1 & -- & -- & -- \\
ORI-P2 & -- & -- & -- \\
ORI-P3 & -- & -- & -- \\
ORI-P4 & -- & -- & -- \\
ORI-P5 & -- & -- & -- \\
ORI-P6 & -- & -- & -- \\ \hline
\end{tabular}
\end{table}

\begin{table}[ht]
\caption{He$n\epsilon$ RRL line fits.}
\begin{tabular}{lccc}
\hline\hline
source & $T_{\mathrm{mb}}$ & $v_{\mathrm{LSR}}$ & $\Delta v_{\mathrm{FWHM}}$ \\
 & $[10^{-3}\mathrm{K}]$ & $[\mathrm{km\,s^{-1}}]$ & $[\mathrm{km\,s^{-1}}]$ \\ \hline
BAR-INSIDE & $4\pm 1$ & $-5.0\pm 0.6$ & $13.1\pm 1.5$ \\
BNKL-PDR & $15\pm 1$ & $-2.6\pm 0.4$ & $16.4\pm 0.8$ \\
TRAP-PDR & $17\pm 1$ & $-3.1\pm 0.6$ & $17.5\pm 1.5$ \\
EAST-PDR & $4\pm 1$ & $-2.7\pm 0.9$ & $19.5\pm 2.1$ \\
ORI-P1 & -- & -- & -- \\
ORI-P2 & -- & -- & -- \\
ORI-P3 & -- & -- & -- \\
ORI-P4 & -- & -- & -- \\
ORI-P5 & -- & -- & -- \\
ORI-P6 & -- & -- & -- \\ \hline
\end{tabular}
\end{table}

\end{appendix}

\end{document}